\documentclass[apj]{emulateapj_aug2010}

\usepackage{graphicx}                 
\usepackage{xspace}                   
\usepackage{natbib}
\usepackage{apjfonts}
\newcommand{\acis}{{ACIS}\xspace}
\newcommand{\acisi}{{ACIS-I}\xspace}
\newcommand{\aciss}{{ACIS-S}\xspace}
\newcommand{\champ}{{ChaMP}\xspace}
\newcommand{\chandra}{{Chandra}\xspace}
\newcommand{\ciao}{{\it CIAO}\xspace}
\newcommand{\CDFS}{{CDFS}\xspace}
\newcommand{\CSC}{{Chandra Source Catalog}\xspace}
\newcommand{\CSCA}{{CSC}\xspace}
\newcommand{\CSCH}{{CSC}-}
\newcommand{\cxo}{{CXO}\xspace}
\newcommand{\hetg}{{HETG}\xspace}
\newcommand{\hetgs}{{HETGS}\xspace}
\newcommand{\hrc}{{HRC}\xspace}
\newcommand{\hrcs}{{HRC-S}\xspace}
\newcommand{\hrci}{{HRC-I}\xspace}
\newcommand{\hrma}{{HRMA}\xspace}
\newcommand{\isis}{{\it ISIS}\xspace}
\newcommand{\letg}{{LETG}\xspace}
\newcommand{\lsf}{{LSF}\xspace}
\newcommand{\marx}{{\tt MARX}\xspace}
\newcommand{\marxh}{{\tt MARX}-}

\newcommand{\psf}{{PSF}\xspace}
\newcommand{\psfs}{{PSFs}\xspace}
\newcommand{\saotrace}{{\tt SAOTrace}\xspace}
\newcommand{\sdss}{{\tt SDSS}\xspace}
\newcommand{\sherpa}{{\it Sherpa}\xspace}
\newcommand{\slang}{{\tt S-lang}\xspace}
\newcommand{\wavdetect}{{\tt wavdetect}\xspace}
\newcommand{\xmm}{{XMM-Newton}\xspace}

\newcommand{\uband}{{\it u}\xspace}
\newcommand{\sband}{{\it s}\xspace}
\newcommand{\mband}{{\it m}\xspace}
\newcommand{\hband}{{\it h}\xspace}
\newcommand{\bband}{{\it b}\xspace}

\newcommand{\sao}{\altaffilmark{1}}
\newcommand{\mki}{\altaffilmark{2}}
\newcommand{\ngis}{\altaffilmark{3}}

\slugcomment{To Appear in The Astrophysical Journal Supplement Series}

\shorttitle{Statistical Characterization of the Chandra Source Catalog}
\shortauthors{Primini et al.}

\begin{document}

\title{Statistical Characterization of the Chandra Source Catalog}

\author{Francis~A.~Primini,\sao\ 
  John~C.~Houck,\mki\ 
  John~E.~Davis,\mki\ 
  Michael~A.~Nowak,\mki\
  Ian~N.~Evans,\sao\ 
  Kenny~J.~Glotfelty,\sao\
  Craig~S.~Anderson,\sao\ Nina~R.~Bonaventura,\sao\ Judy~C.~Chen,\sao\
  Stephen~M.~Doe,\sao\ Janet~D.~Evans,\sao\
  Giuseppina~Fabbiano,\sao\ Elizabeth~C.~Galle,\sao\ Danny~G.~Gibbs~II,\sao\
  John~D.~Grier,\sao\ Roger~M.~Hain,\sao\ Diane~M.~Hall,\ngis\ Peter~N.~Harbo,\sao\
  Xiangqun~(Helen)~He,\sao\ Margarita~Karovska,\sao\
  Vinay~L.~Kashyap,\sao\ Jennifer~Lauer,\sao\ Michael~L.~McCollough,\sao\
  Jonathan~C.~McDowell,\sao\ Joseph~B.~Miller,\sao\ Arik~W.~Mitschang,\sao\
  Douglas~L.~Morgan,\sao\ Amy~E.~Mossman,\sao\ Joy~S.~Nichols,\sao\
  David~A.~Plummer,\sao\ Brian~L.~Refsdal,\sao\
  Arnold~H.~Rots,\sao\ Aneta~Siemiginowska,\sao\ Beth~A.~Sundheim,\sao\
  Michael~S.~Tibbetts,\sao\ David~W.~Van~Stone,\sao\ Sherry~L.~Winkelman,\sao\
  and Panagoula~Zografou\sao}

\email{fap@head.cfa.harvard.edu}

\altaffiltext{1}{Smithsonian Astrophysical Observatory, 60 Garden Street,
  Cambridge, MA 02138}

\altaffiltext{2}{MIT Kavli Institute for Astrophysics and Space Research, 77
  Massachusetts Avenue, Cambridge, MA 02139}

\altaffiltext{3}{Northrop Grumman, 60 Garden Street, Cambridge, MA 02138}

\addtocounter{footnote}{3}

\begin{abstract}
The first release of the \CSC (\CSCA) contains
$\sim$95,000  X-ray sources in a total area of ~0.75\% of the entire
sky, using data from $\sim$3,900 separate ACIS
observations of a multitude of different types of X-ray sources. In
order to maximize the scientific benefit of such a large, heterogeneous
data-set, careful characterization of the statistical properties of the
catalog, i.e., completeness, sensitivity, false source rate, and accuracy of
source properties, is required. Characterization efforts of other, large
\chandra catalogs, such as the ChaMP Point Source Catalog (Kim et al. 2007) or
the 2 Mega-second Deep Field Surveys (Alexander et al. 2003), while
informative, cannot serve this purpose, since the \CSCA analysis procedures are
significantly different and the range of allowable data is much less
restrictive. We describe here the characterization process for the
\CSCA. This process includes both a comparison of real \CSCA results with
those of other, deeper \chandra catalogs of the same targets and
extensive simulations of blank-sky and point source populations. 

\end{abstract}


\section{Introduction}
\label{sec:intro}
The \chandra X-ray Observatory \citep[\cxo;][]{weisskopf:02a} has
observed an extremely diverse range of X-ray emitting astrophysical
sources, ranging from spatially extended diffuse sources such as X-ray
clusters to bright point-like sources such as Galactic black hole
binaries.  Even within the category of X-ray point sources, \chandra
has observed the widest range of source X-ray fluxes of any previously
flown X-ray satellite -- spanning literally more than 10 orders of magnitude from
the $\approx 10^{-18}\,{\rm ergs~cm^{-2}~s^{-1}}$ flux limits
of the \chandra deep fields
\citep{brandt:01a,giacconi:02a,alexander:03a,luo:08a} to the $\approx
10^{-7}\,{\rm ergs~cm^{-2}~s^{-1}}$ of Sco~X-1.  These
observations have occurred in a variety of instrumental arrangements,
determined by whether or not either of the two gratings configurations
(the High Energy Transmission Grating, \hetg,
\citealt{canizares2005}, and the Low Energy Transmission
Grating, \letg, \citealt{brinkman:00a}) was inserted into the optical
path, and by which set of detectors (the Advanced CCD Imaging
Spectrometer, \aciss and \acisi, CCDs, \citealt{garmire:03a}, or the
High Resolution Camera, \hrcs and \hrci, \citealt{murray:00a}) were
placed in the focal plane.  Although nearly all possible
instrument/detector configurations have been used at some point over
the mission lifetime, the majority of \chandra observations have been
conducted with the \acis CCDs inserted into the focal plane and
without the use of any gratings.  For this reason, the first release
of the \CSC \citep[\CSCA;][]{evans09} consists
solely of such observations.

The \CSC follows in the long tradition of using
X-ray satellite observations to create surveys of detected sources,
encompassing both those sources that were the targets of the original
observing proposals and serendipitously discovered sources.  Such
past and present surveys include the Einstein survey \citep[over 800
sources;][]{gioia:90a}, the ROSAT surveys of bright and faint sources
\citep[$\approx 20,000$ sources;][]{voges:99a,voges:00a} and its
counterpart WGACAT \citep[$\approx 45,000$ sources;][]{white:94a}, the
ASCA Medium Sensitivity Survey \citep[$\approx 1,200$
sources;][]{ueda:05a}, and the recent \xmm survey \citep[2XMM, with
$\approx 247,000$ detections from 3,491 observations;][]{watson:2009a}.
What makes the \CSCA unique among these surveys is the unsurpassed (in
the X-ray) spatial resolution of \chandra, which is $\approx 0.5''$
for on-axis sources.  It is anticipated that over a 20 year lifetime,
\chandra will conduct over 20,000 separate \acis and \hrc observations
which will yield over 250,000 significantly detected X-ray sources.
These sources already include a diverse set of objects spanning local
sources within our own solar system to distant clusters of galaxies.
The ultimate goal of the \CSCA is to represent the full diversity of
\chandra observed sources, and to include both point-like and extended
sources.

The initial release of the \CSC limits itself in several ways
\citep{evans09}.  As discussed above, it only considers \acis
observations without any inserted gratings.  (A subset of no-gratings
\hrc observations was included as of release v1.1.  Sources
detected from the zeroth-order images of gratings observations
eventually will be included.)  Furthermore, source detections are
derived from single observations, as opposed to merged observations
from the same field. The \CSC does define ``Master Sources'' as distinct
X-ray sources, which may be observed in more than one
observation. However, Master Source properties such as position and
flux are derived from appropriate combination of the corresponding
properties from spatially coincident sources separately detected in individual
observations. Other Master Source properties, such as 
inter-observation variability, are derived by collating and comparing
properties from contributing sources detected in individual
observations.
Future releases of the \CSCA will include
properties derived from data combined prior to source detection.  
The initial release of
the \CSCA also limits sources 
to (physical and/or instrumental) source extents
$<30^{{\prime}{\prime}}$.  These restrictions of the initially released
\CSCA can 
be compared to those found in a number of other released catalogs
covering \chandra observations.

\begin{figure}[t]
\epsscale{1.0}
\plotone{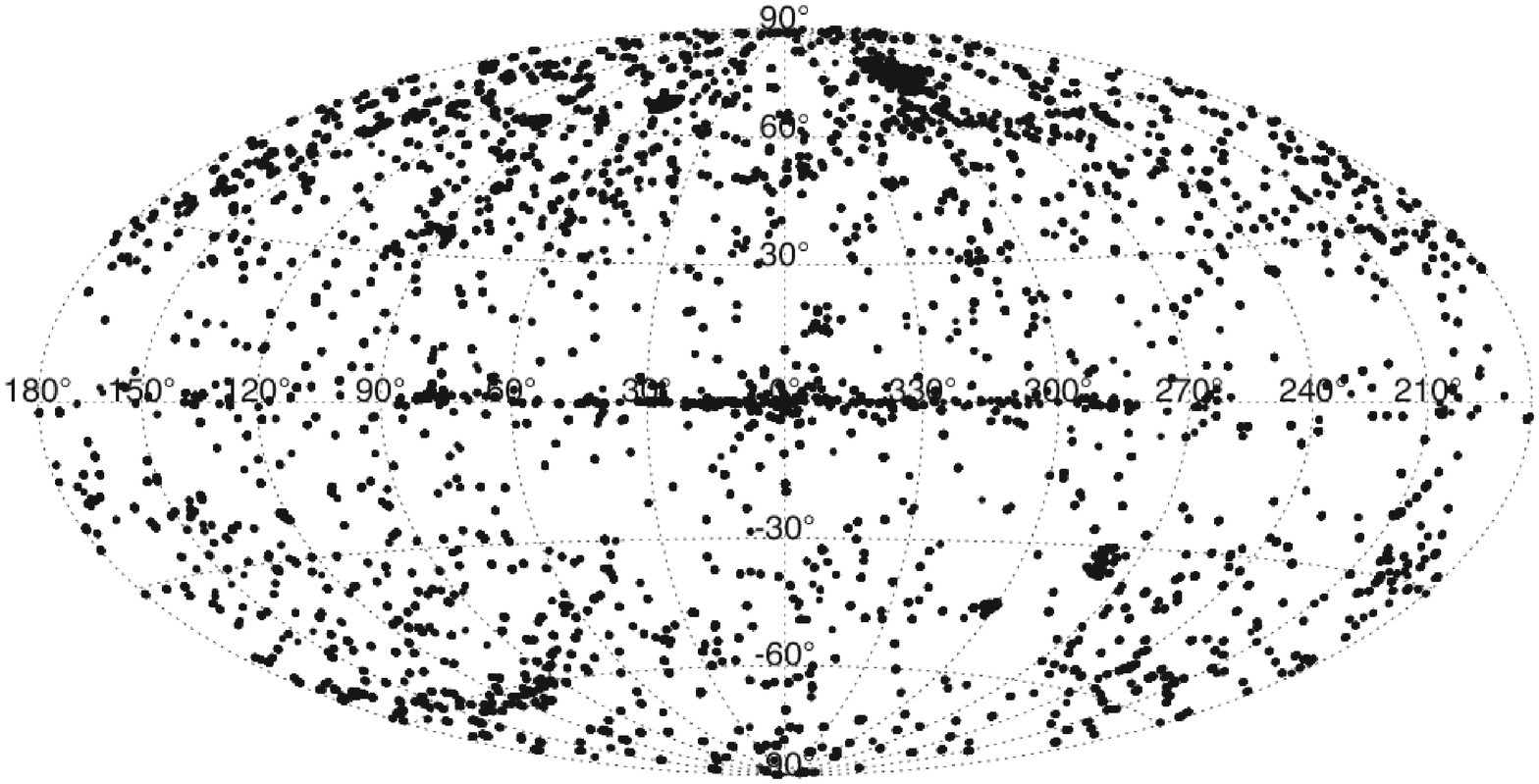}
\caption{\label{skydist} Distribution of \CSCA sources on the sky, in galactic coordinates.}
\vspace*{-0.2cm}
\end{figure}

Numerous such \chandra catalogs already exist.  Prominent among these
are those that deal specifically with a well-defined set of fields of
view.  Examples of such targeted catalogs include the \chandra Deep
Fields North \citep[][now containing over 500 sources]{brandt:01a,
alexander:03a} and South \citep[][with nearly 600 sources when
including the flanking fields]{giacconi:02a,luo:08a}, and the \chandra
Ultra-deep Orion Project \citep[COUP;][with over 1,600
sources]{getman:05a}.  Although these catalogs currently consider
source detections and properties from merged observations, they are
far more restricted in terms of fields of view than the \CSC.  More
general catalogs include the \chandra Multi-wavelength Project
\citep[ChaMP][with nearly 1,000 sources]{kim:04a,kim:04b}; however, it
too does not cover the full scope of fields of view as is covered by
the \CSCA.  Furthermore, these existing catalogs are all driven by the
specific scientific goals of the projects that produced them.  They do
not share commonly defined source properties or analysis procedures.

The \CSC differs from these catalogs in several
important respects.  All data for all observations of a given \chandra
detector are processed in a uniform manner with a uniformly defined
set of source properties.  The \CSCA also aims to be the most inclusive
of any \chandra catalog.  With few exceptions, all data from all
active \acis CCDs were searched for sources \citep[see][for a
description of the criteria by which whole observations, or individual
CCD detectors within an observation, were excluded]{evans09}. The
intended audience for the \CSCA is not limited to X-ray astronomers nor
to any particular sub-field of study within astronomy; it is intended
as a general resource for all astronomers working at any wavelength.

\begin{figure}
\epsscale{0.9}
\plotone{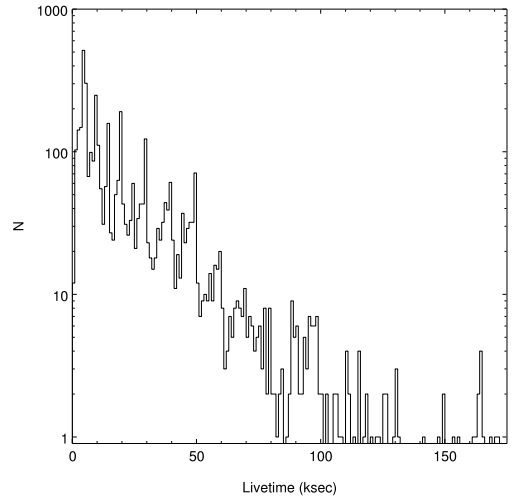}
\caption{\label{expdist} Distribution of livetimes for individual
  observations included in the \CSCA. The median livetime is $\sim14$\,ksec.}
\end{figure}

The \CSC is the product of a series of complex data
processing pipelines.  In order to take full advantage of the \CSCA
products, users must understand the capabilities of both the \chandra
observatory and the \CSCA analysis system.  The \cxo telescope and
detectors have been documented extensively in numerous publications
\citep{weisskopf:02a,garmire:03a,canizares2005,murray:00a,brinkman:00a}.
The \CSCA analysis system and first release products have been
described by \citet{evans09}.  In this work, we describe in more
detail the procedures used to characterize the capabilities of that
analysis system, and the results of this characterization.  The
statistical characterization of the catalog source properties is
accomplished primarily through the use of simulated datasets. These
simulations include both empty fields (blank-sky) and simulated
sources.  For the most part, these simulated datasets are processed by
the catalog pipelines in the exact same fashion as real datasets.  We
present here a summary of those results.

We begin with a summary of the overall properties of the source
catalog.  (See also \citealt{evans09} for further descriptions.)  We
then describe the sky coverage of the first release catalog and
discuss how limiting sensitivities within these fields of view are
determined.  In Section \ref{algorithms} we describe the algorithms used to create
and assess our simulations.  Results of these simulations are then
presented in Section \ref{sec:source_detection} for source detection, 
including the false source rate and the detection efficiency.
Relative and absolute astrometry are discussed in Section
\ref{astrometry}.  Photometry and source colors  (hardness
ratios) are discussed in Sections \ref{sec:photometry} and
\ref{hardness}, respectively. Results of spectral fits for bright sources are
described in Section \ref{specfits}. Estimates of source extents, and errors
on these extents, are presented in Section \ref{extent}.  Section
\ref{variability}  deals with  intra-observation variability within
the catalog.  We end with a summary of the current characterization
efforts, and a discussion of plans for characterization efforts for
future releases of the \CSC.

\section{Overall Properties}
\label{overall}

The first release of the \CSC contains 135,914 individual source entries from
3,912 separate \acis observations available in the Chandra Public
Archive as of to Dec. 31, 2008. Because many Chandra targets were
observed more than once, these individual source entries correspond to
94,676 unique ``master sources''. These include both target and
serendipitous sources. The distribution of sources on the sky, in galactic
coordinates, is shown in Fig.~\ref{skydist}.
Individual observation exposure times
ranged from $\sim0.5-175$ ksec, with a median of $\sim14$
ksec. The observation epochs range from Feb. 3, 2000 (\chandra
MJD 51,577.5) to Dec. 31, 2008 (MJD 54,831.2), with a median 
of Jul. 1 2004 (MJD 53,187.3).

As can be seen in Fig.~\ref{expdist}, the exposure time
distribution exhibits strong peaks at multiples of 5 ksec., reflecting
the inclination of {\it Chandra} Guest Observers to round required
exposure times to these values when requesting observations. This may
seem a trivial point, but it emphasizes an overwhelming dependence of the
\CSCA on a heterogeneous mix of observations with different scientific
objectives and requirements.

\begin{figure}[!b]
\epsscale{1.0}
\plotone{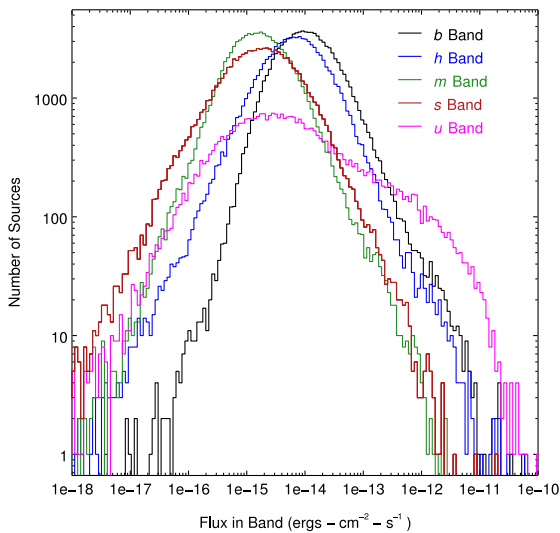}
\caption{\label{fluxdist} Distribution of \CSCA fluxes in the broad
  (black), hard (blue), medium (green), soft (red), and ultrasoft
  (magenta) bands, obtained from the catalog master source table
  {\tt flux\_aper} columns.}
\end{figure}

\CSCA fluxes range from below $\sim10^{-18}$\,erg~cm$^{-2}$~sec$^{-1}$
to $\sim10^{-10}$\,erg~cm$^{-2}$~sec$^{-1}$. Most \CSCA sources have 
fluxes, as shown in Fig.~\ref{fluxdist}, of
$\sim10^{-15}-10^{-13}$erg~cm$^{-2}$~sec$^{-1}$ (\bband band, or
0.5-7.0 keV). We note that the \uband band number-flux distribution 
is much flatter that that observed in the other bands. Since
photoelectric absorption is severe in the \uband band, it is tempting
to attribute the flatter distribution to a population of relatively
near-by sources. However, we caution against assigning
any real astrophysical meaning to the distributions in
Fig.~\ref{fluxdist} because they represent a hetergeneous mixture of sources
of all types included in the \CSCA. The figure is intended
merely to ilustrate the range of fluxes in the catalog. 
Minimum net source counts range from  $\sim10$ for
on-axis sources to $\sim15-30$ for sources with off-axis angle
$\theta\sim10^{\prime}$, depending on exposure.  

\begin{figure}
\epsscale{2.25}
\plottwo{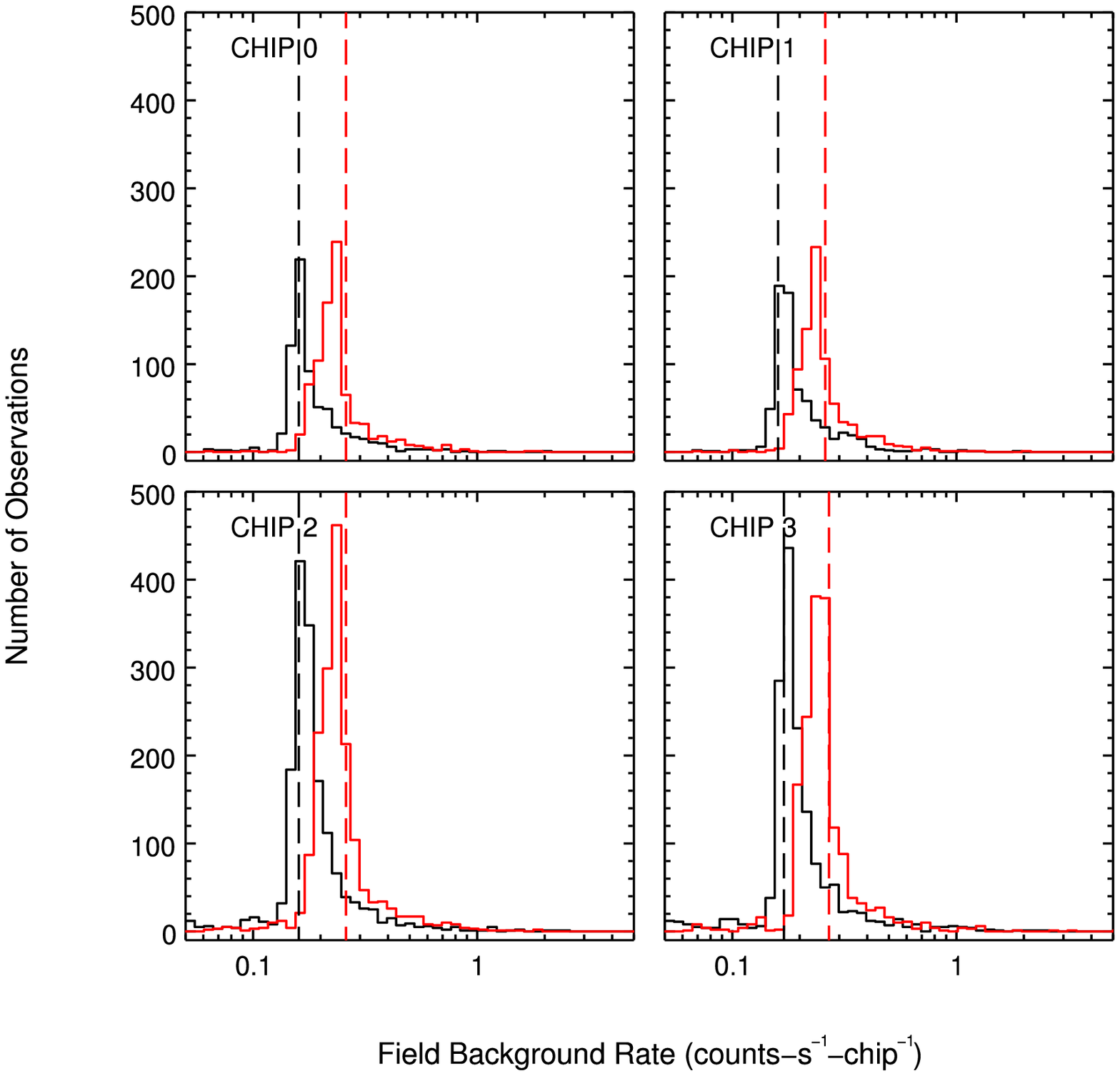}{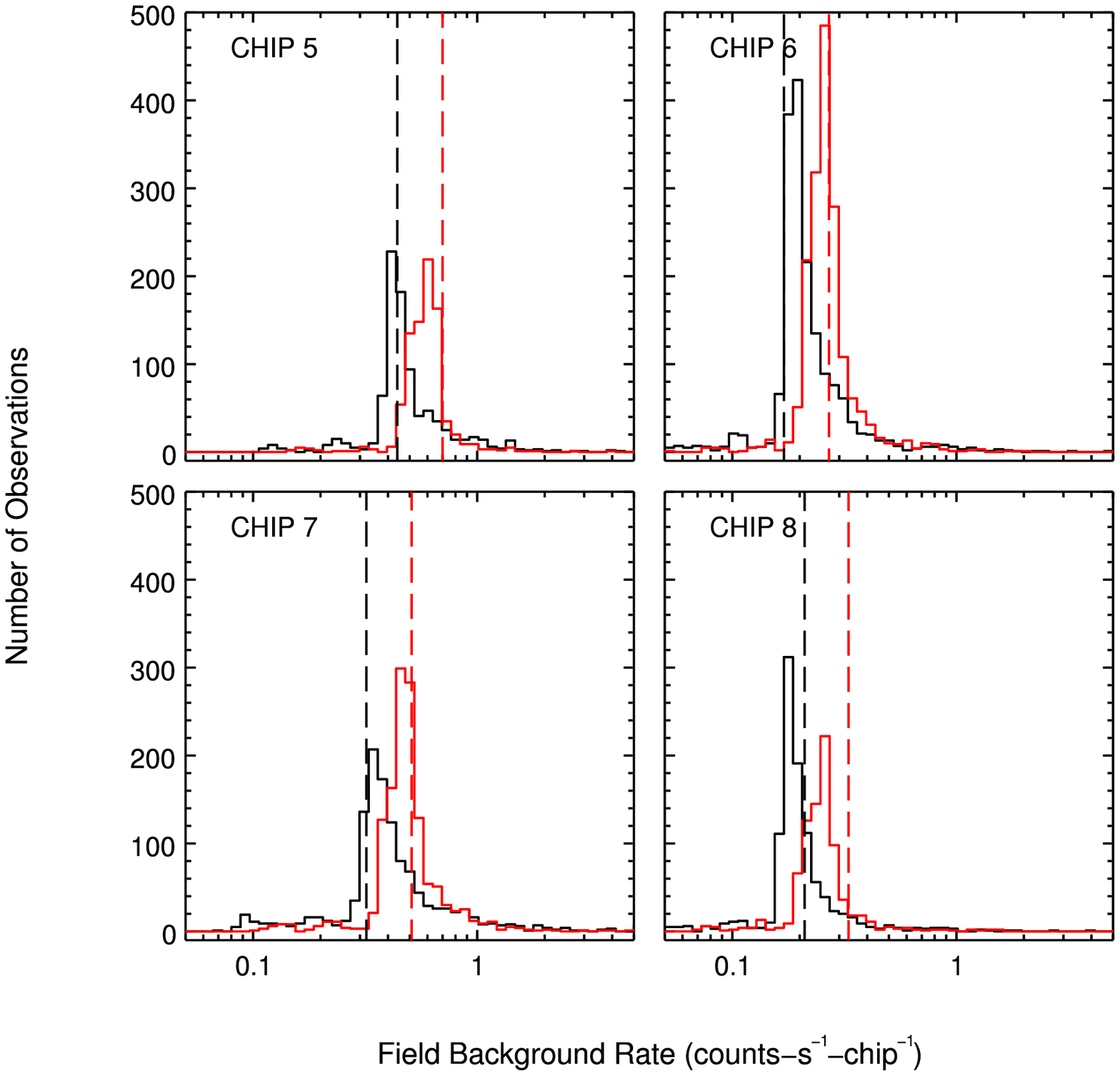}
\caption{\label{brate} Distribution of field background rates for
  commonly used \acis imaging chips. Black (left) histograms refer to
  observations made prior to the median \CSCA epoch of July 1, 2004, and
  red (right) histograms to observations made after that date. Black and red
  vertical lines indicate nominal rates from v. 7 and v. 11 of the
  \chandra Proposers' Observatory Guides, respectively.} 
\end{figure}

\CSCA background rates are in general comparable to those reported in
the \chandra Proposers' Observatory Guide, and reflect the overall
changes in background rate during the lifetime of the mission.
This is illustrated in
Fig.~\ref{brate}, in which we display histograms of background rates
for chips 0--3 and 5--8, using observations taken before (black) and
after (red) the median epoch. 
The background rates were determined by summing all \bband band events in each
chip, subtracting \bband band net counts for \CSCA sources which fell on the
chip, and dividing by the chip livetime. Nominal rates from v. 7 (black)
and v. 11 (red) of the Observatory Guides are also shown.

\begin{figure*}[!t]
\epsscale{1.0}
\plotone{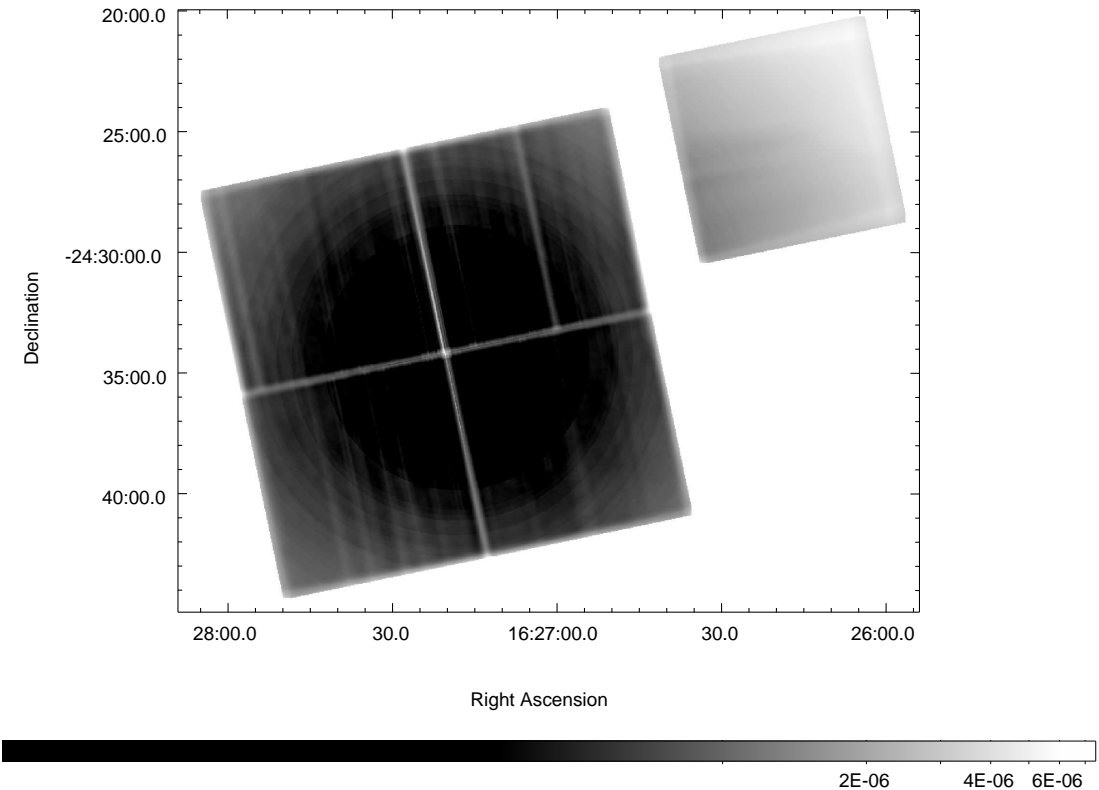}
\caption{\label{lsmap} \bband Band limiting sensitivity map for OBSID
  635. Each pixel value represents the minimum point source photon
  flux needed to yield a flux significance at the catalog inclusion
  limit, at that pixel location. Color bar units are photons-cm$^{-2}$-s$^{-1}$.}
\end{figure*}

\vskip 20pt
\section{Limiting Sensitivity and Sky Coverage}
\label{limiting}

A limiting sensitivity map is computed for each Observation Id (OBSID)
that contributes to the \CSC, in each of the 5 science energy
bands. The maps are derived from the \CSCA model background maps for
the OBSID. Statistical noise appropriate to the observation is
introduced by randomly sampling from Poisson distributions whose means
are equal to the model background values in each map pixel. Each
sensitivity map pixel represents the minimum point source photon flux
needed to yield a flux significance greater than or equal to the
catalog inclusion limit ($3\sigma$) at that location, when background
is obtained from a region in the randomized background map appropriate
to background apertures at that pixel location. The algorithm is
described in detail in \citet{evans09}. An example sensitivity map is
shown in Fig.~\ref{lsmap}.

\begin{figure}
\epsscale{1.2}
\plotone{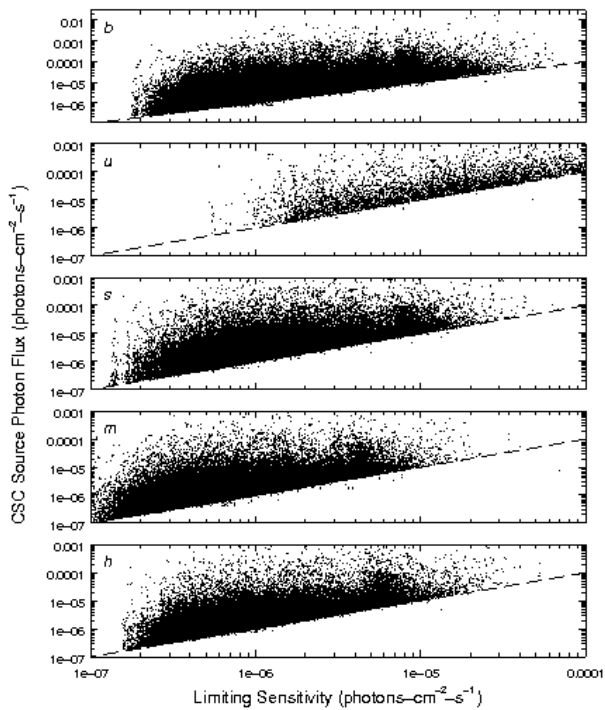}
\caption{\label{flux.vs.ls} Comparison of photon fluxes and limiting
  sensitivities in each band for sources with flux significances $\geq$3.0 in
  that band. Fluxes for reported sources should all fall on or above
  the dashed lines, for which flux and sensitivity are equal. 
}
\end{figure}

Because the limiting sensitivity maps are derived from model
background maps, and not directly from the event data used to compute
individual photon fluxes, it is important to demonstrate that they are
consistent with the fluxes of sources included in the \CSC.
We compare the photon fluxes of sources reported in
individual OBSIDs in the \CSCA to the values of those OBSIDs' sensitivity maps at
the corresponding source locations. Photon fluxes for detected sources should
all be greater than or equal to the corresponding limiting sensitivity
values. The results for all bands are shown in Fig.~\ref{flux.vs.ls}. To
simplify our procedure for matching source fluxes to limiting
sensitivity, we have limited our sample of OBSIDs to those which
included only a single Observation Interval (OBI). We  find 120,230
sources with \bband band flux significances $\geq$ 3.0 in our sample,
of which 464 ($\sim0.4\%)$ have photon fluxes less than the expected
limiting sensitivity value. The corresponding numbers for the \uband,
\sband, \mband, and \hband bands are  112/4,552 ($\sim2.5\%$),
538/50,052 ($\sim1.1\%$), 595/57,480 ($\sim1\%$), and 252/49,360
($\sim0.5\%$), respectively.   

\begin{figure}
\epsscale{1.2}
\plotone{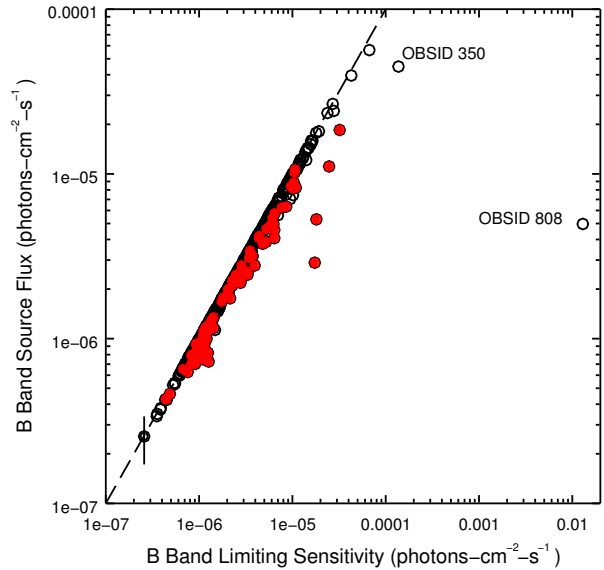}
\caption{\label{badsens} 
\bband band photon fluxes and sensitivities for sources for which the photon flux
is less than the corresponding limiting sensitivity. The dashed line
represents the threshold at which fluxes and sensitivities are
equal. One-sigma error bars are indicated for the faintest source, and are
typical of the errors for all the sources. Red (halftone in paper
edition) filled circles denote those sources
whose fluxes are in error due to a bug in computing source exposure
(see the discussion in Section \ref{limiting}). Labeled sources were
observed in OBSIDs with anomalous chip configurations (see text). 
}
\end{figure}

Although these percentages are small, it is worth examining the  sources
contributing to them in more detail.  In Fig.~\ref{badsens}, we show
the 464 sources whose \bband band flux is less than the 
corresponding sensitivity. Of these, all but 21 are consistent with
the threshold (dashed line) at which fluxes and sensitivities are
equal, when flux errors are
taken into account.
Seventeen of these twenty-one
are members of a set of CSC sources for which
incorrect exposure times were used in calculating fluxes. The entire
set includes 93 of the 464 sources in Fig. ~\ref{badsens}, shown in
red, and $\sim2,200$ sources in $\sim160$ OBSIDs in the entire CSC. For these
sources, exposure times for chips other than the source chip
were used, 
leading to errors of $\sim 3\% $ or more in photon fluxes. Properties for
these sources have been revised in Release 1.1 of the catalog. Two
of the twenty-one  are inconsistent with the sensitivity
limit when 68\% confidence bounds on flux are considered, but are
consistent at the 90\% level. 
For the remaining two sources, labeled by OBSID in Fig.~\ref{badsens}, we find anomalous
chip configurations. For 
OBSID 350, the target chip (chip 7) contained significant extended emission
and was dropped from analysis; the source in question was located at the
interface of chips 6 and 7. For OBSID 808, a subarray was used and the entire
chip active area contained extended emission. In such cases, the background
map algorithm fails and hence limiting sensitivity results are
suspect. Similar results apply to the small percentages of failed sources in
the other bands. We conclude that apart from these exceptional cases, the
limiting sensitivities cited in the catalog are consistent with the actual
distribution of measured source fluxes.

\begin{figure*}
\epsscale{1.0}
\plotone{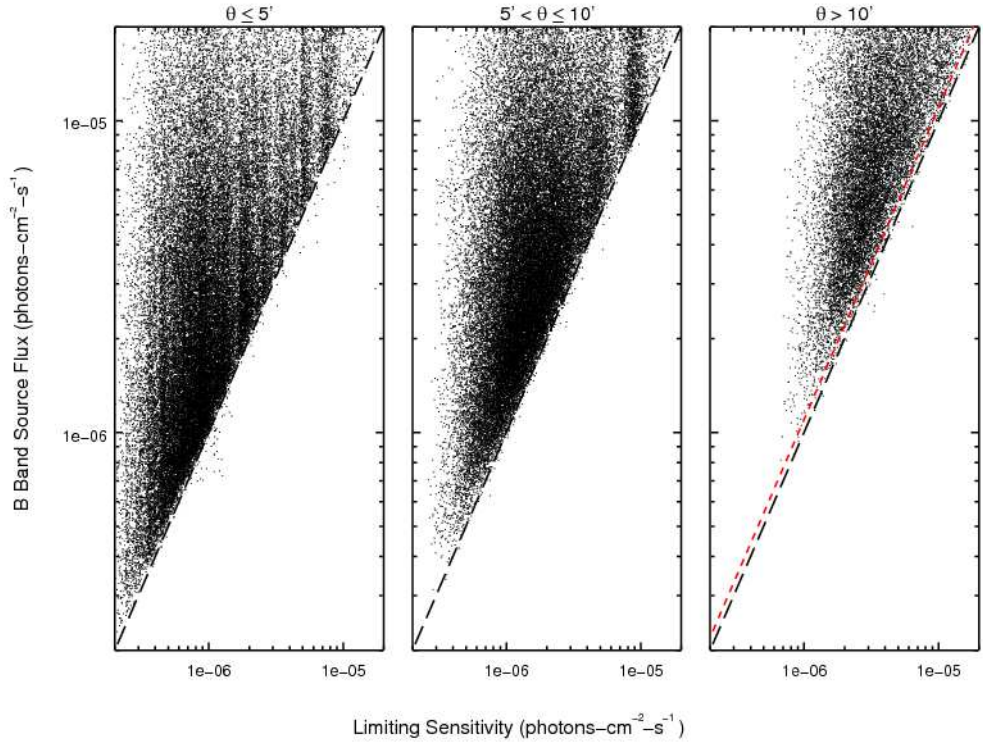}
\caption{\label{lstheta} Comparison of \bband band fluxes and sensitivities for
  sources in different  ranges of off-axis angle. In each panel, the black (longdash)
  line  represents the threshold at which fluxes and sensitivities are
  equal. For $\theta \geq 10^{\prime}$, the distribution of fluxes does not
  extend to this threshold, as indicated by the red (shortdash) line $flux = 1.1 \times
  sensitivity$. This indicates that either fluxes are over-estimated by $\sim
  10\%$, or sensitivities are underestimated by a similar amount.
}
\end{figure*}

Finally, we examine the behavior of limiting sensitivities with off-axis angle
$\theta$. In Fig.~\ref{lstheta} we reproduce the top panel (\bband band) of
Fig.~\ref{flux.vs.ls}, but now displaying different ranges of $\theta$
separately. We find that for $\theta \leq 10^{\prime}$, the distribution of
photon fluxes is consistent with the $flux=sensitivity$ threshold. However,
for $\theta > 10^{\prime}$, the flux distribution does not extend down to the
threshold (Fig.~\ref{lstheta}, right panel). The differences amount to
$\sim 10\%$, as indicated by the dashed red line at
$flux=1.1~\times~sensitivity$, and may be interpreted as either an
overestimate of fluxes or underestimate of sensitivities by this amount. Since
there is some evidence from simulations for a slight overestimate of fluxes in
this range of $\theta$, we consider the former possibility to be the
most likely case here.

\begin{figure*}
\epsscale{1.0}
\plotone{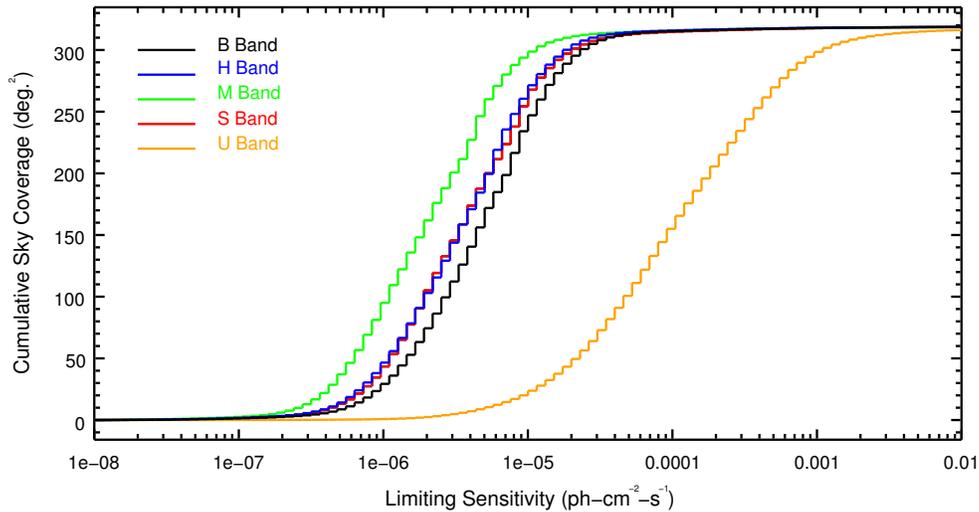}
\caption{\label{skycover} \CSCA Sky Coverage for each science band. The value at
  each flux {\it F}  represents the total \CSCA area sensitive to point sources
  with fluxes $\geq F$.}
\end{figure*}

The sky coverage represents the total area in the \CSCA sensitive to
point sources greater than a given flux, as a function of flux. We
estimate sky coverage by assigning all non-zero limiting sensitivity
map values to all-sky pixels, using the HEALPix projection
\citep{gorski:2005a}, keeping only the most sensitive (i.e., 
lowest) value in each all-sky pixel. To reduce computational load and
size of the projections (i.e., the number of HEALPix pixels), we
rebinned the sensitivity maps to block 64
($\sim31.5^{{\prime}{\prime}}\times\sim31.5^{{\prime}{\prime}}$), used
$\sim25.8^{{\prime}{\prime}}$ HEALPix pixels, and assigned rebinned
sensitivity map pixels to the nearest HEALPix pixel, ignoring
spillover. The resulting sky coverage function for the all bands is shown
in Fig.~\ref{skycover}. Total \bband band sky coverage is $\sim320$
deg.$^2$.

\begin{figure*}
\epsscale{1.0}
\plotone{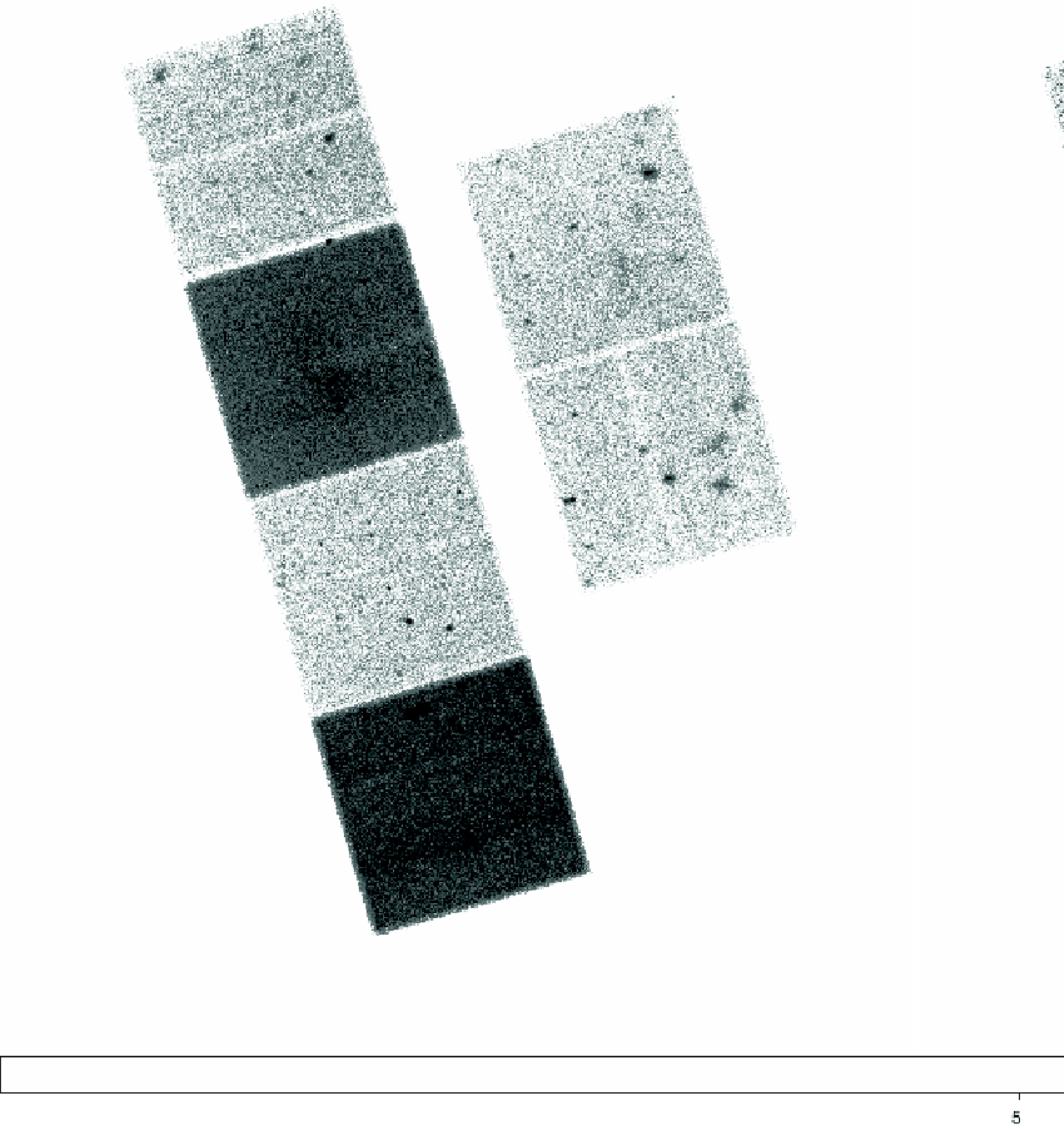}
\caption{\label{fig:4613sim}Images of seed event list (left) and corresponding
empty-field simulated event list (right) for 118 ksec \aciss observation 4613.}
\end{figure*}

\section{Simulation Algorithms}
\label{algorithms}

We use simulations of empty fields to estimate the number of false 
source detections in the catalog as a function of exposure, chip location,
and detector configuration. We then inject simulated sources into these empty
fields to investigate source properties such as position, flux, and extent.

In all cases except for variability studies, we start with actual observations
that have been processed through the \CSC calibration pipeline. We selected
four ``seed'' observations that span a wide range of exposures, for both
\acisi and \aciss aimpoints. The set of seed observations is shown in
Table \ref{table:seed-observations}.  We then replace the actual event
lists with simulated lists that share the same metadata, such as exposure,
attitude, and detector configuration. These simulated event lists are then
processed through the \CSCA source detection and properties pipelines.

We felt it necessary to adopt this ``cuckoo's egg'' approach because of the
complexity of the \CSCA software pipelines, in which multiple inputs
to multiple
programs could affect source detection or properties. We therefore treat the
entire source detection and properties pipeline as a ``black box''
experimental apparatus, to be calibrated by 
studying its response to various artificial inputs. The exception to this
approach is the characterization of source variability. In this
case, it is simpler to simulate the variability analysis outside of
the pipeline (see below).

\begin{deluxetable}{cccc}
\tablecaption{Simulation Seed Observations\label{table:seed-observations}}
\tablehead{\colhead{OBSID} & \colhead{Aimpoint} & \colhead{Exposure (ksec)} &  \colhead{Chip Configuration}}
\startdata
  379        &	ACIS-I	&	9	                 &	0,1,2,3,6,7\\
  1934	&	ACIS-I	&	29	                 &	0,1,2,3,6,7\\
  4497	&	ACIS-I	&	68	                 &	0,1,2,3,6,7\\
  927	&	ACIS-I	&	125	                 &	0,1,2,3,6,7\\
  5337	&	ACIS-S	&	10	                 &	2,3,5,6,7,8\\
  4404	&	ACIS-S	&	30	                 &	2,3,5,6,7,8\\
  7078	&	ACIS-S	&	51	                 &	2,3,5,6,7,8\\
  4613	&	ACIS-S	&	118	                 &	2,3,5,6,7,8\\
\enddata

\tablenotetext{}{
  Seed observations for empty-field and point source simulations. Outputs from
  the \CSCA Calibration Pipeline for these observations were used in the
  simulation tests, with the event list replaced by simulated event lists that
  matched the metadata of the seed observations.
}
\end{deluxetable}

\subsection{Empty Field Simulations}
\label{sec:emptyfieldsims}
To simulate event lists containing background only, we start with the \acis
blank-sky data in the \chandra calibration data base. For each seed event
list, we determine the appropriate blank-sky data sets for the active chips,
using the \ciao tool {\tt acis\_bkgrnd\_lookup}.  The \chandra
blank-sky datasets were adequate for all chips except chip 4 (S0),
chip 8 (S4), and chip 9 (S5). For chip 8 we were unable to match the
horizontal streaks in \CSCA data due to the different destreaking
processing applied to the blank-sky datasets and the \CSCA event
lists. For this chip, we constructed our own blank-sky dataset from
\CSCA event lists of several long exposures that contained no bright
sources in chip 8. Chip 4 and chip 9 have only one blank sky dataset
at a focal plane temperature of -110 C.  Given that they are very far
off axis, and are not typically used in \aciss \emph{imaging}
observations, we have not included blank sky simulations for these chips. 
We expect that their characterization should be similar to other
front-illuminated chips at large off-axis angles.

We estimate the expected number of background events for each chip
from the chip nominal field background rate and observation on-time,
and compute the ratio of this quantity to the number of events in the
corresponding blank-sky dataset. For each chip column, we then
determine the number of events by randomly sampling from a Poisson
distribution whose mean is the number of events in that column in the
blank-sky dataset, scaled by the event ratio. Row positions for these 
events are determined by randomly sampling from a normalized cumulative
distribution derived from the row positions of events in the corresponding
column of the blank-sky dataset.

\begin{figure*}
\epsscale{1.0}
\plotone{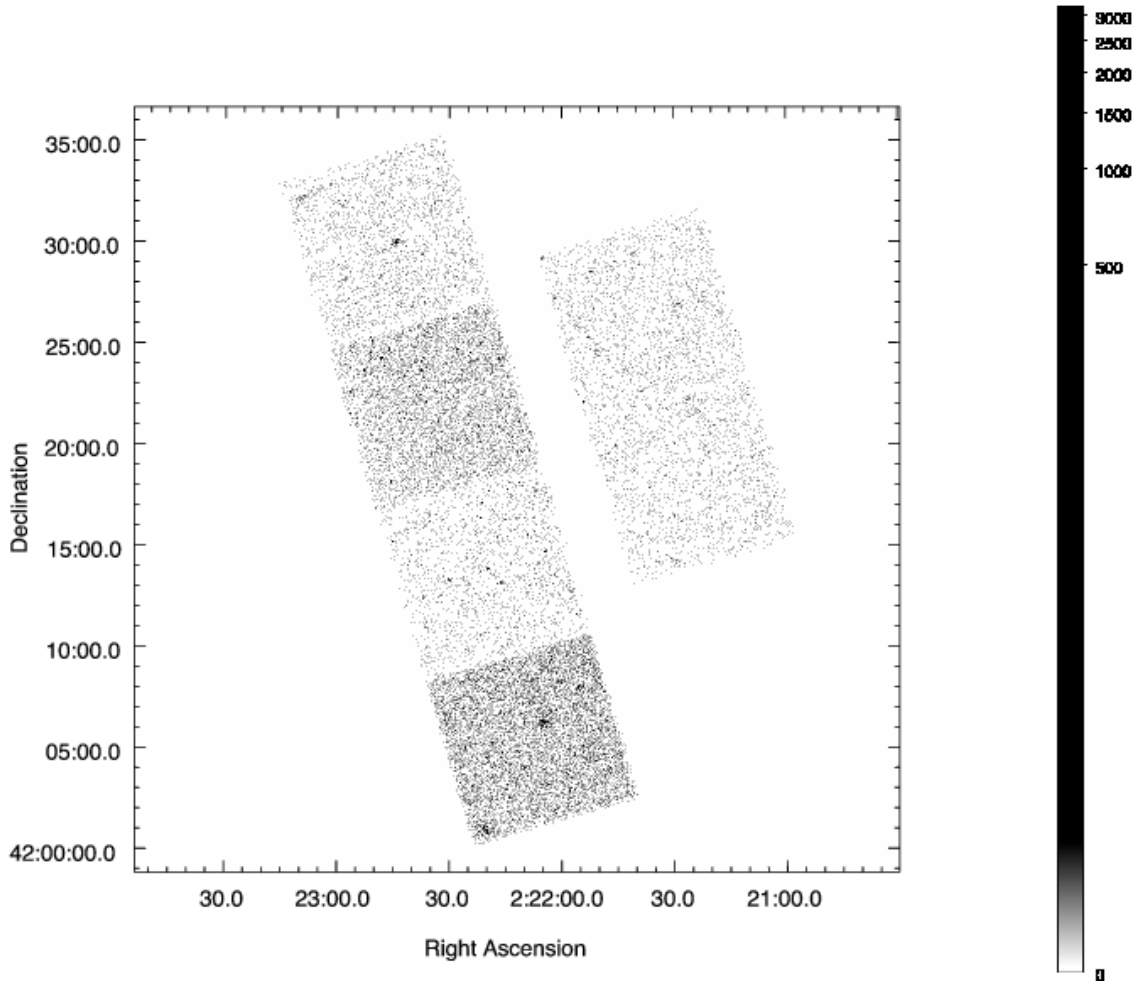}
\caption{\label{fig:4613simsrc}An
empty-field event list for \aciss observation 4613 with simulated sources inserted.}
\end{figure*}

We simulate numbers of events and their positions in this fashion in order to
preserve the column-to-column variations due to detector defects such as bad
columns, and variations in quantum efficiency.  The simpler technique of
setting pixel values in simulated images to random samples
from Poisson distributions whose means are the corresponding pixel values 
in the seed blank-sky images cannot be used because at the desired
resolution the seed 
images contain zero-valued pixels. Since zero is an invalid mean for a
Poisson distribution, appropriate random samples cannot be generated
for such pixels, and simply setting the corresponding pixel values in
the simulated images to zero would introduce unwanted statistical
correlations in the set of simulated images for each seed obsid.

We approximated the nominal field background rates for each chip by
values cited in the \chandra Proposers' Observatory Guide, except for
the longer \aciss observations (OBSIDs 7078 and 4613) which include
chip 8. Here, since we were using an input blank-sky dataset derived
from \CSCA event lists, we estimated the field background rates
directly from source-free regions of the \CSCA event list for the
longest exposure OBSID 4613. We found the rates to be $\sim$67\% of
the corresponding values from the Observatory Guide for chips 2, 3, 5,
6, and 7, and scaled the POG values by this amount. We attribute these
differences to the more rigorous data screening in the \CSCA
processing.

Finally, we distribute event times randomly within the good time intervals
available for each chip, and re-compute the sky coordinates for the chip
with the \ciao tool {\tt reproject\_events}, using the actual aspect
solution from the seed observation. The final chip event lists are
re-assembled into a single event list with the  \ciao tool {\tt
  dmmerge}. An example of a simulated event list for seed OBSID 4613 is shown
in Fig. \ref{fig:4613sim}. Approximately 50 empty-field simulations were
generated for each seed OBSID.

\subsection{Point-Source Simulations}
\label{pointsourcesims}
Simulated point sources were generated using \marx-4.3.  A user-defined
source model was input to \marx to generate X-ray photons incident from a
spatially uniform random distribution of point sources, all having the
same spectral shape of either a power-law (photon index $\Gamma=1.7$)
or a blackbody ($kT=3.0$\,keV), and with an absorbing column of
$N_{\rm H} =3\times 10^{20}$\,cm$^{-2}$. 

More specifically, input source positions were generated by sampling
from uniform random distributions of rotations about orthogonal axes
aligned with directions of increasing 
Right Ascension and Declination, and offset from the observation
aimpoint.  These angular offsets were then
converted to unit vectors in this coordinate system for input to
\marx. They were also converted to Right Ascension and
Declination using the coordinates of the aimpoint.
The mean spatial density of randomly generated source positions was about 
1.2\,arcmin$^{-2}$.  This source density was a compromise aimed at
limiting source confusion and reducing the total number of simulations
required to derive useful statistics on the performance of the
software pipeline. A different random sequence was used to generate
each simulated source population.

The source photon fluxes
were drawn from a powerlaw distribution in which the number of
sources, $N(f)df$ with photon flux between $f$ and $f+df$ is $N(f)df
\propto (f/f_0)^{-\alpha}df$ with $\alpha = 1.5$.  For a simulation
based on an OBSID with exposure time $t$ in seconds, the minimum
photon flux was $f_0 = (0.003/A)(10^5/t)^{1/2}$\,photons
s$^{-1}$~cm$^{-2}$, 
where $A=2,269.55$~cm$^{-2}$ is the geometric area of the mirrors.  

The effect of photon pileup (i.e., when two or more photons are
recorded in a single 
CCD pixel in a single readout frame, and are either misinterpreted as a
single event or discarded as a ``bad'' event) was included by
post-processing each 
simulation with {\tt marxpileup}. The effect of observation-specific bad
pixels was included by post-processing each simulation with
{\tt acis\_process\_events}; events falling on bad pixels were flagged
appropriately.  Because the source and background components were created and
processed separately and then combined only in the final step, we did not
include the (negligible) effect of pileup due to coincidence between source
and background photons.

To simulate an \acis imaging observation based on a particular \chandra OBSID,
two separate \marx simulations were usually required, one for the \acisi chips
and one for the \aciss chips.  Each simulation used the observation-specific
aspect solution ({\tt asol} file), detector position ({\tt SIM\_Z}), start
time ({\tt TSTART}), and exposure time ({\tt EXPOSURE}).

The source events from the two \marx simulations were merged with the
simulated background events, discarding all \marx-simulated source events on
unused CCDs.  After quantizing the background event arrival times to match the
frame times of the relevant CCDs, the full set of event arrival times was
sorted in ascending order. A table containing the coordinates of each
simulated source and the associated flux in each spectral band was appended to
the merged event file.

\begin{figure*}
\epsscale{1.0}
\plotone{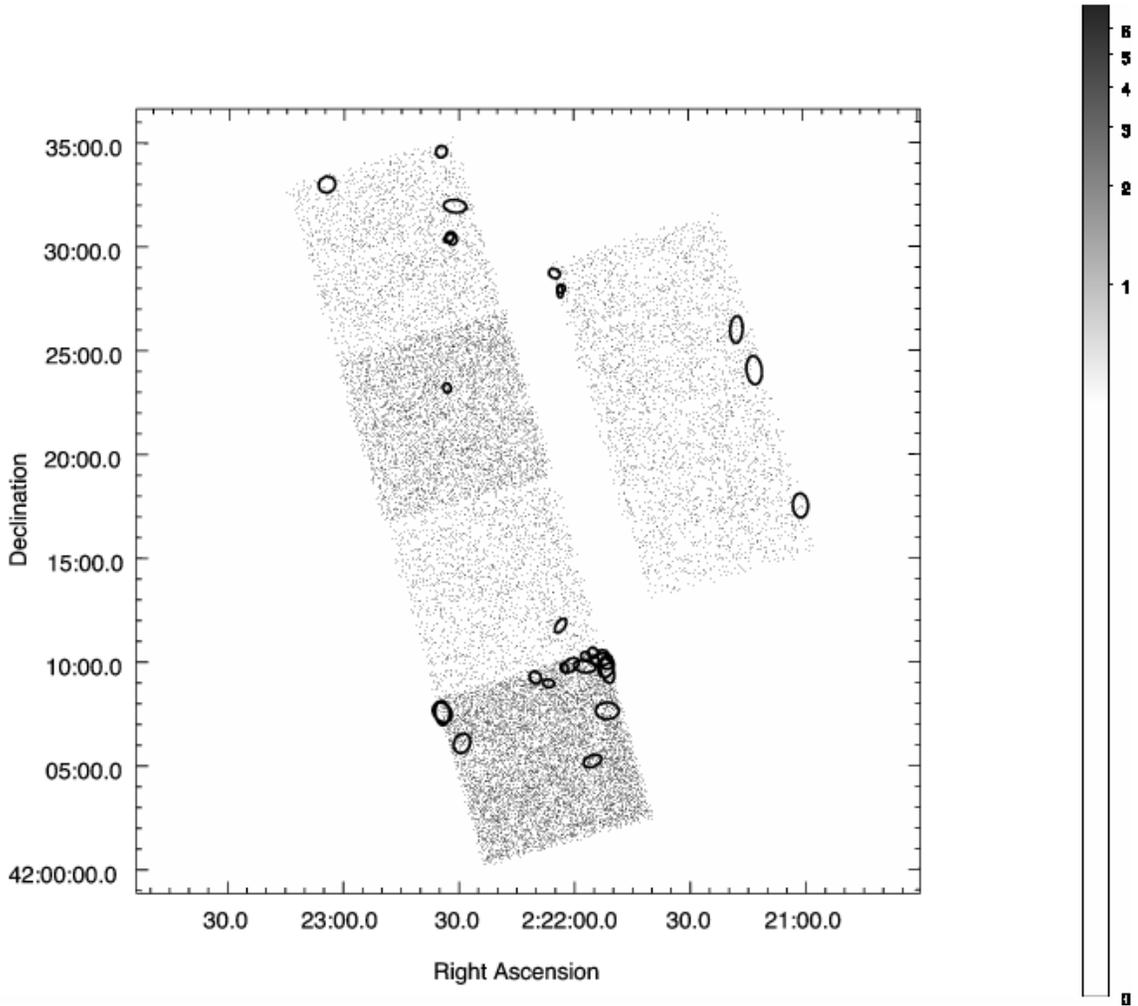}
\caption{\label{blanksky} An example simulated event list using the
  metadata for OBSID 4613. A total of 25 simulation runs were
  performed for this OBSID, yielding 30 source detections that passed
  \CSCA inclusion criteria. These detections are shown as black ellipses.}
\end{figure*}

An example of an event list for seed OBSID 4613 with simulated sources
inserted is shown in Fig. \ref{fig:4613simsrc}. Approximately 20 point-source
simulations were generated for each seed OBSID, for each input spectrum, with
$\sim500-600$ sources per simulation. It should be noted that the distribution
of fluxes for these simulated sources extends well below the
anticipated \CSCA detection limit; the actual number of detected
sources available for characterization analysis is approximately half
the total number.

\subsection{Variability Simulation Algorithms}\label{sec:var_sim_alg}

To assess intra-observation variability, the \CSC employs three
variability tests, described below, to assess whether event arrival
times are consistent with the expectations for a steady source.
Detected count rate variations for a steady source should be dictated
solely by Poisson statistics and the time variable response of the
spacecraft detectors.  The latter is driven primarily by
the effects of spacecraft dither.  The pointing direction of the
\chandra spacecraft is varied in a Lissajous pattern with typical
periods of 1,000 and 707 seconds in perpendicular directions when
observing with the \acis detectors.  Thus a source chip position can
dither beyond the edges of the CCDs, or over detector locations with 
different responses or with different numbers of bad pixels, etc.  

The algorithms for creating background simulations described
in \ref{sec:emptyfieldsims} reproduce very well the \emph{time
averaged} background with the proper counting statistics.  The \marx
simulations used to create the discrete source simulations
(Section~\ref{pointsourcesims})  essentially yield lightcurves that have the
proper counting statistics for a steady source (i.e., white noise)
dithering in a realistic time-dependent manner across the detector.
The final simulations used to assess the \CSCA pipeline, however, are
a combination of these time averaged and time-dependent
components. Although these simulations are suitable for assessment of
source detection, flux, and size algorithms, they are not suitable for
detailed assessment of the source variability detection
algorithms. This is especially true near chip edges where the effects
of dither are expected to be the most significant.  We plan to address
these simulation shortcomings with future updates of the \CSCA
characterization.  

For this initial characterization we perform a series of lightcurve
simulations and variability tests outside of both the \marx package
and the \CSCA pipeline.  These simulations thus lack detector details
such as the CCD response and the spacecraft dither motion; however,
they otherwise have been designed to mimic some properties of
real \chandra lightcurves.  The simulations have discrete time bins
with 3.24104\,sec resolution (the 41.04\,ms \acis readout deadtime is
not included in the simulations), total lengths ranging from
1--150\,ksec, and count rates ranging from 0.0006--0.03\,cps
(corresponding to 0.002--0.1 counts per readout frame).  The goals of
the simulations were to determine the rate of false positives for pure
``white noise'' simulations and to determine the sensitivity of the
tests to real variability for ``red noise'' simulations.

The three intra-observation variability tests performed in the \CSCA
pipeline are the Kolmogorov-Smirnov (K-S) test (essentially as
described and implemented by \citealt{press:2007a}), its variant the
Kuiper test (\citealt{kuiper:1960a}; also based upon the
implementation of \citealt{press:2007a}), and the Gregory-Loredo
variability test \citep{gregory:1992a}.  Statistical properties and
sensitivity of the first two of these tests are described
by \citet{stephens:1974a}.  Essentially one is comparing the
cumulative fraction of all lightcurve events that occur between the
start of the observation and some given time, $t$, to the
theoretically expected cumulative fraction also at time $t$.  For a
steady source, the latter is a curve that rises from 0 to 1 in direct
proportion to the detector area-weighted ``good time'' that has
elapsed.  The K-S and Kuiper tests assess the significance of the
maximum deviations of the measured cumulative fraction curve compared
to the theoretical one.  It is straightforward to incorporate
time-dependent changes in detector efficiency into both of these
tests.

\begin{deluxetable*}{ccccc}
\tablecaption{CSC False Source Rates\label{tabblanksky}} 
\tablewidth{0pt}
\tablehead{\colhead{OBSID}&\colhead{ACIS Configuration}&\colhead{Exposure (ksec)}&\colhead{\#Sources (\#Runs)}&\colhead{False Source Rate}}
\startdata
379	&	ACIS-I	&	9	&	0 (50)	&	0.0	\\
1934	&	ACIS-I	&	29	&	0 (50)	&	0.0 \\
4497	&	ACIS-I	&	68	&	11 (50)	&	0.22 \\
927	&	ACIS-I	&	125	&	64 (50)	&	1.28 \\
5337	&	ACIS-S	&	10	&	1 (50)	&	0.02 \\
4404	&	ACIS-S	&	30	&	5 (50)	&	0.12 \\
7078	&	ACIS-S	&	51	&	5 (24)	&	0.21 \\
4613	&	ACIS-S	&	118	&	30 (25)	&	1.2 \\\hline
\enddata
\tablenotetext{}{
  False Source Rates derived from blank-sky simulations. Column 1:
  OBSID from which observation metadata were chosen; column 2:
  detector configuration; active chips for \acisi were 0, 1, 2, 3, 5,
  6; those for \aciss were 2, 3, 5, 6, 7, 8; column 3: observation
  livetime; column 4: numbers of source detections and runs; column 5:
  mean false source rate (sources per field per run). For the OBSID 4404
  simulations, background data for chip 8 were
  unavailable and the false source rate was renormalized to account for
  this missing chip.
}

\end{deluxetable*}

The Gregory-Loredo test is a Bayesian algorithm that takes a given
lightcurve and successively divides it into a greater number of
uniformly spaced time bins.  It then compares the Poisson likelihood
that these uniformly binned lightcurves are a more probable description
than the single bin lightcurve \citep{gregory:1992a}.  The algorithm
also returns a ``best estimate'' of the time-dependent lightcurve.
Time-dependent detector variations can be incorporated into this test,
but only in an approximate way.  The algorithm implicity assumes that
there is no correlation between the intrinsic variability time scales
of the source and the variability time scales of the detector
efficiency.  Additionally, the Gregory-Loredo algorithm is testing a
more specific hypothesis than the K-S and Kuiper tests.  The latter
tests are assessing the significance of \emph{any} deviations from the
expectations for a steady source.  The Gregory-Loredo test is
specifically examining the significance of \emph{uniformly binned}
lightcurves.  These differences will be discussed further in Section
\ref{variability}.

In our simulations, all three of the above tests were implemented
as \slang\footnote{http://www.jedsoft.org/slang/} 
scripts run via \isis \citep{houck:2000a}. The scripts for
the K-S and Kuiper tests were the same as those run in the \CSCA
pipeline, whereas the script for the Gregory-Loredo test was an
independent version from the C-code implementation used in the
pipeline.  The \slang script, however, was extensively tested against
the C-code and found to give nearly identical results in all cases.  

Lightcurve simulations were also performed with 
\slang scripts run
under \isis.  Two types of simulations were performed: ``white noise''
and ``red noise'' simulations.  For the latter, we followed the Power
Density Spectrum (PDS) based approach outlined
by \citet{timmer:1995a}.  Essentially, one creates an instance of a
lightcurve using the mean PDS profile, where the PDS is normalized
such that its integral over Fourier frequency is the lightcurve mean
square variability. For each Fourier frequency bin, one draws a
Fourier amplitude that is distributed as $\chi^2$ with two degrees of
freedom times the square root of the PDS amplitude.  The Fourier phase
in each bin is independently and uniformly distributed between
0--2$\pi$.  The Fourier spectrum is then inverted to create the
lightcurve, and the lightcurve mean is normalized to a desired level.
(\citet{vaughan:2007a} refer to simulations of this type as following
the ``Davies-Harte'' method, after \citet{davies:1987a}, and discuss
how this method can be generalized to include even more complex
statistical properties.)  For the case of a red noise lightcurve, the
mean PDS was $\propto f^{-1}$ between $1/T$ and $f_{Ny} \equiv
(2 \Delta t)^{-1}$, where $f$ is the Fourier frequency, $T$ is the
total lightcurve length, $f_{Ny}$ is the Nyquist frequency defined by
the bin size of the lightcurve, $\Delta t$.  The root mean square
(rms) variability was also defined by the integral between those two
frequencies.

Once the lightcurve was created, any time bins that fell below zero
were truncated at zero.  (This was required only for a few bins in
each lightcurve for rms variabilities $>10\%$.)  The lightcurve
amplitude in \emph{each} time bin was then used to draw a Poisson
variable for that time bin, which was used as the counts for the time
bin.  Note that the simulation process for the white noise lightcurves
began at this point.  Time bins with multiple counts were considered
to be potentially subject to the effects of pileup, following the
simple pileup model of \citet{davis:2001a}. For each
count in a single time bin in such cases, we assigned a 0.95 chance
that it fell 
within the central ``piled region'', and then drew a random variable
(to be compared to the binomial distribution) to determine how many of
the events were within this region.  Once that number, $n$, was
determined, a probability $\alpha^{n-1}$ was assigned to all the piled
region events being read as a single event, with $1-\alpha^{n-1}$
being the probability that no counts would be registered for the piled
region.  This procedure then yielded the final lightcurves to which
each of the above three variability tests was applied.

\section{Source Detection}
\label{sec:source_detection}

\subsection{False Source Rate}
\label{FSRsection}
To estimate false source rates, we conducted a series of blank-sky
simulations at exposures of $\sim10$, $\sim30$, $\sim60$, and
$\sim120$ ksec, for typical \acisi and \aciss chip
configurations, as discussed in Section \ref{sec:emptyfieldsims}.
Each simulated event list was then processed using the standard \CSCA
source detection and properties software, and the resulting source
detections that would have been included in the catalog were
tabulated. The results are shown in Table \ref{tabblanksky}, and an
example simulated observation is shown in Fig.~\ref{blanksky}.

\begin{figure}
\epsscale{1.3}
\plotone{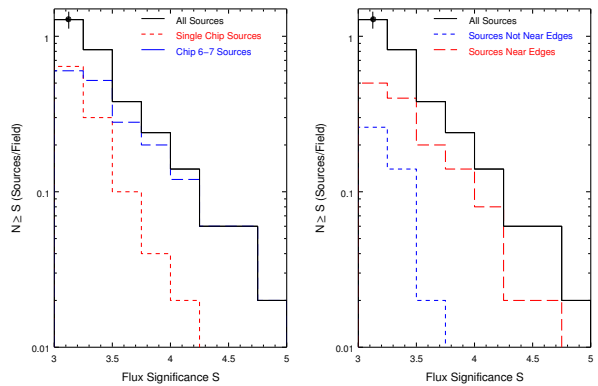}
\caption{\label{927fsr} False source rates as a function of flux
  significance for OBSID 927. The maximum flux significance of all
  science bands is used. Left: Single-chip sources are those whose
  source regions cover only a single chip, as indicated by the
  {\tt multi\_chip\_code}. Chip 6-7 sources are those whose source regions
  dither across chips 6 and 7. Right: Sources near edges are those
  whose source regions dither off a chip edge during the observation.}
\end{figure}

\begin{figure}
\epsscale{1.25}
\plotone{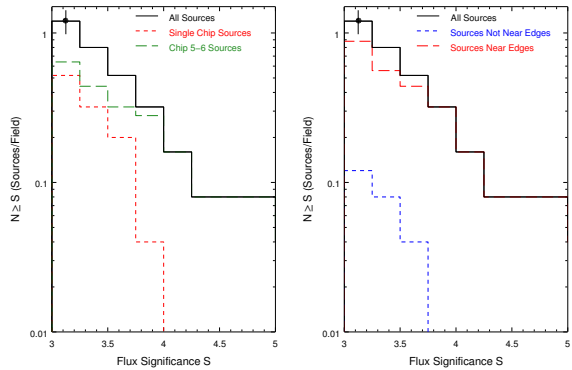}
\caption{\label{4613fsr} False source rates as a function of flux
  significance for OBSID 4613. The definitions for different subsets
  are the same as in Fig.~\ref{927fsr}. }
\end{figure}

As can be seen in Table \ref{tabblanksky}, the false source rate is
appreciable only for exposures longer than $\sim$50 ksec. There
is also some evidence for a clustering of false source detections near
chip edges and between the back- and front-illuminated chips. To
investigate these effects further, we considered the longest \acisi
and \aciss simulation sets, and examined the false source rate
separately near chip edges and interfaces. The results for OBSID 927
are shown in Fig.~\ref{927fsr} and for OBSID 4613 in
Fig.~\ref{4613fsr}, and demonstrate that false source rates are
enhanced in these regions.

\begin{figure*}
\begin{center}
\epsscale{0.96}
\plotone{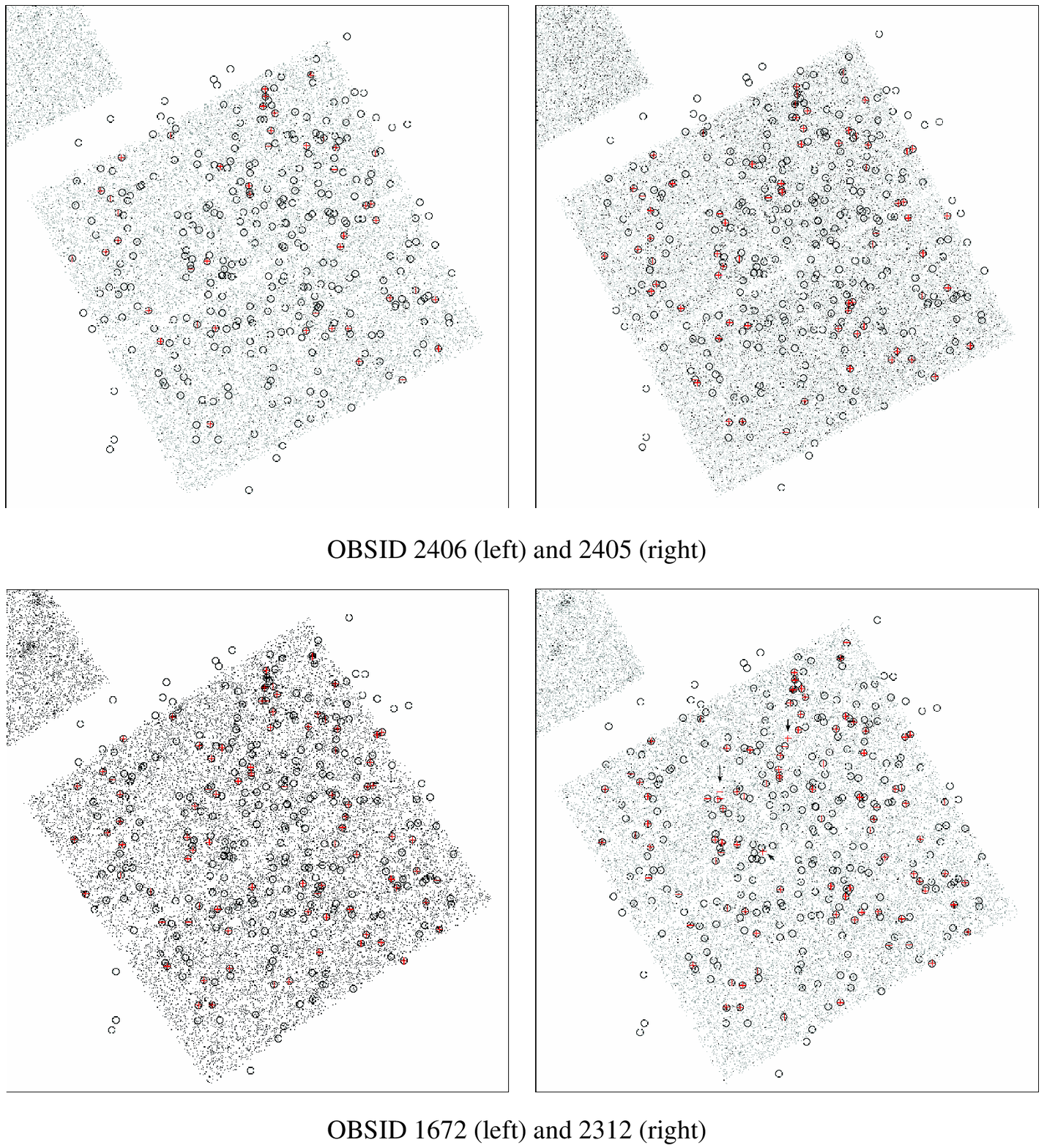}
\caption{\label{cdfs_falsesrc} \CSCA (crosses) and \CDFS (circles)
  sources in four \CDFS OBSIDs of $\sim30$, $\sim60$, $\sim95$, and
  $\sim124$ ksec. False sources, indicated by black arrows, are
  evident only for the longest exposure.}
  \vspace{-0.1cm}
\end{center}
\end{figure*}

We can verify the conclusions of our simulation studies by examining
\CSCA sources detected in individual observations that are themselves
parts of longer-exposure observing programs. We use the \chandra
Deep Field South (\CDFS) Catalog of \citet{alexander:03a}, which
contains 326 sources in a total exposure of $\sim940$ ksec,
comprising 11 separate \acisi observations with similar
aimpoints. Since source detection is performed on the deeper,
combined \CDFS images, we assume the \CDFS catalog is complete at the
level of individual component observations, and that therefore any
\CSCA sources detected in individual \CDFS observations that do not
match sources in the \CDFS catalog are likely to be false sources. We
are implicitly ignoring the possibility of long term variability,
where a real source is marginally detected in a single observation,
but falls below the detection level for the combined observations.

In Fig.~\ref{cdfs_falsesrc} we show \CSCA sources detected in
individual \CDFS OBSIDs 2406 (30 ksec), 2405 (60 ksec), 1672 (95 ksec)
and 2312 (124 ksec), together with sources in the \CDFS catalog.  For
OBSIDs 2406, 2405, and 1672, all \CSCA sources match \CDFS sources,
consistent with false source rates of $<1$ per field shown in Table
\ref{tabblanksky}. For OBSID 2312, three \CSCA sources do not match
sources in the \CDFS catalog. The mean rate from Table
\ref{tabblanksky} is 1.28 for an \acisi observation of this length. If
we assume a Poisson statistical model for the false source
distribution, the probability of finding three or more false sources
is $\sim14$\%. We conclude that the false source rates determined from
real \chandra observations are consistent with those derived from
our simulations. 

\begin{figure*}
\begin{center}
\includegraphics[scale=0.95]{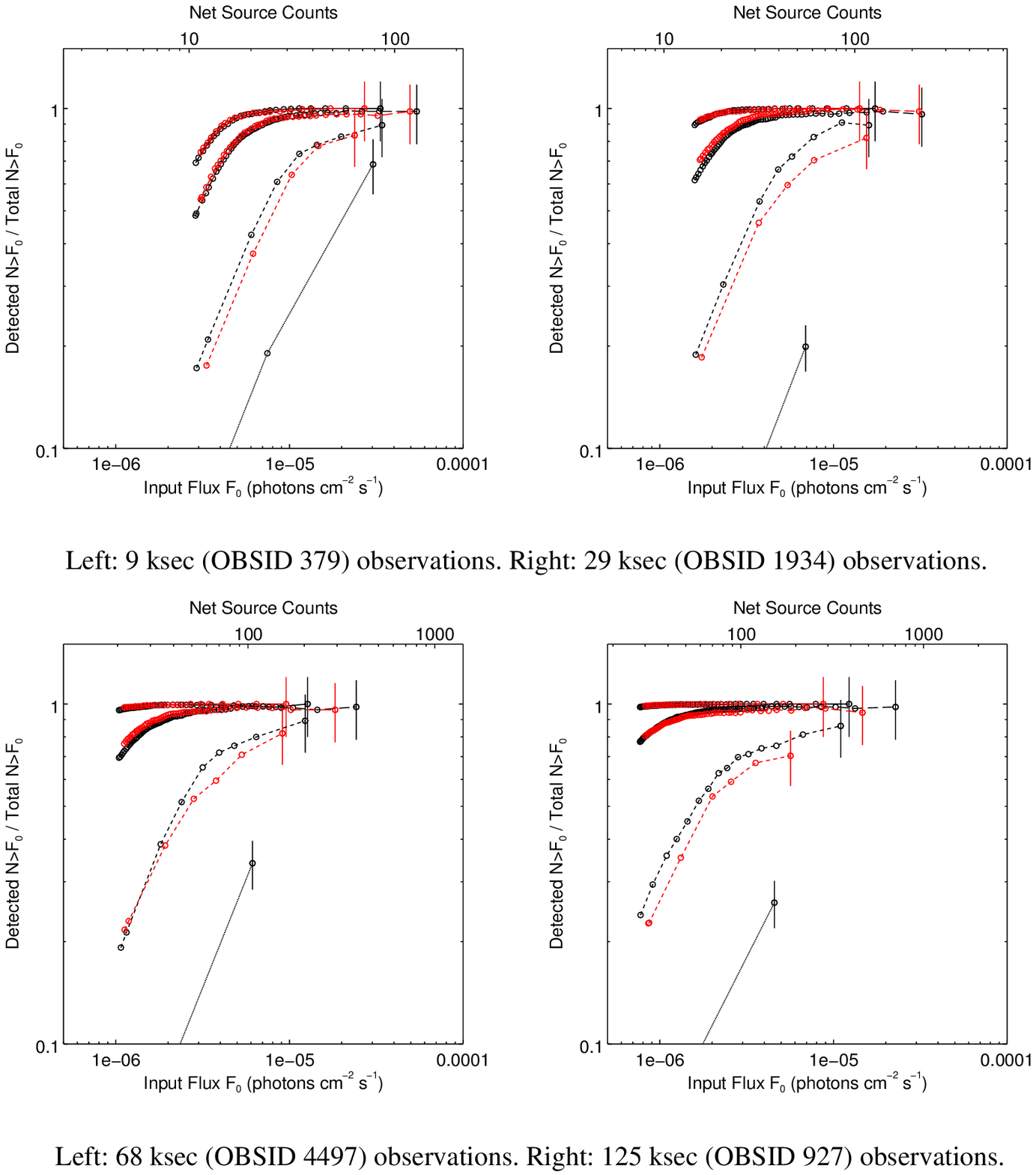}
\end{center}
\vspace{-0.4cm}
\caption{\label{deteffacisi} Detection Efficiencies for simulated \acisi sources
 with powerlaw (black, left curve) spectra and blackbody (red, right curve)
  spectra, for sources with off-axis angle
  $\theta\leq5^{\prime}$ (solid lines),
  $5^{\prime}<\theta\leq10^{\prime}$ (long dash),
  $10^{\prime}<\theta\leq15^{\prime}$ (short dash) and
  $15^{\prime}<\theta\leq20^{\prime}$ (dot).
  Simple statistical error bars (i.e., $\sqrt{N}$) for the last bin are shown.}
\vspace{-0.2cm}
\end{figure*}

\begin{figure*}
\begin{center}
\includegraphics[scale=0.93]{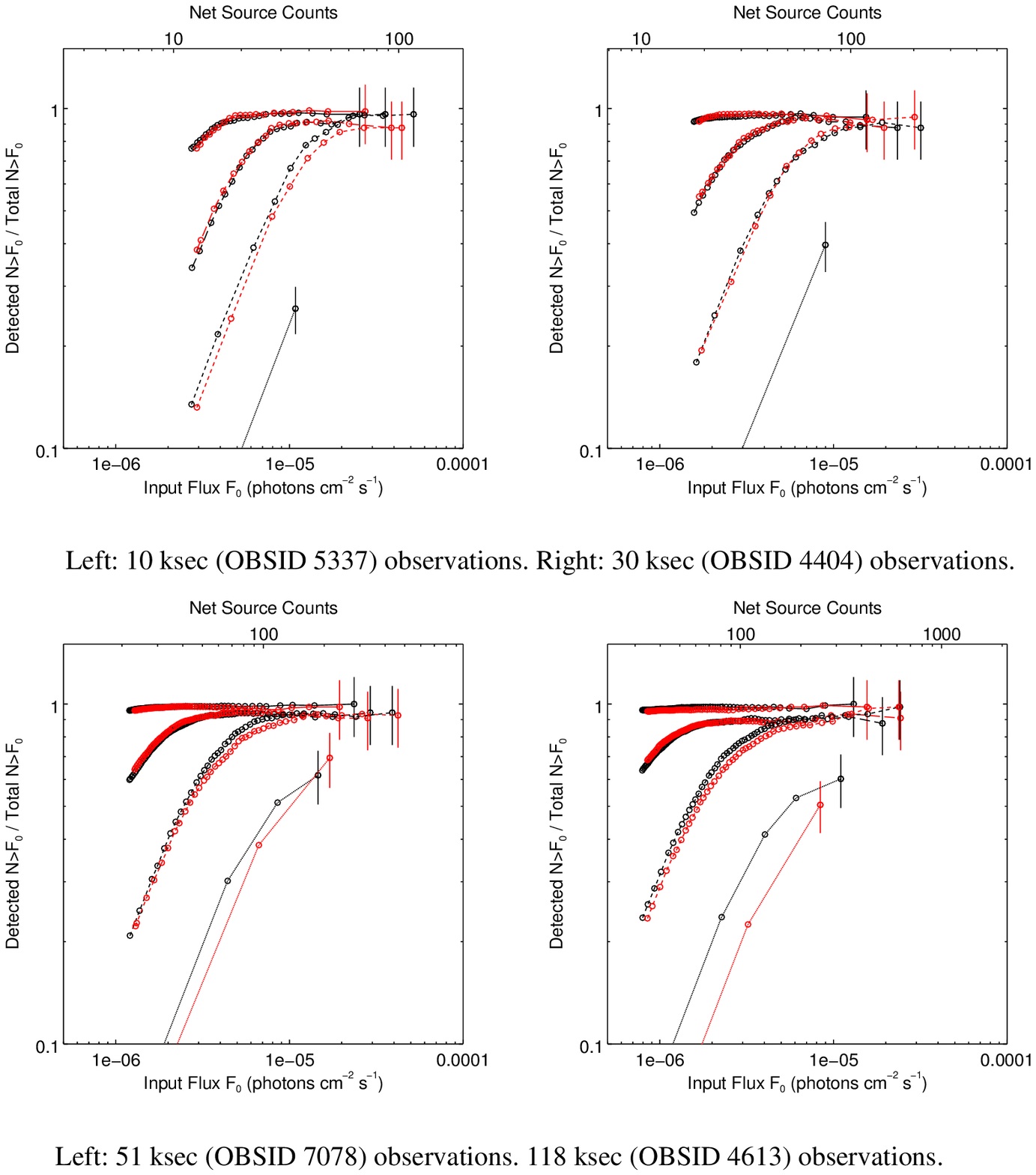}
\end{center}
\vspace{-0.5cm}
\caption{\label{deteffaciss} Detection Efficiencies for simulated \aciss
  sources (see Fig.~\ref{deteffacisi} for a description of the various components).}
\vspace{-0.1cm}
\end{figure*}

\begin{figure}[!h]
\epsscale{1.2}
\plotone{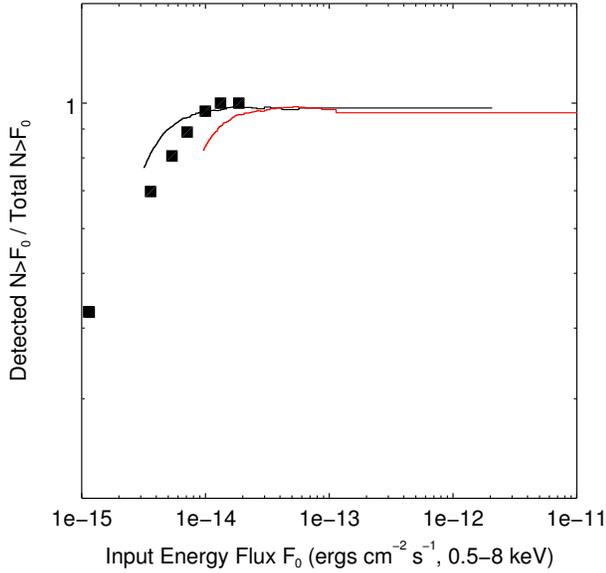}
\caption{\label{csc_cdfs_eff} Detection Efficiency for OBSID 2405, derived
  from sources detected in the \CDFS catalog
  \citep{alexander:03a}. Efficiencies for powerlaw (black) and blackbody
  (red, or halftone in paper version of the article) sources in simulated \acisi observations of comparable exposure are included.
}
\end{figure}

\begin{figure}
\begin{center}
\includegraphics[width=3.25in]{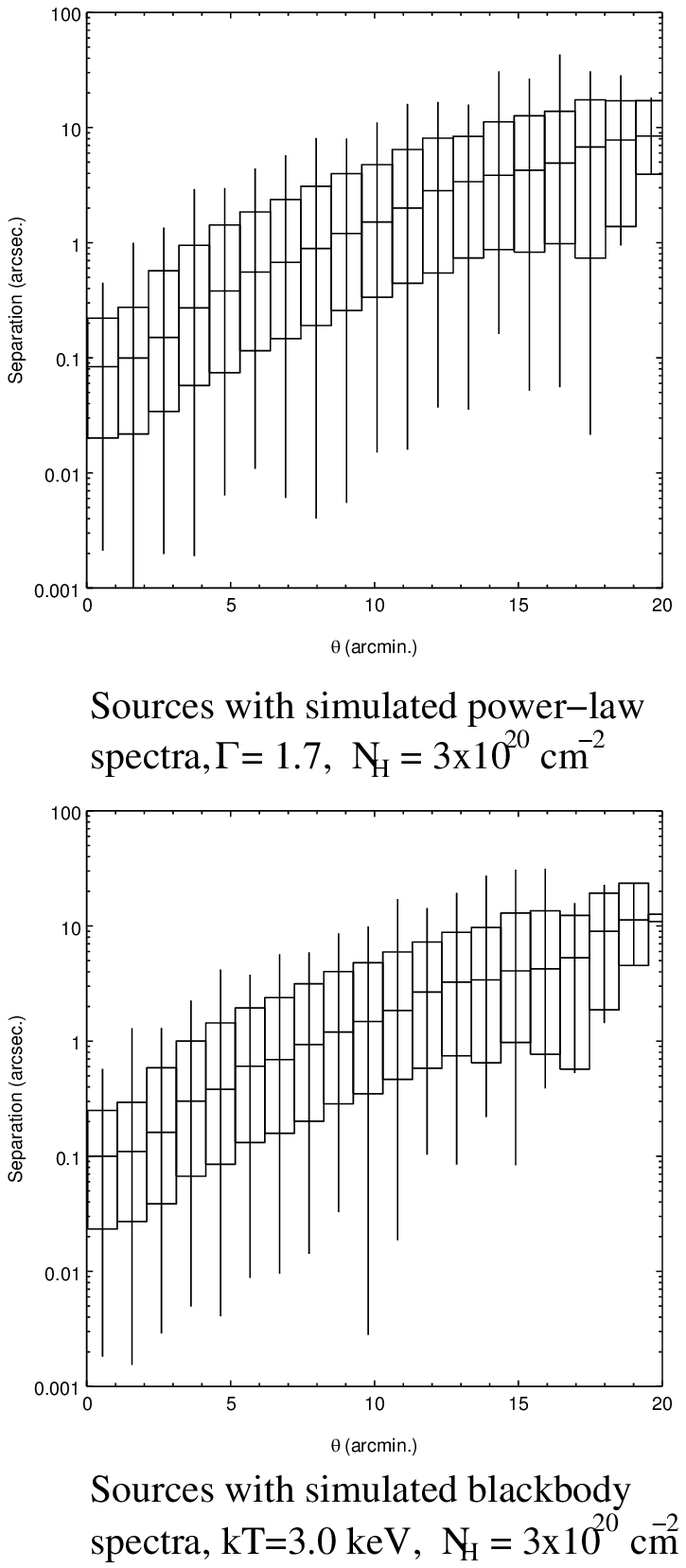}
\end{center}
\caption{\label{relastro_1} Distribution of angular separations between input and
  measured source positions, as a function of source off-axis angle
  $\theta$. Median separations are indicated by horizontal
  lines. Boxes   indicate the 95\% (upper) and 5\% (lower) percentiles
  of the distribution in each bin, and vertical lines indicate extreme values.}
\end{figure}

\subsection{Detection Efficiency}
\label{deteffsection}

We use the point-source simulations described in Section
\ref{pointsourcesims} to estimate detection efficiency as a function
of exposure time for observations with \acisi and \aciss
aimpoints. Sources with simulated powerlaw and blackbody spectra were
analyzed separately; results were similar for both spectral models.
Approximately 214,000 simulated sources were available for analysis, of
which approximately half were detected by the \CSCA source detection
pipeline and passed the quality assurance and flux significance
criteria for inclusion in the catalog\footnote{We emphasize that for
  the remainder of this section, the term ``detected'' refers to such
  sources, while the term ``undetected'' refers to sources which
  failed either the source detection, quality assurance, or flux
  significance criteria for catalog inclusion.}. 

For each seed OBSID in Table \ref{table:seed-observations} we
constructed histograms of input \bband band photon fluxes for both
detected and undetected sources, choosing bin boundaries such that
there were 50 detected sources in each flux bin. We then constructed
cumulative $N>S$ distributions from each histogram. The ratio of the
distribution for detected sources to that for all sources represents
the detection efficiency, i.e., the fraction of input sources brighter
than a given incident flux that are actually detected. Results for the
\bband band detections for the \acisi and \aciss simulation sets are
shown in Figs.~\ref{deteffacisi} and \ref{deteffaciss}. Efficiencies
are plotted against both input photon flux and net source counts. The
latter are based on a linear regression between net counts and input
flux for detected sources and are only intended to provide an
approximate counts scale for the plots. 

These curves are in general similar to those derived for the ChaMP
Point Source Catalog \citep{Kim:2007}, but are  presented separately
for standard \acisi and \aciss chip configurations, since the
different chips sampled in each configuration may result in different
efficiencies for certain ranges of off-axis angle $\theta$. For example, in the
range $5^{\prime}<\theta\leq10^{\prime}$, \acisi observations sample
the relatively low-background, front-illuminated chips 0-3, while
\aciss observations sample both the high-background, back-illuminated
chip 7 and  the badly-streaked chip 8. As indicated in
Figs.~\ref{deteffacisi} and \ref{deteffaciss}, the detection
efficiencies for the \aciss observations are systematically lower than
those for the \acisi observations of comparable exposure in this range
of off-axis angle. 

Finally, we compare the detection efficiencies derived from our
simulations to those measured from real \chandra observations, again
using \CSCA sources detected in OBSID 2405 and the \CDFS Catalog
\citep{alexander:03a}. The \CSCA includes 72 sources with \bband band
energy fluxes above $\sim1.3\times10^{-15}~{\rm ergs~cm^{-2}~s^{-1}}$
in \acis chips 0-3 (those covered by  \CDFS) in OBSID 2405. All have
counterparts in the \CDFS catalog, which includes an additional 228
sources in the same field-of-view, with fluxes above
$\sim9\times10^{-17}~{\rm ergs~cm^{-2}~s^{-1}}$ in the energy band
from 0.5 to 8.0 keV. We use the \CDFS fluxes in this energy band for
both  detected and undetected sources, to compute detection
efficiency, using the procedure described previously. We chose bin
boundaries to include 10 detected sources in each flux bin.  To
compare to the efficiencies from our simulations, we convert the input
photon fluxes of our simulated sources to \CDFS energy fluxes, using
\sherpa \citep{Freeman:2001,Doe:2007} and our powerlaw and blackbody spectral models. We find
conversion factors of   $3.03\times10^{-9}~{\rm erg~photon^{-1}}$ for
sources with powerlaw spectra and $8.56\times10^{-9}~{\rm
  erg~photon^{-1}}$ for sources with blackbody spectra. We then
computed detection efficiencies for simulated sources within
$10^{\prime}$ of the aimpoint in \acisi OBSID 4497,  which has an
exposure time comparable to that of OBSID 2405. We do not divide the
data into ranges of off-axis angle since \CDFS sources typically
contain contributions from multiple off-axis angles. 

Our results are shown in Fig.~\ref{csc_cdfs_eff} and indicate
general agreement. We note that the \CDFS sources exhibit a range of
spectra, and their efficiency is bracketed by those derived from our
two spectral models.

\begin{figure*}
\epsscale{1.1}
\plotone{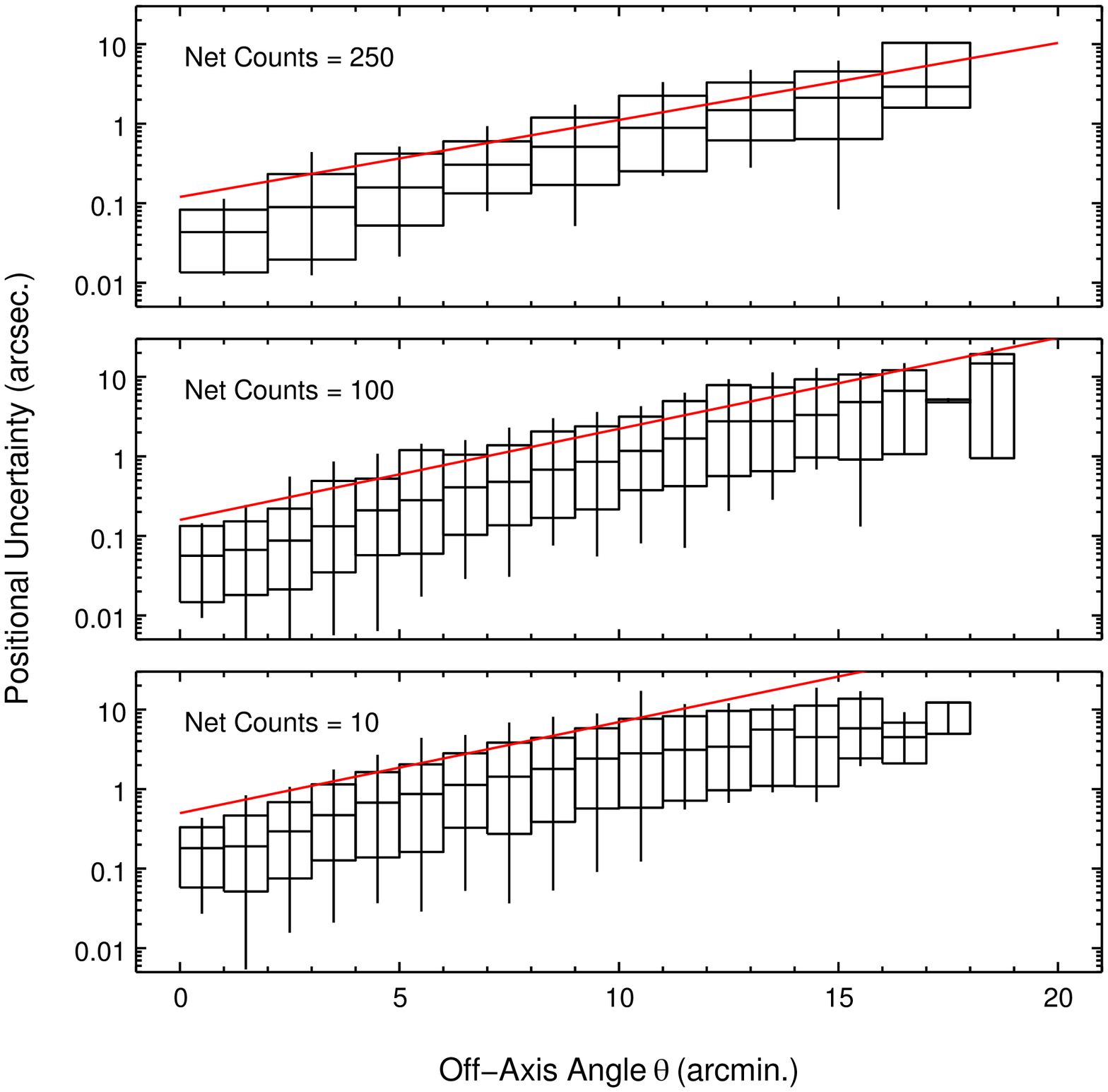}
\caption{\label{relastro_2} Distribution of angular separations between input and
  measured source positions, as a function of source off-axis angle $\theta$,
  for three values of net counts. Red straight lines indicate the \champ 95\%
  positional uncertainties, as reported by \citet{Kim:2007}.}
\end{figure*}

\begin{figure*}
\epsscale{1.0}
\plotone{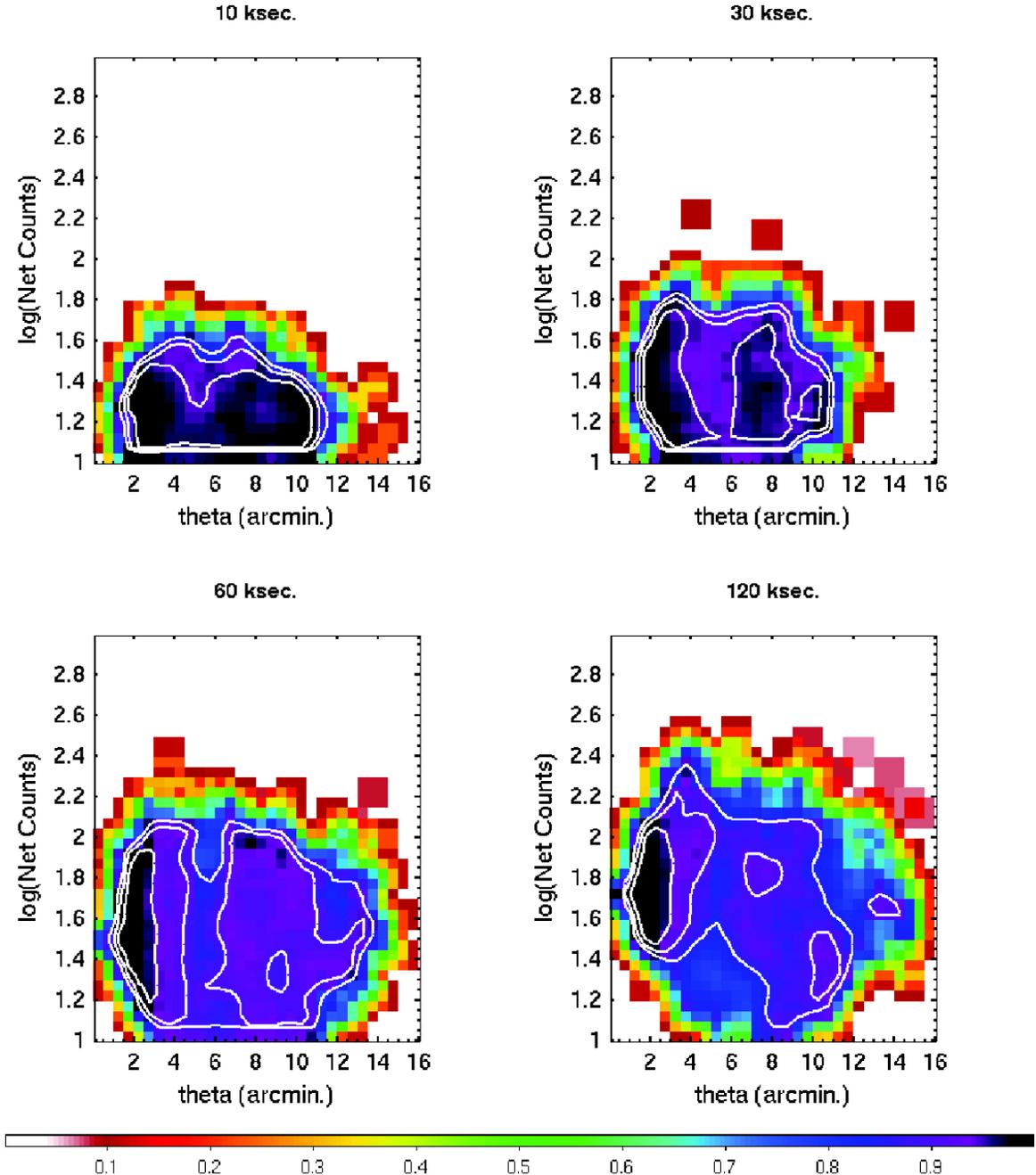}
\caption{\label{relastro_3} Fraction of simulated sources with
  position errors less than \champ 95\% uncertainties, as a function of
  off-axis angle $\theta$,   and net counts, for four exposure times
  used in the point-source simulations. Contours for fractions of
  0.85, 0.9, and 0.95 are indicated.} 
\end{figure*}

\begin{figure}
\epsscale{1.0}
\plotone{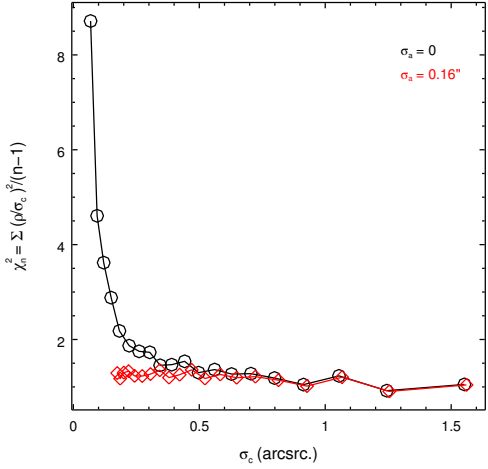}
\caption{\label{relastro_4} Reduced $\chi^{2}$ vs. combined \CSCH
  \sdss position
  error, for no assumed systematic astrometric error (black circles) and for a
  systematic error of $0.16^{{\prime}{\prime}}$ (red diamonds).}
\end{figure}

\begin{figure}
\epsscale{1.0}
\plotone{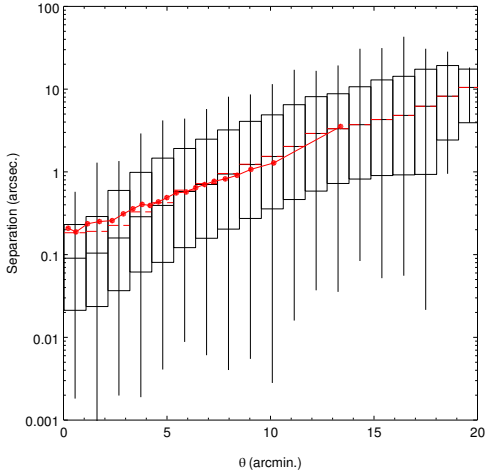}
\caption{\label{relastro_5} Distribution of separations between input and
  source positions for all simulated sources (see Fig.~\ref{relastro_1} for an
  explanation of the meaning of various plot components). Also plotted
  as red filled circles
  are the average separations from the \CSCH \sdss cross-match
  catalog. Dashed red horizontal lines are the medians in each bin
  with the astrometric systematic error added in quadrature. }
\end{figure}

\begin{figure}
\epsscale{1.0}
\plotone{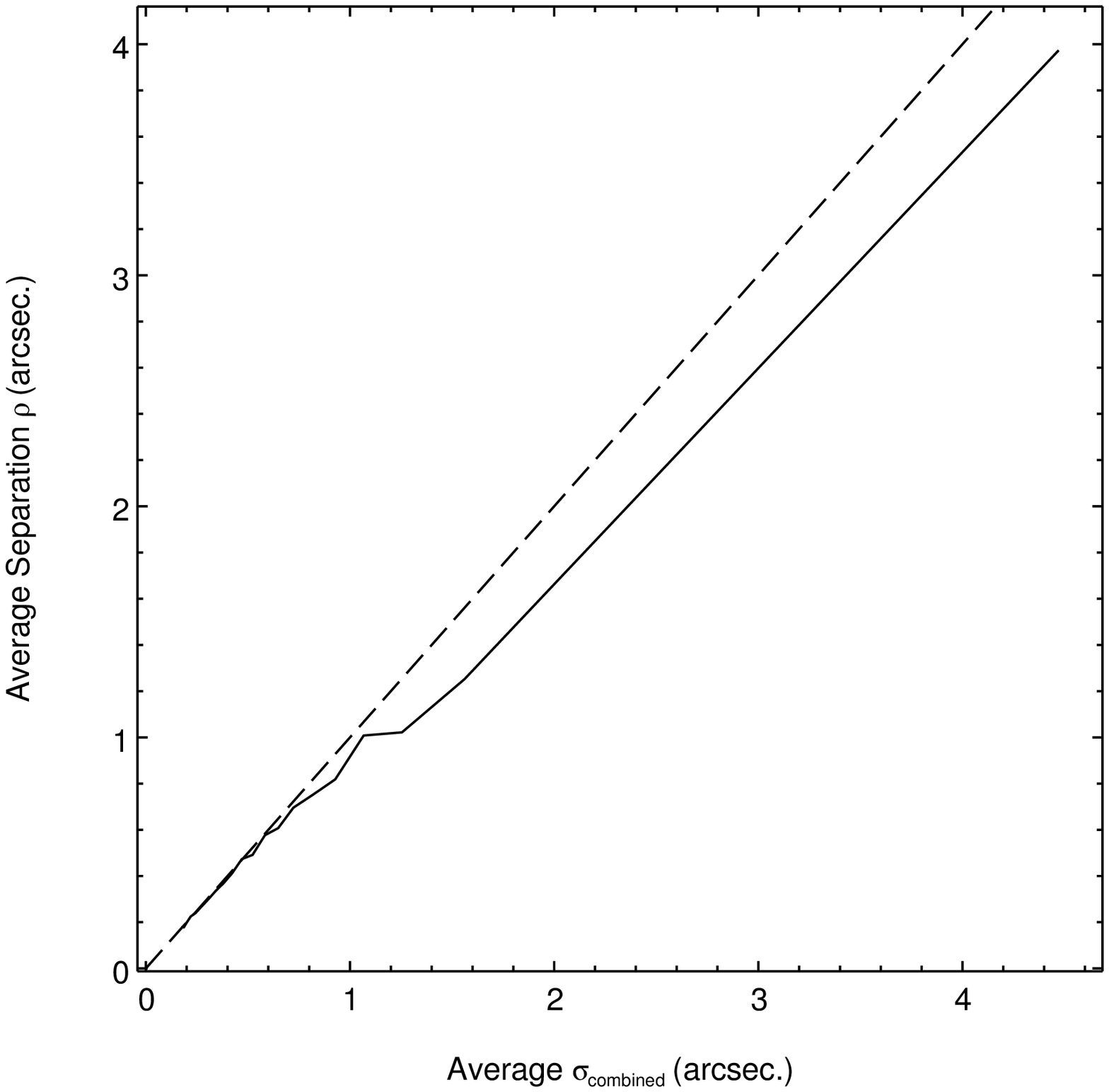}
\caption{\label{relastro_6} Average \CSCH \sdss separations vs. average combined
  error for cross-match pairs in the bins used in
  Fig.~\ref{relastro_4}.  The combined errors include the
  $0.16^{{\prime}{\prime}}$ systematic astrometric 
  error. The dashed line has a slope of 1.}
\end{figure}

\section{Astrometry}
\label{astrometry}

\CSC source positions in individual observations are derived from
centroids of events found in source apertures \citep{evans09}; their 
uncertainties are characterized by error circles  whose sizes were
determined from simulations generated by the \champ project
\citep{Kim:2007} and verified in an earlier, limited set of \CSCA
simulations.  In the case of multiple detections of the same source,
an error ellipse is derived from a combination of the error circles
associated with the individual  detections \citep{evans09}. To
characterize the astrometric properties of the \CSCA, we first
consider the accuracy with which we can locate sources in the frame of
the observation, using simulated point sources. This can  provide  a
good measure of the statistical uncertainty of the source
position in the frame of the observation,  but does not address any
systematic errors in the absolute astrometry. To investigate these
errors, we consider a subset of \CSCA sources with known counterparts
of high astrometric quality, obtained from cross-matching \CSCA
positions with positions from Data Release 7 of the Sloan Digital Sky
Survey \citep{Abazajian:2009}.  

\subsection{Statistical Uncertainties}\label{sec:astrometry_1}
To estimate the relative astrometric precision of the \CSCA, we use
the point source simulations described in Section \ref{pointsourcesims},
and compare input and detected source positions.  To be explicit,
simulated sources are distributed in sky coordinates and rays are
propagated onto chip coordinates using the \marx internal mirror and
detector models. These simulations are passed through the \CSCA
pipeline, where detected source positions are assigned to sky
positions via knowledge of the spacecraft geometry.  Thus the detected
positions of the simulated sources are both a measure of the
accuracy of the pipeline algorithms, as well as a measure of the
fidelity of the \marx simulations.  The correspondence between the
\marx simulations and the true spacecraft geometry is explicitly
discussed in the Appendix, and it is found to be excellent.

Approximately 90,000 simulated sources were identified by the \CSCA
detection pipeline and meet the criteria for inclusion in the
catalog. For these sources we have tabulated input source position and
flux, detected source position and net counts from the \CSCA detection
pipeline, and final source properties from the \CSCA properties
pipeline.  Distributions of angular separation between input and
detected positions as a function of off-axis angle $\theta$ are shown
in Fig.~\ref{relastro_1}. Median separations range from
$\sim0.1^{{\prime}{\prime}}$ on-axis to $\sim4^{{\prime}{\prime}}$ at
$\sim15^{\prime}$ off-axis. We find little difference in the results
for the different input spectra, and so combine results from both in
subsequent analysis.  

We use these results to revisit the question of the suitability of the
\champ error relations for the \CSCA. The \champ error relations are
essentially functions of net counts and $\theta$ fit to particular
percentiles of measured position error distributions at certain values
of net counts and $\theta$. To examine how well they describe \CSCA
position errors, we compare them to percentiles of \CSCA error
distributions from our simulations, for appropriate values of net
counts and $\theta$. In Fig.~\ref{relastro_2} we show three plots
similar to those in Fig.~\ref{relastro_1}, but now limited to
sources with net counts within 10\% of 10, 100, and 250 counts.  The
net counts used here are the quantities reported by \wavdetect in
the \CSCA source detection pipeline; these are the same quantities
used to derive the \champ positional uncertainty relations and to
calculate the error circles in the \CSCA pipeline. They differ
slightly from, but are well-correlated with, the net counts determined
from aperture photometry and reported in the catalog.  The number of
sources in each set are 2,341, 1,534, and 430, respectively. Also
plotted are curves for the \champ 95\% positional uncertainties from
eq. 12 of \citet{Kim:2007}, for sources with 10, 100, and 250 net
counts.  For all three values of net counts, the \champ relations lie
above the observed 95\% percentiles (upper edges of boxes) for
positional error distributions  for $\theta \lesssim 3^{\prime}$.
We conclude that the \champ uncertainties and hence the \CSCA
uncertainties slightly overestimate the actual positional errors in
this range. Similarly, for net counts=100 and 250, the \champ
uncertainties appear to underestimate the true errors for  $\theta \gtrsim
8^{\prime}$.  

We investigate this result in more detail by constructing
two-dimensional histograms in net counts and $\theta$, and computing
the fraction of sources in each bin for which the separation between
input and detected position is less than the \champ 95\% positional
uncertainty for that source. We divide our data into four subsets,
corresponding to simulation exposures of $\sim10$, $\sim30$, $\sim60$,
and $\sim120$ ksec (see Table
\ref{table:seed-observations}). The number of sources in each subset
are $\sim$13,000, 16,000, 29,000, and 32,000,  respectively. If the \champ
relations are always and everywhere a good measure of the \CSCA
statistical position uncertainties, all histogram values should be
$\sim$0.95. Images of the histograms are shown in
Fig.~\ref{relastro_3}, where we have lightly smoothed the histograms
by a simple $3\times3$ boxcar kernel, to aid in constructing
contours. Only histogram bins containing more than 10 sources are
shown. For exposures $\lesssim30$ ksec, the \champ
uncertainties are greater than the 95\% percentiles of the actual
position error distributions for net counts $\lesssim$40 and for most
values of $\theta$ for which there are sufficient data. For higher
exposures, the \champ uncertainties overestimate the actual  95\%
percentiles for low values of $\theta$, and underestimate the 95\%
percentiles at larger values, as suggested by
Fig.~\ref{relastro_2}. For all exposures, the \champ uncertainties 
approximate error distribution percentiles of $\gtrsim$80\% for most
of the range of net counts and $\theta$ for which we have sufficient
data.   

\subsection{Absolute Astrometry}
We have cross-matched the \CSCA with the \sdss DR-7 catalog
\citep{Abazajian:2009}, using the probabilistic cross-match algorithm
of \citet{budavari2008}. We selected objects with a cross-match
probability greater than $90\%$ and which were classified as stars in
the \sdss catalog. The resulting cross-match catalog contained 6,310
\CSCH \sdss pairs, corresponding to 9,476 sources detected in
individual \CSCA observations, since many objects were observed
several times by Chandra.  We use the combined spatial error estimate
of each object pair in this catalog as the independent variable and
analyze the statistical distribution of the measured \CSCH \sdss
separations, $\rho$,  to derive the value of any unknown \CSCA 
astrometric error. \CSCA provides a 95\% error circle radius, while
the \sdss provides independent 1-$\sigma$ errors in Right Ascension  and
declination \citep{pier:2003}.  The combined error is derived by adding the geometric
means of the major and minor axes for \sdss in quadrature with the
\CSCA error and any unknown astrometric error, namely,
$\sigma_{combined}=\sqrt{\sigma_{RA}\sigma_{Dec}+(0.4085\sigma_{\CSCA})^2+\sigma_{a}^2}$,
where the numerical constant $0.4085$ is used to convert from a 95\%
to a 1-$\sigma$ error\footnote{For a two-dimensional, circularly
symmetric Gaussian distribution, the 95\% error radius $R_{95}$ is
given by the solution to the integral equation
$(2\pi\sigma^{2})^{-1}\int_{0}^{R_{95}}\,e^{-\frac{r^2}{2\sigma^2}}\,2\pi\,r\,dr\,=\,0.95$,
or $R_{95}\,=\,2.448\,\sigma$.}. 
The RA error bar is a true angular error bar
in that a factor of $\cos (Dec)$ has been incorporated into it.

We sorted the cross-match pairs in increasing order of
$\sigma_{combined}$ into bins containing $n=$100, 200, 300, and 400
sources for the first 4 bins, and 500 sources thereafter (the last bin
contained 476 sources). We used smaller numbers in the first few bins
since we assume that any unknown astrometric error, $\sigma_{a}$, is
relatively small compared to the \CSCA uncertainties, especially
off-axis, and that it therefore affects mainly those pairs with small
combined errors. The statistical distribution of the separations will
therefore change more rapidly for lower values of
$\sigma_{combined}$. We characterized the statistical distribution of
separations in each bin in terms of the reduced $\chi^2$ of the
normalized separations 
$\rho_{N}=\rho/\sigma_{combined}$ 
\begin{equation} 
\chi^{2}_{n}=\frac{\displaystyle\sum^{n}_{i=1} \rho_{N,i}^2}{n-1},
\end{equation}
and examined the behavior of $\chi^{2}_{n}$ vs. the mean value of
$\sigma_{combined}$ in the bins, for different assumed values of an
unknown $\sigma_a$. As can be seen in Fig.~\ref{relastro_4}, for
$\sigma_{a}=0, \chi^{2}_{n}\sim1$ for
$\sigma_{combined}\gtrsim0.25^{{\prime}{\prime}}$ but rises steeply
below this value,  validating our assumption that a systematic
astrometric error dominates at small values of combined error. A value
of $\sigma_{a}\sim0.16^{{\prime}{\prime}}$ yields reasonable values of 
$\chi^{2}_{n}$ for all values of $\sigma_{combined}$, and we adopt
this as our estimate for the \CSCA systematic astrometric error. Note,
this value should be added in quadrature to all \CSCA 1-$\sigma$
positional uncertainties in Release 1.0.1 of the catalog. (This
additional error is already incorporated into later catalog releases.) 

We can use the \CSCH \sdss cross-match catalog to verify the simulation
results derived in Section \ref{sec:astrometry_1}. We show in
Fig.~\ref{relastro_5} a plot similar to that in Fig.~\ref{relastro_1},
but now  combining results from both powerlaw and blackbody
sources. We also plot the average \CSCH \sdss separations in various
bins in $\theta$. The \CSCH \sdss separations agree well with the
simulation results for $\theta\gtrsim5^{\prime}$, but exceed the
median simulation separations for smaller $\theta$. This result is to
be expected since the simulation results do not include a  systematic
astrometric error, which dominates the \CSCH \sdss results for the
small separations prevalent at small $\theta$. When the systematic
uncertainty is added (as indicated by the horizontal red lines), 
the results are in good agreement. 

Finally, we use the \CSCH \sdss results to investigate the suitability
of the \champ errors, as in Section \ref{sec:astrometry_1}. In
Fig.~\ref{relastro_6}, we show the average \CSCH \sdss separations as
a function of  $\sigma_{combined}$ for the data in the bins used to 
compute reduced $\chi^2$ above. For values of  separation $\lesssim
0.7^{{\prime}{\prime}}$ (corresponding to $\theta\lesssim7-8^{\prime}$
in Fig.~\ref{relastro_5}) , the two agree well,  but at larger
values ,  $\sigma_{combined}$ becomes increasingly larger than the
average separation, indicating that the \champ errors overestimate the
true errors for $\theta\gtrsim7-8^{\prime}$. This is roughly
consistent with the results in Section \ref{sec:astrometry_1},
especially for exposures $\lesssim$30 ksec.  We note the median
exposure in \CSCA observations is $\sim$13 ksec.  

\begin{figure}
\begin{center}
\includegraphics[width=3.4in]{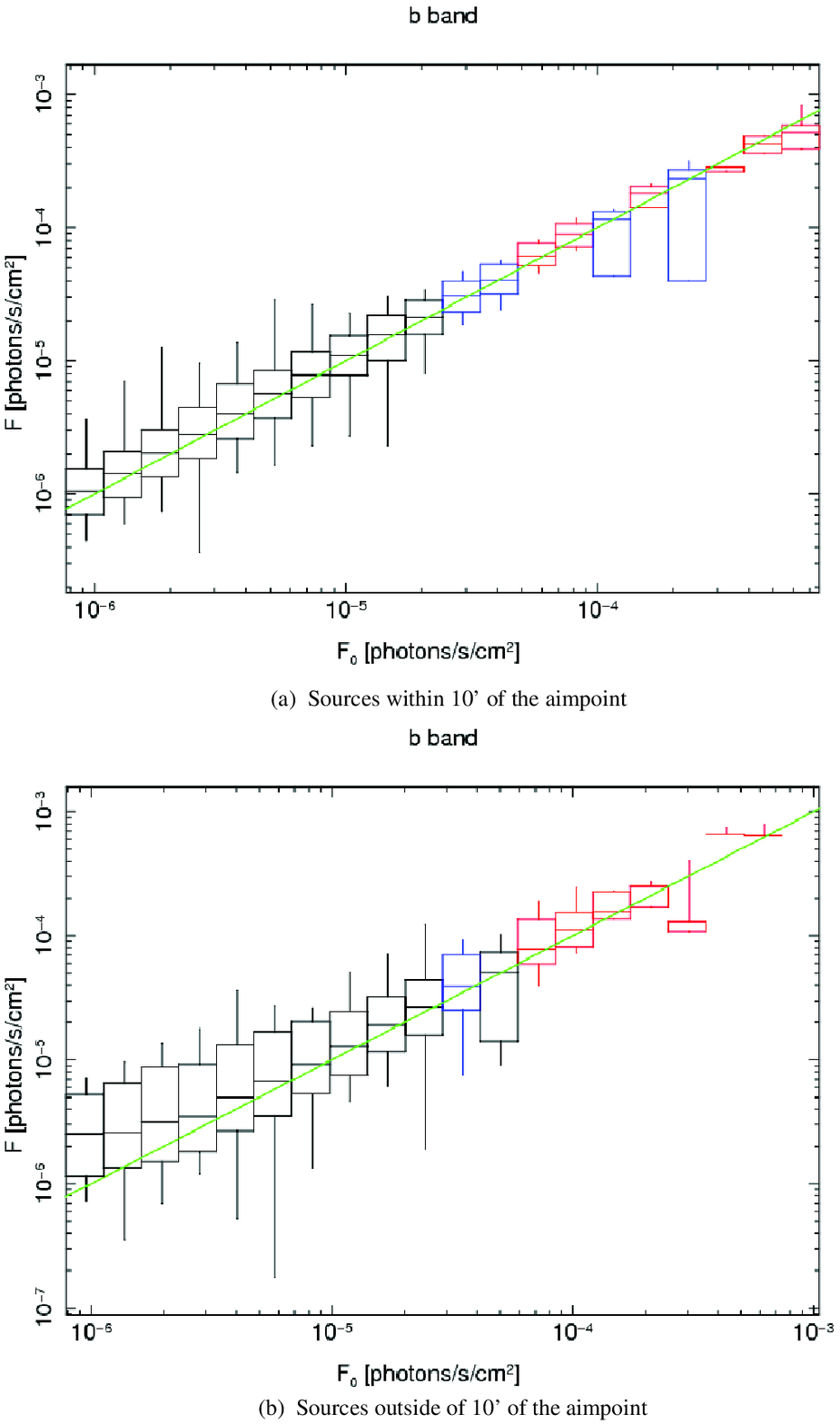}
\end{center}
\vspace{-0.2cm}
\caption{\label{plflux} Comparison of input and measured \bband band fluxes
  for sources with powerlaw spectra. Bins in red contain fewer than
  100 measurements; bins in blue contain 100-400 measurements; bins in
  black contain more than 400 measurements.} 
\vspace{-0.2cm}
\end{figure}

\begin{figure}
\begin{center}
\includegraphics[width=3.5in]{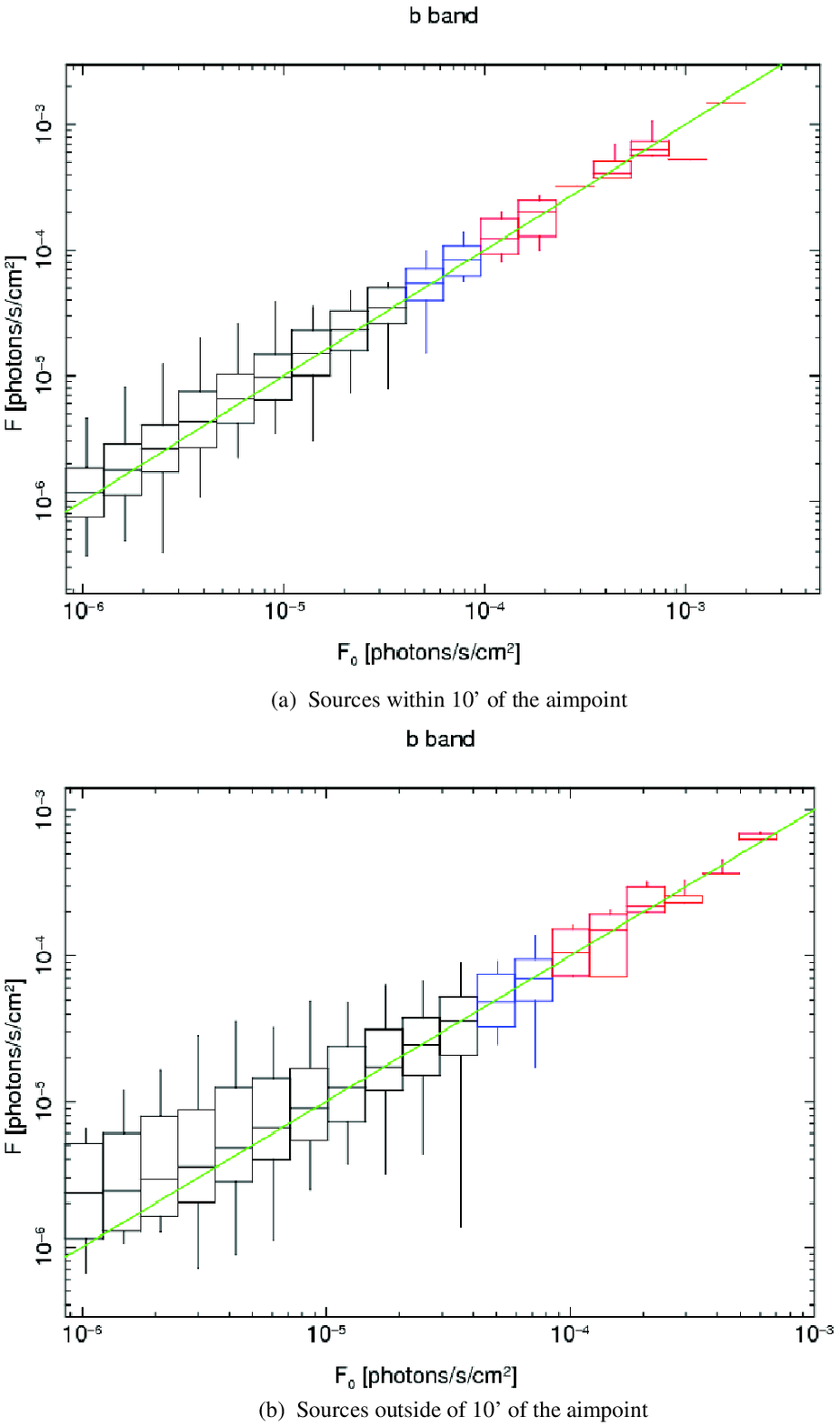}
\end{center}
\vspace{-0.2cm}
\caption{\label{bbflux} Comparison of input and measured \bband band fluxes
  for sources with blackbody spectra. Bins in red contain fewer than
  100 measurements; bins in blue contain 100-400 measurements; bins in
  black contain more than 400 measurements.}
\vspace{-0.2cm}
\end{figure}

\begin{figure}
\begin{center}
\includegraphics[width=3.5in]{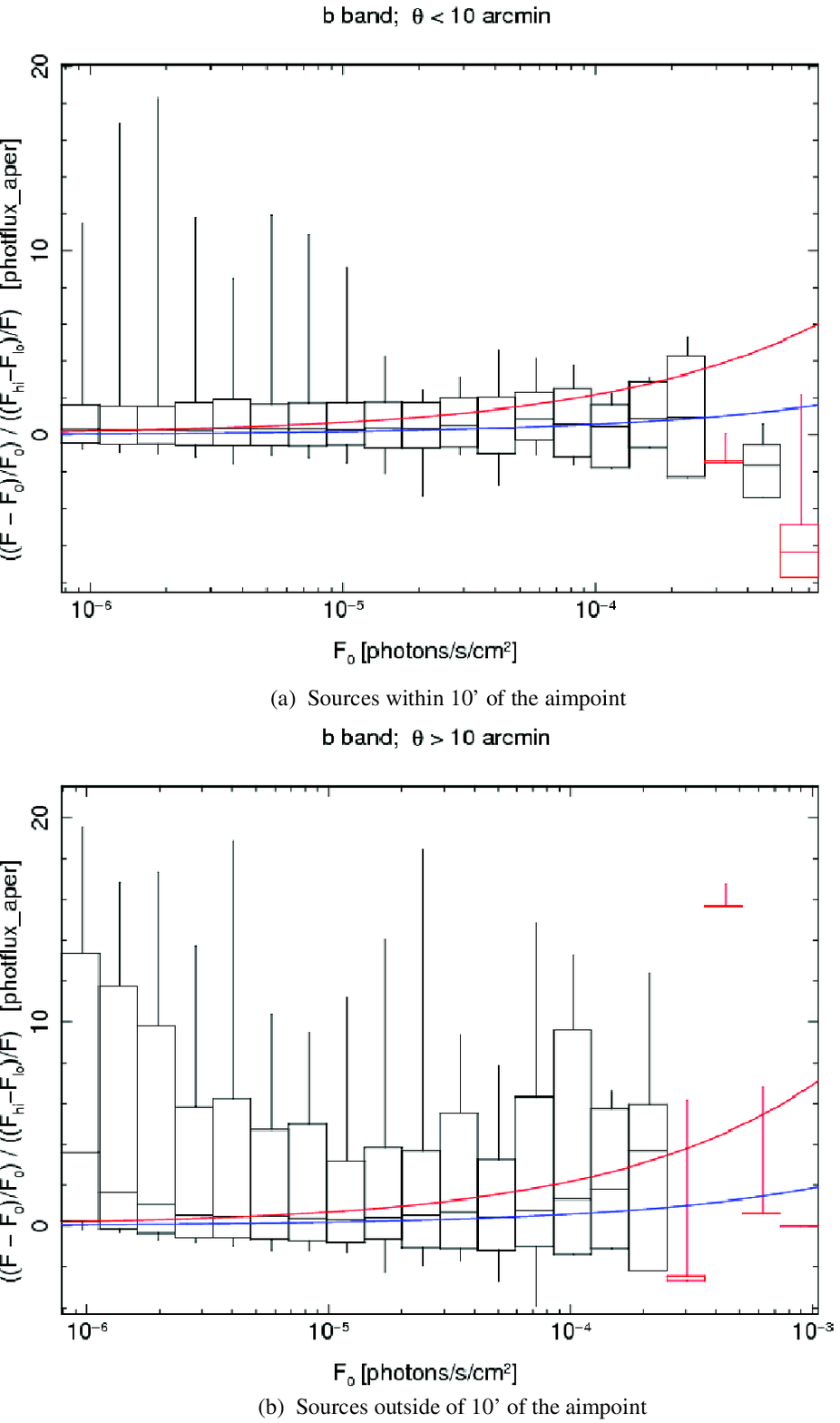}
\end{center}
\caption{\label{plnormb} Fractional difference between input and
  measured fluxes, normalized by measured fractional error, for
  sources with powerlaw spectra, in the \bband band. The smooth curves show
  the predicted systematic error for exposure times of 9\,ksec (blue,
  lower curve)
  and 125\,ksec (red, upper curve).}
\end{figure}

\begin{figure}
\begin{center}
\includegraphics[width=3.4in]{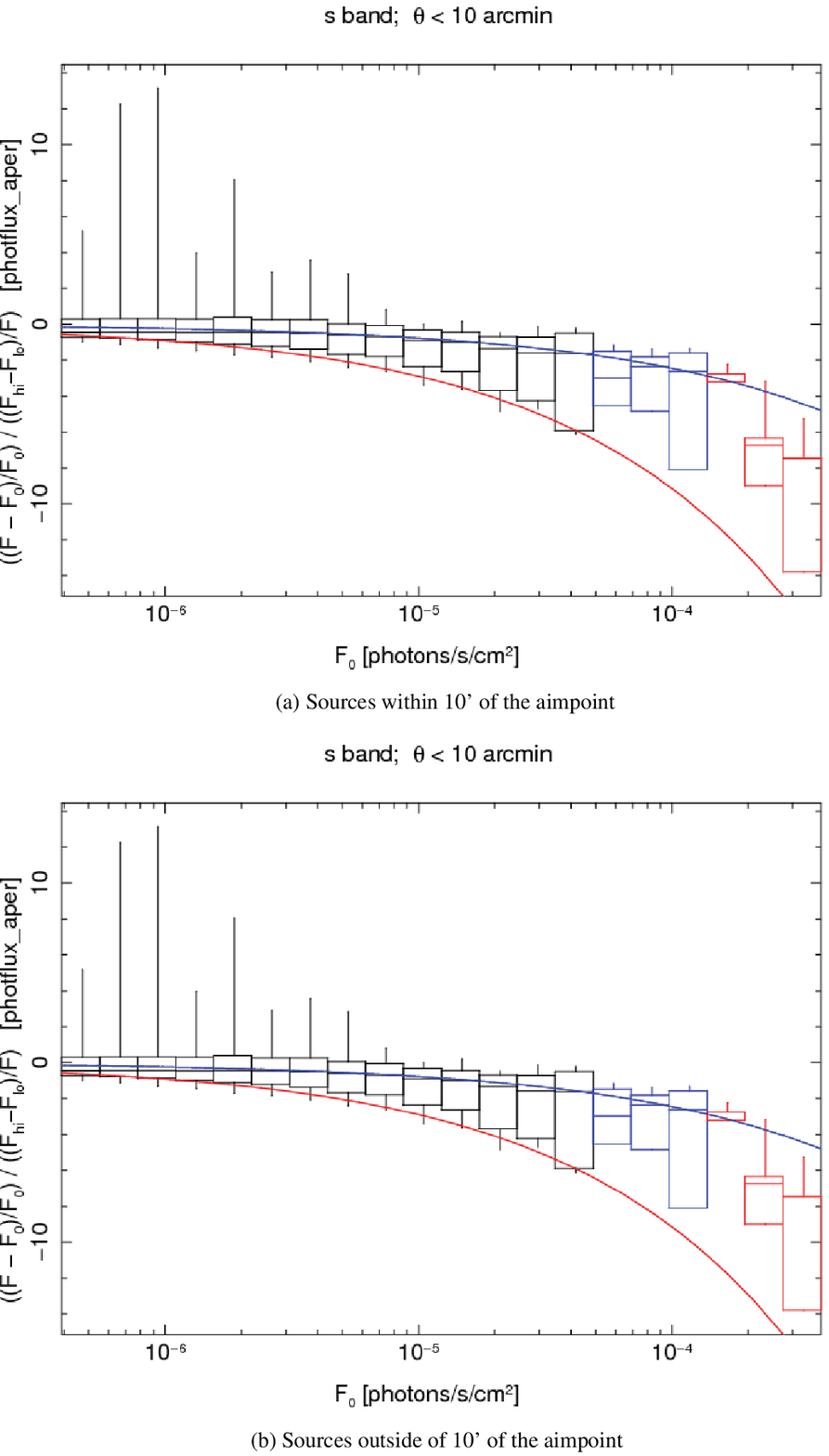}
\end{center}
\vspace{-0.2cm}
\caption{\label{plnorms}Fractional difference between input and
  measured fluxes, normalized by measured fractional error, for
  sources with powerlaw spectra, in the \sband band. The smooth curves show
  the predicted systematic error for exposure times of 9\,ksec (blue,
  upper curve)
  and 125\,ksec (red, lower curve). } 
\vspace{-0.2cm}
\end{figure}

\begin{figure*}
\begin{center}
\includegraphics[width=4.3in]{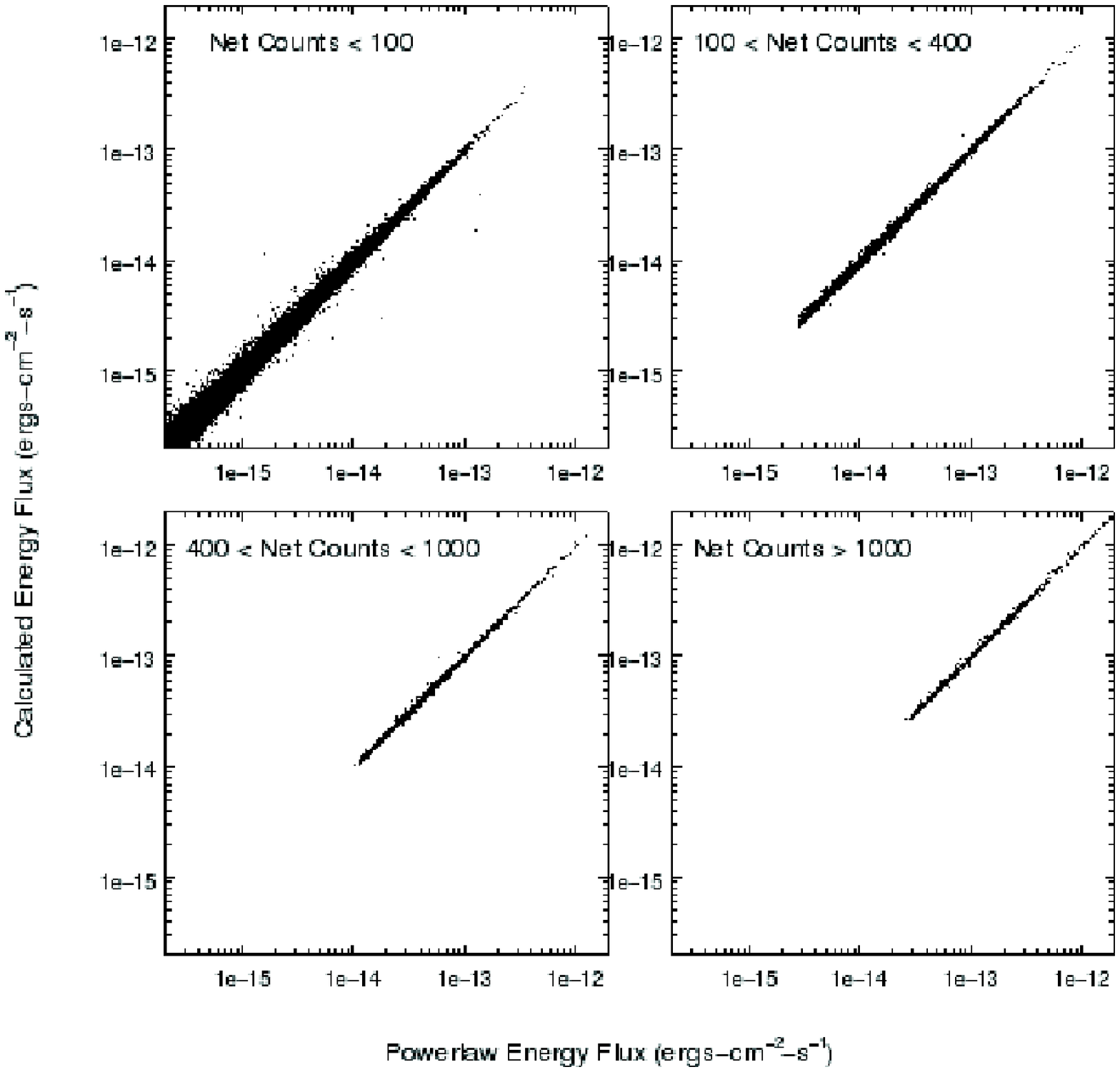}
\end{center}
\vspace*{-0.24in}
\caption{\label{flux_vs_cts_m} 
Comparison of energy fluxes calculated from individual event energies
and fluxes calculated assuming a powerlaw spectrum in the \mband band, for
sources with 4 different ranges of \mband band net counts.                                
}
\vspace*{-0.30in}
\end{figure*}

\begin{figure*}
\begin{center}
\includegraphics[width=4.3in]{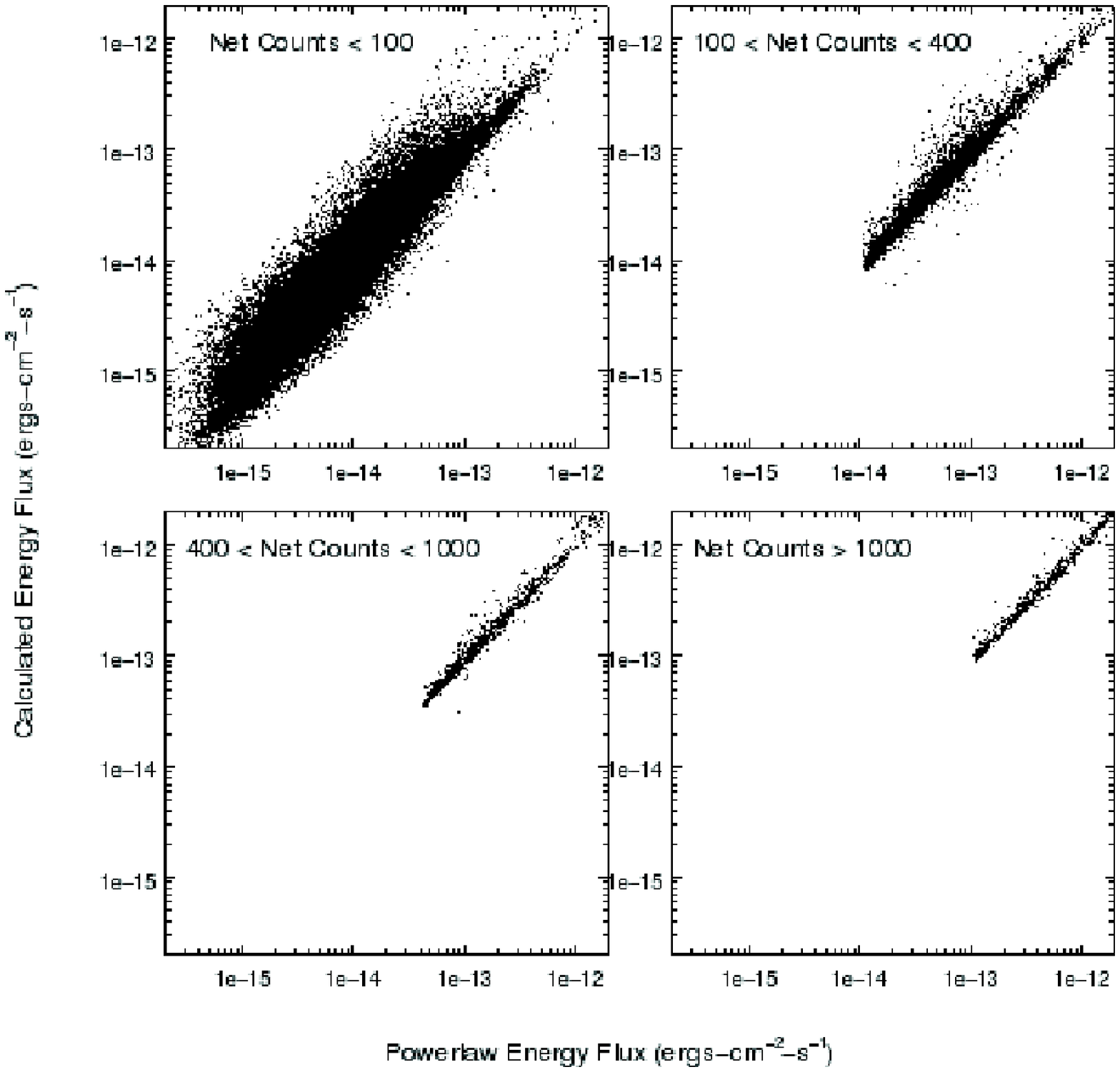}
\end{center}
\vspace*{-0.24in}
\caption{\label{flux_vs_cts_h} 
Comparison of energy fluxes calculated from individual event energies
and fluxes calculated assuming a powerlaw spectrum in the \hband band,
for sources with 4 different ranges of \hband band net counts.                               
}
\end{figure*}

\begin{figure*}
\begin{center}
\includegraphics[width=4.3in]{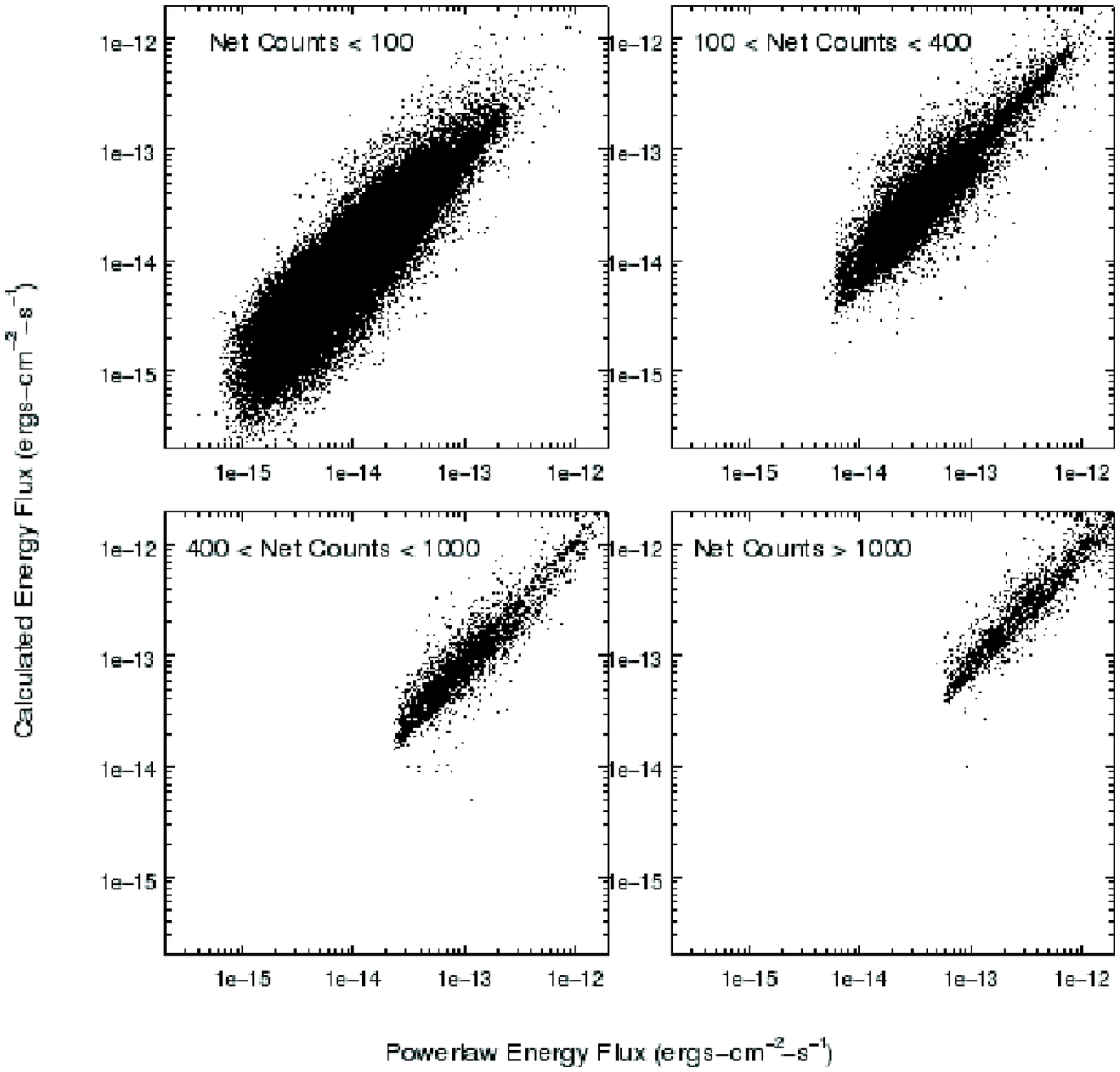}
\end{center}
\vspace*{-0.24in}
\caption{\label{flux_vs_cts_b} 
Comparison of energy fluxes calculated from individual event energies
and fluxes calculated assuming a powerlaw spectrum in the \bband band,
for sources with 4 different ranges of \bband band net counts.                               
}
\vspace*{-0.30in}
\end{figure*}

\begin{figure*}
\begin{center}
\includegraphics[width=4.3in]{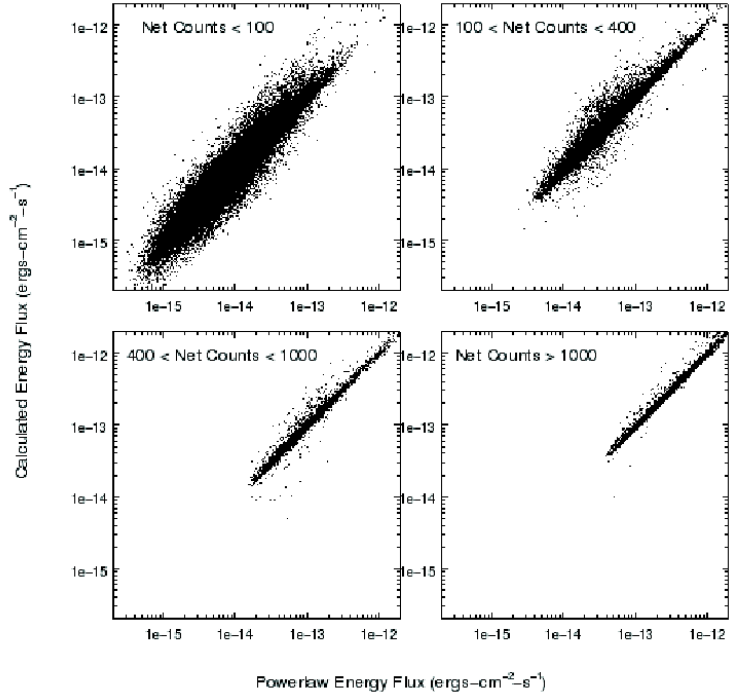}
\end{center}
\vspace*{-0.24in}
\caption{\label{flux_vs_cts_b_bandsum} 
Comparison of energy fluxes calculated from individual event energies
and fluxes calculated from the sum of the powerlaw spectrum fluxes in
the \sband, \mband, and \hband bands. 
}
\end{figure*}

\begin{figure}
\begin{center}
\epsscale{1.0}
\plotone{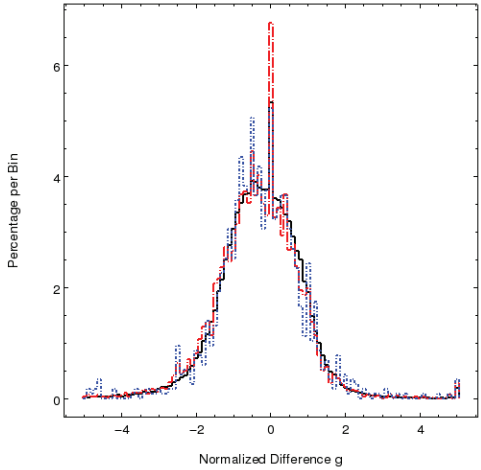}
\end{center}
\caption{\label{flux_norm_diff} 
Histogram of normalized differences between calculated and model
\hband band energy fluxes for source with \hband band net counts less
than 100 (black), between 100 and 400 (red, longdash) and between 400 and 1,000
(blue, shortdash). All histograms are normalized to sum to 100\%. 
}
\end{figure}

\begin{figure}
\epsscale{1.3}
\plottwo{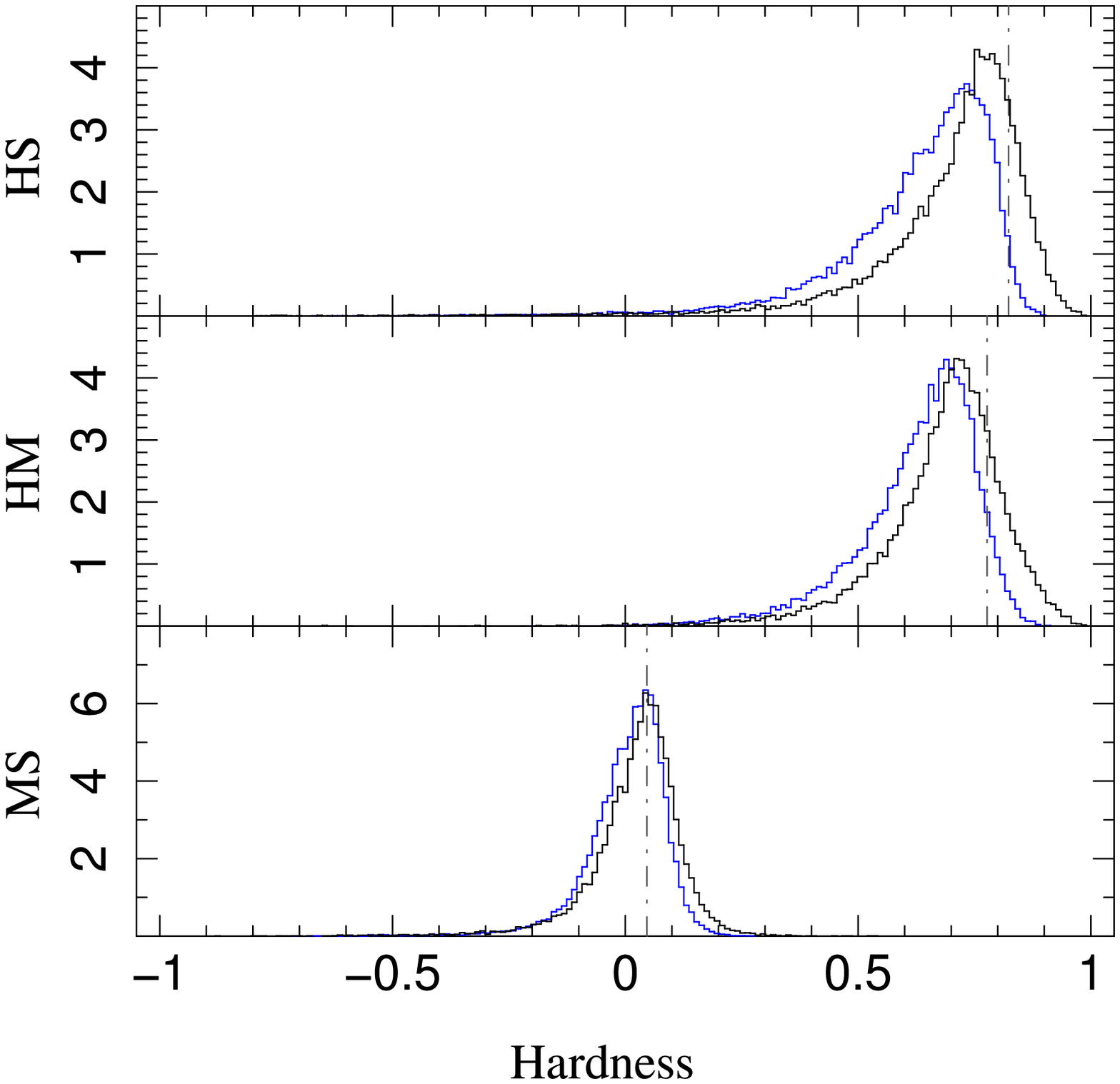}{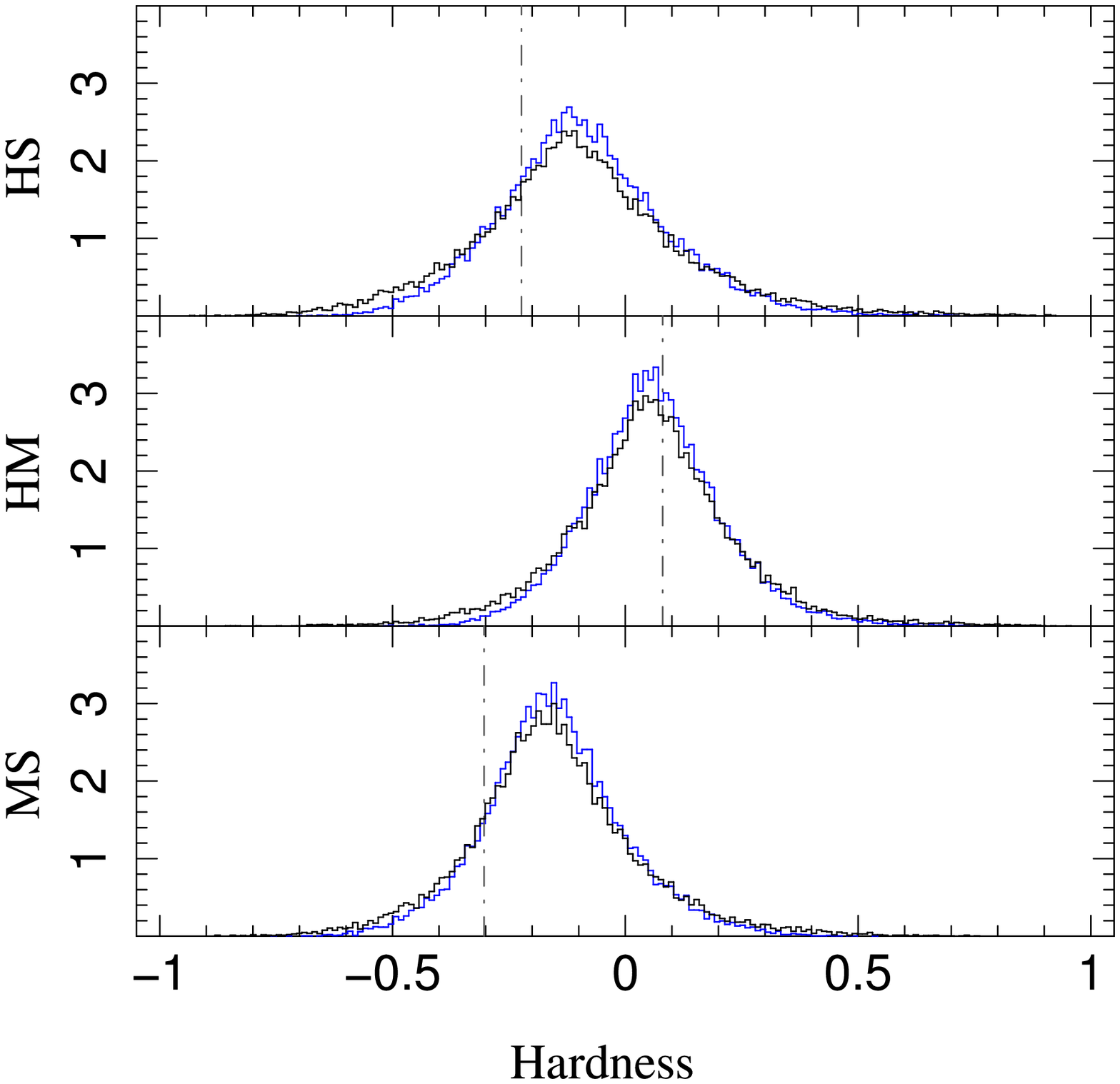}
\caption{Normalized histograms of catalog pipeline-derived hardnesses
  for simulated blackbody (top) and powerlaw 
  (bottom) sources.  HS represents the
  hard vs. soft bands, HM represents the hard vs. medium bands, and MS
  represents the medium vs. soft bands. Blue histograms are the
  hardnesses as calculated by the \CSCA implementation of the
  \protect{\citet{park:06a}} algorithm.  Black histograms are the
  hardnesses calculated from the catalog derived aperture photon
  fluxes.  The vertical lines are the theoretical source colors for
  the ideal input models (i.e., using true model fluxes in a given
  band, not monochromatic estimated fluxes).} \label{fig:color}
\end{figure}

\begin{figure*}
\centerline{ 
   \includegraphics[width=0.3\textwidth]{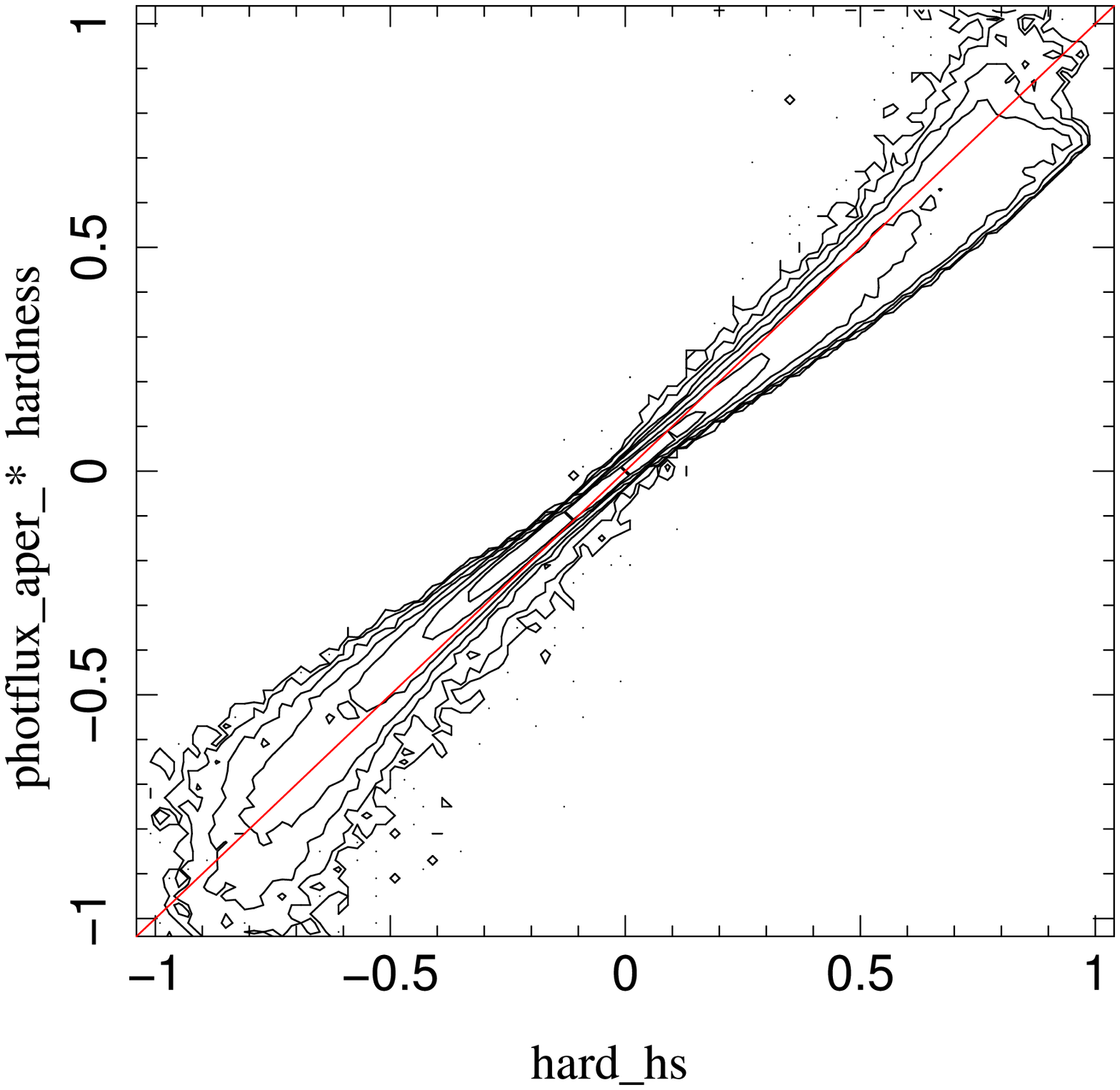} 
   \includegraphics[width=0.3\textwidth]{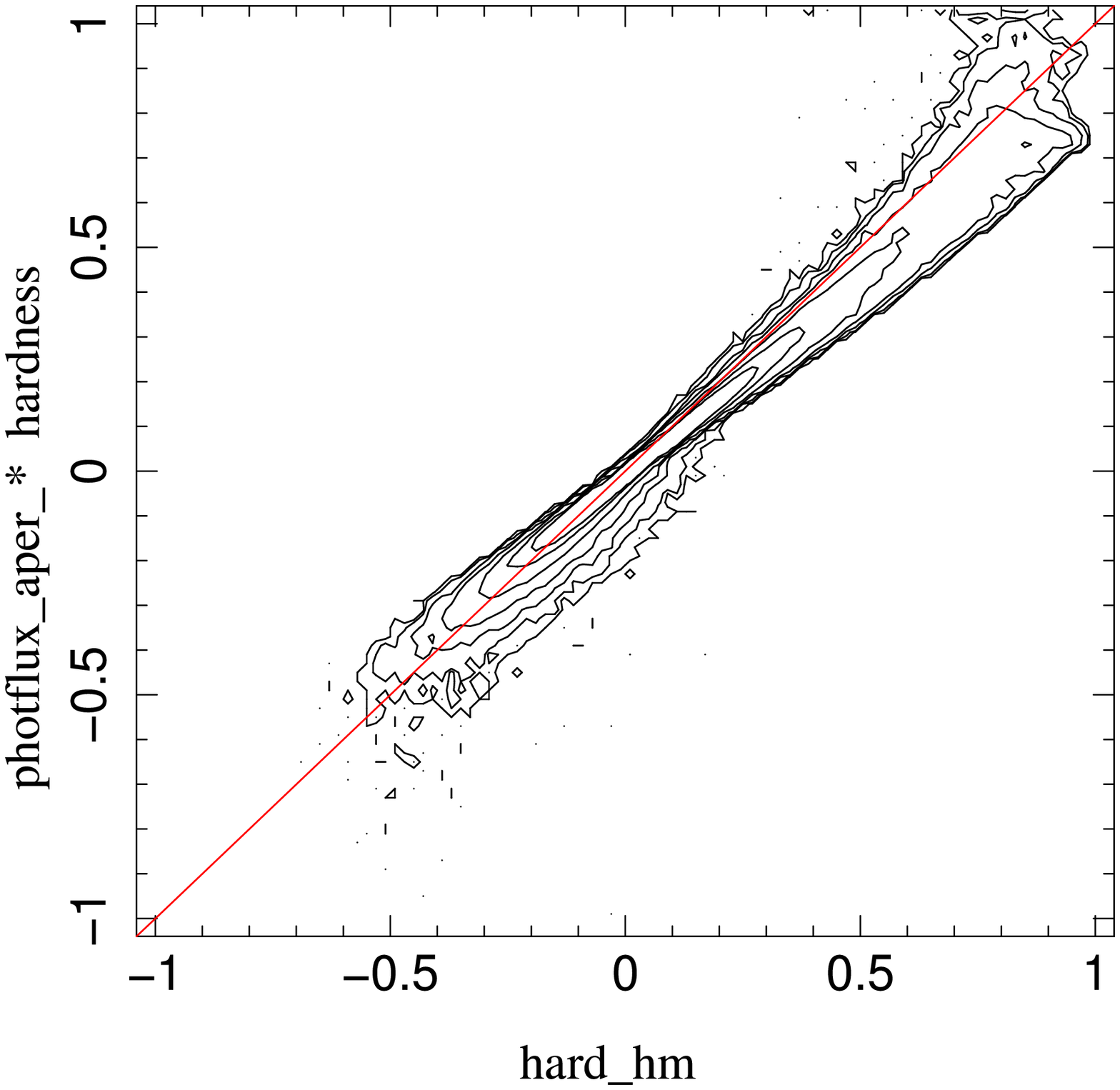} 
   \includegraphics[width=0.3\textwidth]{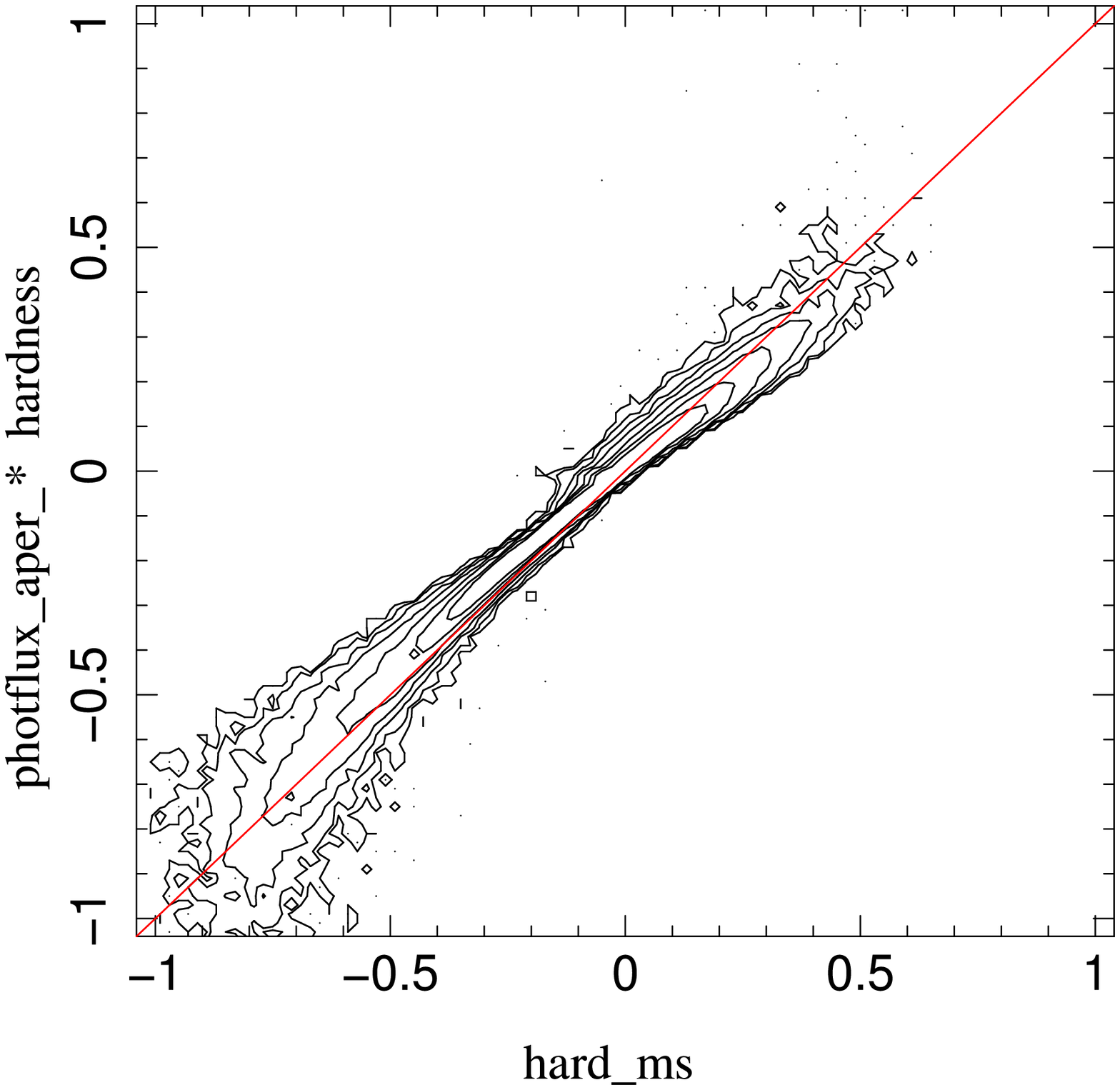} 
}
\caption{Contours derived from two dimensional histograms comparing
  the \CSCA calculated hardnesses (horizontal axes) to the hardness
  directly calculated from the aperture fluxes (vertical axes). The
  left figure is for the hard vs. soft channel, the middle figure is
  for the hard vs. medium channel, and the right figure is for the
  medium vs. soft channel.} \label{fig:cat_color}
\end{figure*}

\begin{figure}
\epsscale{1.23}
\plottwo{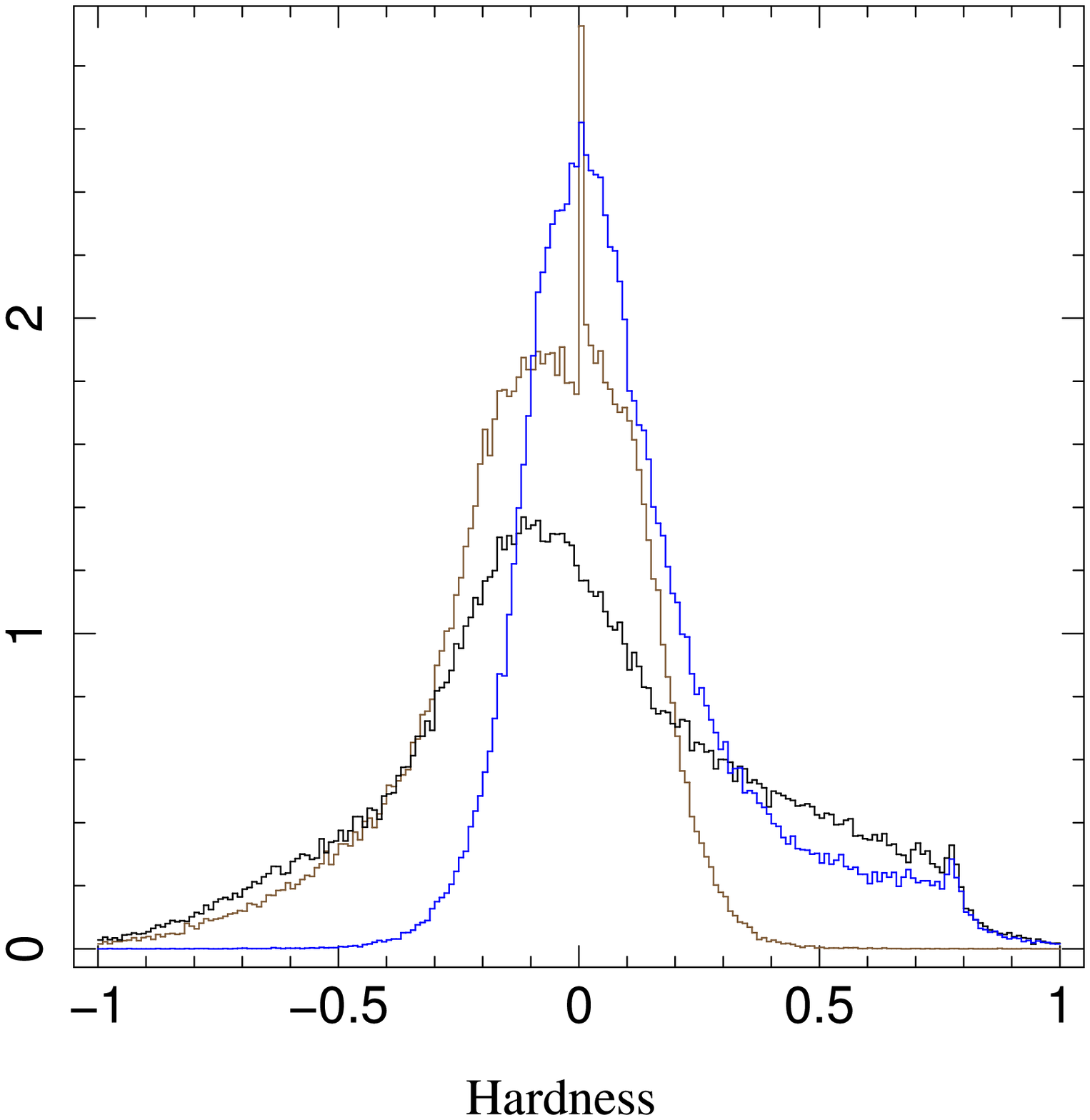}{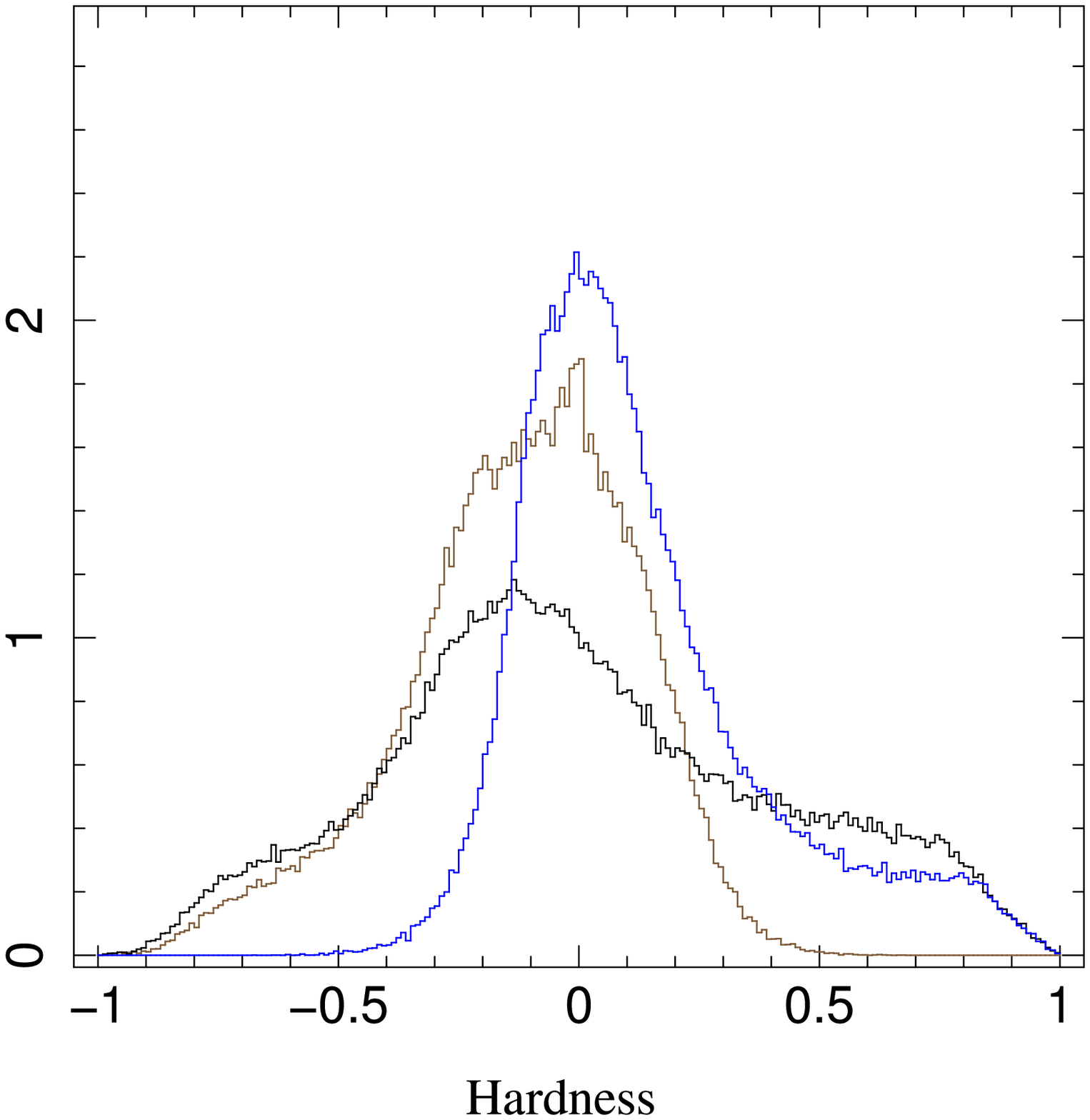} 
\vspace{-0.2cm}
\caption{Top: Normalized histograms of colors calculated directly
  from the aperture photon fluxes taken from the \CSCA v1.0.1.
  Bottom: Normalized histograms of the {\tt hard\_*} hardness values
  taken from the \CSCA v1.0.1 catalog.  For both figures, the brown
  histogram is for the medium vs. soft bands, the blue histogram is
  for the hard vs. medium bands, and the black histogram is for the
  hard vs. soft bands.} \label{fig:cat_flux_color}
\vspace{-0.2cm}
\end{figure}

\section{Photometry}
\label{sec:photometry}

To assess the accuracy of \CSC source fluxes, we compare the input and
measured fluxes of the simulated sources. We use fluxes derived from data in
\CSCA source regions ({\it photflux\_aper}). Fluxes derived from data in
regions  enclosing 90\% of the local point response functions  
({\it photflux\_aper90}) are, in general, similar.
Results for the powerlaw and
blackbody simulation sets are shown in Figs.~\ref{plflux} and
\ref{bbflux} for the \bband band and indicate good agreement for sources
within $10^\prime$ of the aimpoint. For sources beyond $10^\prime$,
there appears to be a systematic overestimate of a factor of $\sim2$
for sources fainter than $\sim3\times10^{-6}$ ph-cm$^{-2}$-s$^{-1}$.
We note, from Figs.~\ref{deteffacisi} and \ref{deteffaciss}, that
detection efficiency for this range of off-axis angle is low and
falling rapidly as flux decreases, and suggest that the flux
overestimates are the result of an Eddington bias
\citep{1940MNRAS.100..354E}, in which more sources with positive than
negative statistical fluctuations in counts are detected near
detection threshold. 
We have attempted to correct for
the bias using the technique of 
\citet{2009ApJS..180..102L}, but are able to account for only $\sim10-20\%$
of the overestimate using their Equation 3. We note, however, that we
use a different
likelihood 
function to explicitly account for
source contamination in 
background apertures (see Section~3.7 of \citet{evans09}). This may
account for the differences, although we cannot exclude the
possibility of other systematic errors. Additional
work is in progress to understand this effect.
  
We also examined the
fractional difference between input and measured fluxes $(F-F_0)/F_0$, 
normalized by
the fractional errors in measured fluxes, $(F_{hi}-F_{lo})/F$. Here,
$F_0$ and $F$ are the simulated and measured fluxes, and
$F_{lo}$ and $F_{hi}$ are the lower and upper confidence bounds for
the measured flux.
Representative plots of this
quantity are shown in Figs.~\ref{plnormb}--\ref{plnorms} and indicate
the presence of additional systematic errors at high flux limits,
even for
sources within $10^\prime$ of the aimpoint.
The effect is more prominent in the \sband band
(Fig.~\ref{plnorms}). 

Preliminary analysis indicates the effect is due to the
assumption of a monochromatic exposure map in computing source
fluxes. This assumption can lead to systematic errors
because it ignores the energy dependence of the telescope response.
The size of the systematic error depends on both the telescope response and the
shape of the incident spectrum, $S(E)$.  For example, in the limit of perfect
background subtraction in spectral band $X$, the ratio of the estimated photon
flux, $F$, to the true photon flux, $F_0$, in that band is
\begin{equation}
\phi_X \equiv \left. \frac{F}{F_0} \right|_X =
  \frac{(A(\overline{E})T)^{-1}\sum_{h \in X}C(h)}{\int_{X} S(E)\,dE},
\label{eq:flux-syserr}
\end{equation}
where the number of counts in each narrow pulse-height bin is
\begin{equation}
  C(h) \equiv T \int_{\Delta E_h} R(h,E)A(E)S(E)\,dE,
\label{eq:model-counts}
\end{equation}
$R(h,E)$ is the redistribution matrix, $T$ is the exposure time, $A(E)$ is the effective area, and
$A(\overline{E})$ is the effective area at energy $\overline{E}$ used to estimate the photon flux in the
band of interest (which includes $\overline{E}$).  In equation \ref{eq:flux-syserr}, the integral in the
denominator spans the incident photon energies, $E \in X$, while the integral
in the equation \ref{eq:model-counts} spans all incident photon energies that
contribute counts to the narrow pulse height bin, $E \in \Delta E_h$ .

To estimate the size of the systematic error defined by equation
\ref{eq:flux-syserr}, we selected from CSC release 1.1 the response
functions for 282 catalog sources
with \verb|flux_significance_b|$>5$ in the obsids listed in Table 1. 
These obsids were observed between May 2000 and July 2006 and
represent a reasonable sample of the time-dependent ACIS detector
contamination in the \CSCA.
For each
source in this arbitrary sample, we computed $\phi_X$ in each band for
both the powerlaw and blackbody spectral models from \S 9, using the
\CSCA -archived response functions.  Within this sample, the systematic
errors from the \mband and 
\hband bands have no significant time dependence because those bands are relatively
unaffected by the increasing amount of detector contamination; for this
sample, $\phi_m =0.94-1.04$ and $\phi_h=0.79-0.90$ for both powerlaw and
blackbody spectra. The increasing detector contamination has a more noticeable
effect on the s- and b-bands, introducing a weak time-dependence within the
range $\phi_s=0.62-0.78$, $\phi_b=0.90-1.25$ for powerlaw sources and
$\phi_s=0.90-1.0$, $\phi_b=1.12-1.28$ for blackbody sources.  Flux
measurements in the u-band are subject to large systematic errors for some
spectral shapes; for the powerlaw spectrum, $\phi_u=0.80-2.4$, but for the
blackbody spectrum, $\phi_u=1-25$.

The smooth curves  in Figs.~\ref{plnormb}--\ref{plnorms}  illustrate
the effect as a function of $F_0$. To generate these curves we used
the \isis~{\it fakeit} command to simulate noise- and background-free
powerlaw spectra 
for a range of $F_0$ and exposure times of 9 and 125 ksec, using
canonical {\it Chandra} response functions. From these spectra we
computed counts in the \bband and \sband bands, and their
``statistical'' ($\sqrt n$) errors and converted to ``measured'' flux
and flux errors by dividing by exposure and $A(\overline{E})$ for the
band. Although the resulting curves ignore contributions due to
background subtraction and variations in {\it Chandra} response
functions with time and detector, they do reproduce the general
behavior of the observed values and add confidence to our explanation
for the systematic errors at high fluxes. 

As \citet{evans09} note, the method of calculating CSC energy fluxes
by applying quantum efficiency and effective area corrections to
individual event energies can be inaccurate for sources with few
counts in energy bands where the \chandra effective area is small and
changing rapidly. We have investigated this effect by comparing the
energy fluxes calculated in this fashion with model fluxes calculated
assuming our canonical power-law spectrum. Our results are shown in
Figs.~\ref{flux_vs_cts_m} and \ref{flux_vs_cts_h}, respectively, and
indicate good agreement for \mband band fluxes for all sources, but
considerable scatter for sources with fewer than 100 counts in the
\hband band. Results for the \sband and \uband bands are similar to
those in the \hband band. For the \bband band, as indicated in
Fig.~\ref{flux_vs_cts_b}, the fluxes show appreciable scatter even for
sources with more than 100 net counts. We attribute this to the fact
that some source spectra cannot be adequately approximated by a single
power law in the \bband band. We note that when we compare calculated
\bband band fluxes to the sum of powerlaw fluxes in the \sband,
\mband, and \hband bands, the scatter is significantly reduced (see
Fig.~\ref{flux_vs_cts_b_bandsum}). 

To quantify our results, we compute a normalized difference
\begin{equation}
g = (f - p)/\sigma
\end{equation}
where $f$ is the energy flux calculated from individual event energies
and effective areas, $p$ is the flux calculated using our canonical
powerlaw spectrum, and $\sigma$ is defined as: 
\begin{equation}
\sigma = \left\{ \begin{array}{rl}
 f - f_{lo} &\mbox{ if $f \ge p$} \\
  f_{hi} - f &\mbox{otherwise}
       \end{array} \right.
\end{equation}
Here, $f_{lo}$ and $f_{hi}$ are the lower and upper bounds for the $1
\sigma$ credible region for f\footnote{The bounds are determined using
Bayesian methodology \citep{evans09} and hence define a ``credible
region'' in the terminology of Bayesian statistics.}. In Fig.~\ref{flux_norm_diff}, we show
histograms of $g$ for \hband band fluxes in three separate ranges of
net \hband band counts. In all three histograms, the  percentage of
sources with $|g|\le2$ is $\sim90$\%, compared with an expected
$\sim95$\% for a Gaussian distribution. 

Finally, we consider sources with zero counts or only an upper limit
to the flux in one of the narrow bands. We examined events in the
source regions of ~7,000 discrepant sources with fewer than 20 counts,  
extracting the highest-flux photon in the broad band. For only
$\sim10$\% of these sources did this photon contribute more than
$\sim$50\% of the total energy flux in the band; $\sim$3\% percent had
a single photon with $\sim$80\% of the flux. This corresponds to only
$\sim$0.2\% of the entire catalog. The effect is reduced even further
when  background is accounted for. In several of the cases that we
investigated in detail, the highest flux photon was actually
compensated by a large subtracted background flux in that energy
band. We conclude that  $\sim5$\% of \CSCA sources may have
underestimated energy fluxes or errors, but the number of cases in
which a combination of a single photon and low background yield
egregious flux estimates is negligible. 

\section{Hardness Ratios and Colors}
\label{hardness}

The \CSC defines source hardness ratios that are meant to reflect the
ratios of the aperture source photon fluxes ({\tt photflux\_aper\_*}, in
terms of the source properties columns).  That is, in the high
statistics limit, the source hardnesses are of the form
\begin{equation}
  H_{xy} = \frac{ F^\gamma_x - F^\gamma_y }{F^\gamma_b} ~~,
\end{equation}
where $F^\gamma_x$ is the aperture photon flux in band $x$,
$F^\gamma_y$ is the aperture photon flux in band $y$, and $F^\gamma_b$
is the aperture flux in the broad band\footnote{Note that Table 1 of
  \citet{evans09} incorrectly states that the hardness ratios are
  calculated from energy fluxes.  The description within the text of
  \citet{evans09}, and that given here, based upon estimated photon
  fluxes is in fact the definition used in the catalog.}.  The concept
behind the colors reflecting the values of the aperture photon fluxes
is to partially normalize out variations induced by spatially and
temporally dependent detector responses.  Chief among these
dependencies are the differing soft X-ray responses between the
frontside and backside illuminated \acis CCDs, as well as the time-
and position-dependent \acis contamination that has led to a decrease
of the soft X-ray effective area over the lifetime of the mission.  By
using hardnesses related to aperture photon flux rather than solely
counts or count rate, it is hoped that sources with the same intrinsic
colors will yield similar estimated hardnesses regardless of observing
epoch or detector position. Note that also as defined above, we expect
hardnesses to be bounded between $-1$ and $1$.

In reality, the source hardnesses are calculated from the \emph{total}
counts (source plus background) in the aperture source region, the
\emph{total} counts in the background region, and scaling factors to
convert from net source counts in the source region to aperture photon
flux.  The intrinsic hardness to be estimated is defined as
\begin{equation}
H^i_{xy} ~\equiv~ \frac{f_x x_i - f_y y_i}{f_s s_i + f_m m_i + f_h h_i} ~~,
\label{eq:hard}
\end{equation} 
where $x_i$, $y_i$, are the intrinsic \emph{source} counts in bands
$x$ and $y$, i.e., the soft, \sband, medium, \mband, or hard, \hband
bands, and the broad band in this case is the sum of the individual
bands\footnote{This is to be contrasted to the broad band flux being
  derived separately from the defined broad band source
  properties. For example, the broad band has its own monochromatic
  conversion factor from net broad band counts to broad band photon
  flux.}.  The factors $f_*$ are the conversion factors to transform
from net source counts in the source region to source photon flux.
These factors incorporate estimates of the detector effective area and
exposure time in the given band, as well as the fraction of the point
spread function within the source region.

The detected total counts will include a contribution from background
counts that must be estimated.  Furthermore, given the excellent
sensitivity of \chandra to extremely faint sources, many faint \CSCA
sources have zero net counts in one or two bands.  The catalog
estimates of hardnesses must account for these effects. To this end,
the \CSCA employs an implementation of the Bayesian algorithm of
\citet{park:06a}.  This algorithm, derived by considering the Poisson
nature of the detected counts in both the source and background
regions, is designed to be applicable even when no counts are detected
in a given band.  Furthermore, it is designed to yield a probability
distribution for the hardness ratio that is properly bounded between
$-1$ and $1$.  Confidence limits are derived from this probability
distribution, and thus never exceed an absolute value of 1.  (This
would not be guaranteed to be true if the hardnesses were determined,
for example, by a Gaussian statistics approximation.)

To assess the success of the \CSCA implementation of the
\citet{park:06a} algorithm, we have compared the calculated hardnesses
for the simulated blackbody and powerlaw sources described in Section
\ref{sec:source_detection} to both the ideal expectations based upon
the model input spectra, as well as to hardnesses directly calculated
from the catalog aperture photon fluxes.  These results are presented
in Fig.~\ref{fig:color}.  As can be seen from these figures, whereas
the distribution of estimated hardnesses peak near the ideal model
input hardnesses, there are biases in the hardness.  Furthermore,
these biases have the opposite sense for the blackbody vs. the
powerlaw simulated spectra.  The blackbody spectra are biased towards
calculated colors that are too soft for hardnesses involving the hard
channel.  Conversely, the powerlaw spectra are biased towards
calculated colors that are too hard for hardnesses involving the soft
channel.

We have previously noted the biases in the estimated photon fluxes in
Section \ref{sec:photometry}, and they have also been described in
Section 2.5.2 of \citet{evans09}.  These biases predominantly arise from the
assumption of a monochromatic energy band when computing the
conversion factor from counts to photon flux.  The form of
eq.~(\ref{eq:hard}), however, requires such a single conversion factor
in each band, in contrast to a conversion factor \emph{per event} as
is used in the calculation of the aperture energy fluxes.  In general
we expect that the fidelity between the ``true'' hardness and the
estimated hardness will be spectrum and possibly detector-dependent.

The simulations show, however, that although the colors are biased,
there is a very good agreement between hardness estimates whether they
are taken from the catalog pipeline or whether they are calculated
directly from the aperture photon fluxes.  When looking at the results
for the \CSCA as a whole, we find for the actual sources in the v1.0.1
catalog that this overall agreement between hardnesses derived from
these two methods holds.  In Fig.~\ref{fig:cat_color} we plot contours
of 2-D histograms comparing the \CSCA results for these two estimates.
The contours are tightly gathered around a unity-correspondence.  This
opens up the possibility for a catalog user to calculate the expected
bias in the hardnesses from a hypothesized spectrum in a few test
cases, and then using these calculated biases to inform an acceptable
set of hardness filtering criteria.

\begin{figure}
\epsscale{0.98}
\plotone{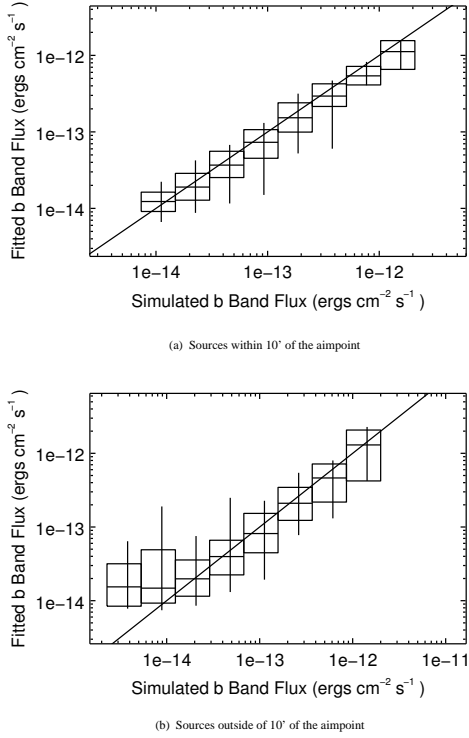}
\caption{\label{plfitflux} Comparison of input and fitted \bband band energy fluxes
  for sources with simulated power-law spectra.}
\end{figure}

\begin{figure}
\epsscale{0.98}
\plotone{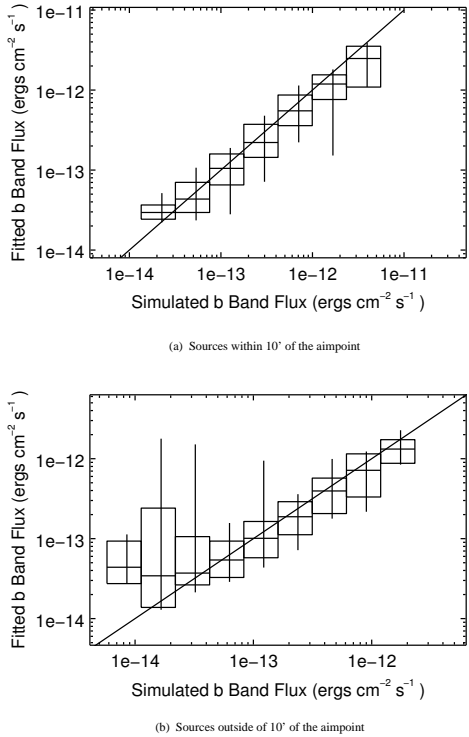}
\caption{\label{bbfitflux} Comparison of input and fitted \bband band energy fluxes
  for sources with simulated blackbody spectra.}
\end{figure}

\begin{figure}
\epsscale{1.0}
\begin{center}
\plotone{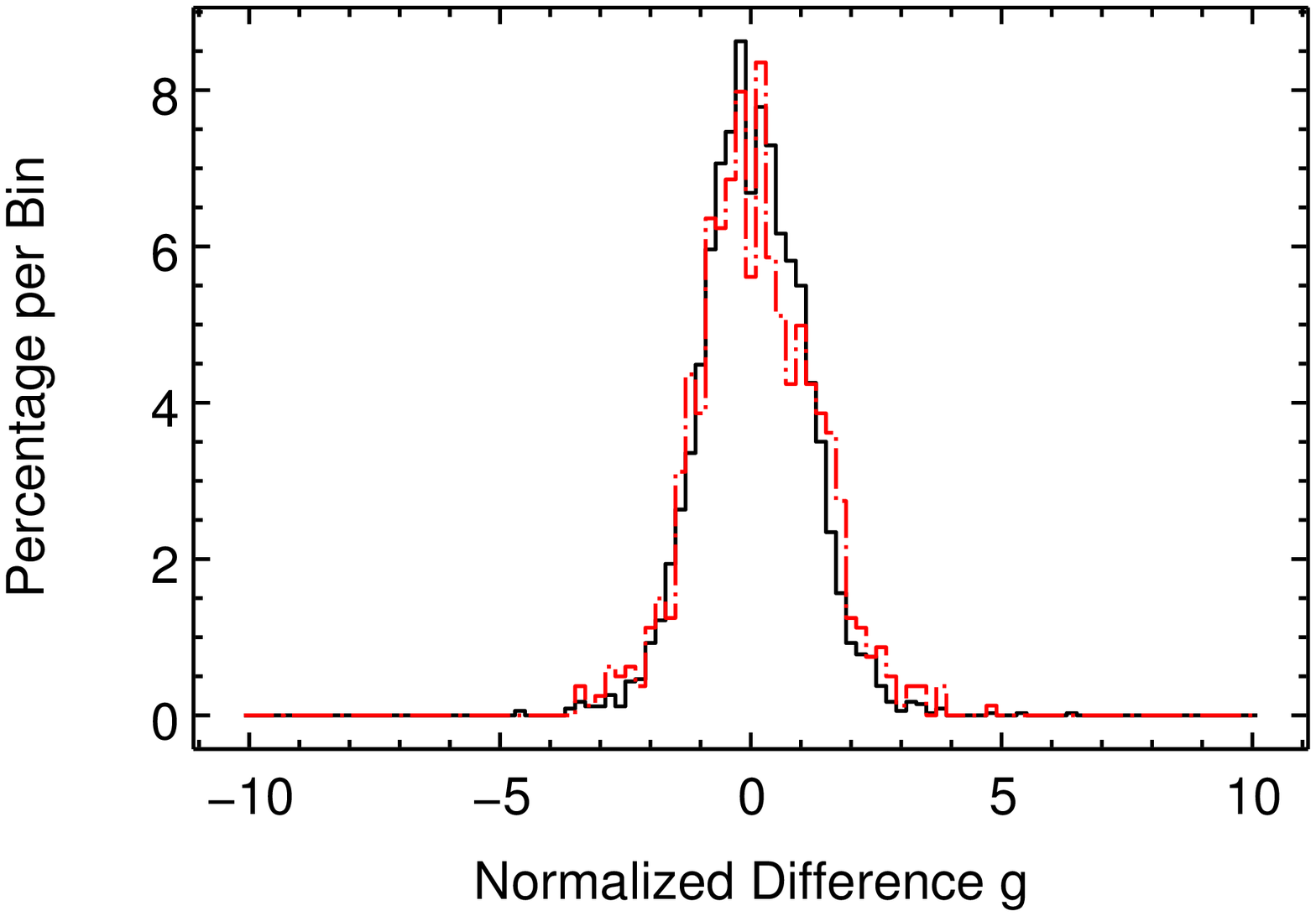}
\vspace{-0.2cm}
\plotone{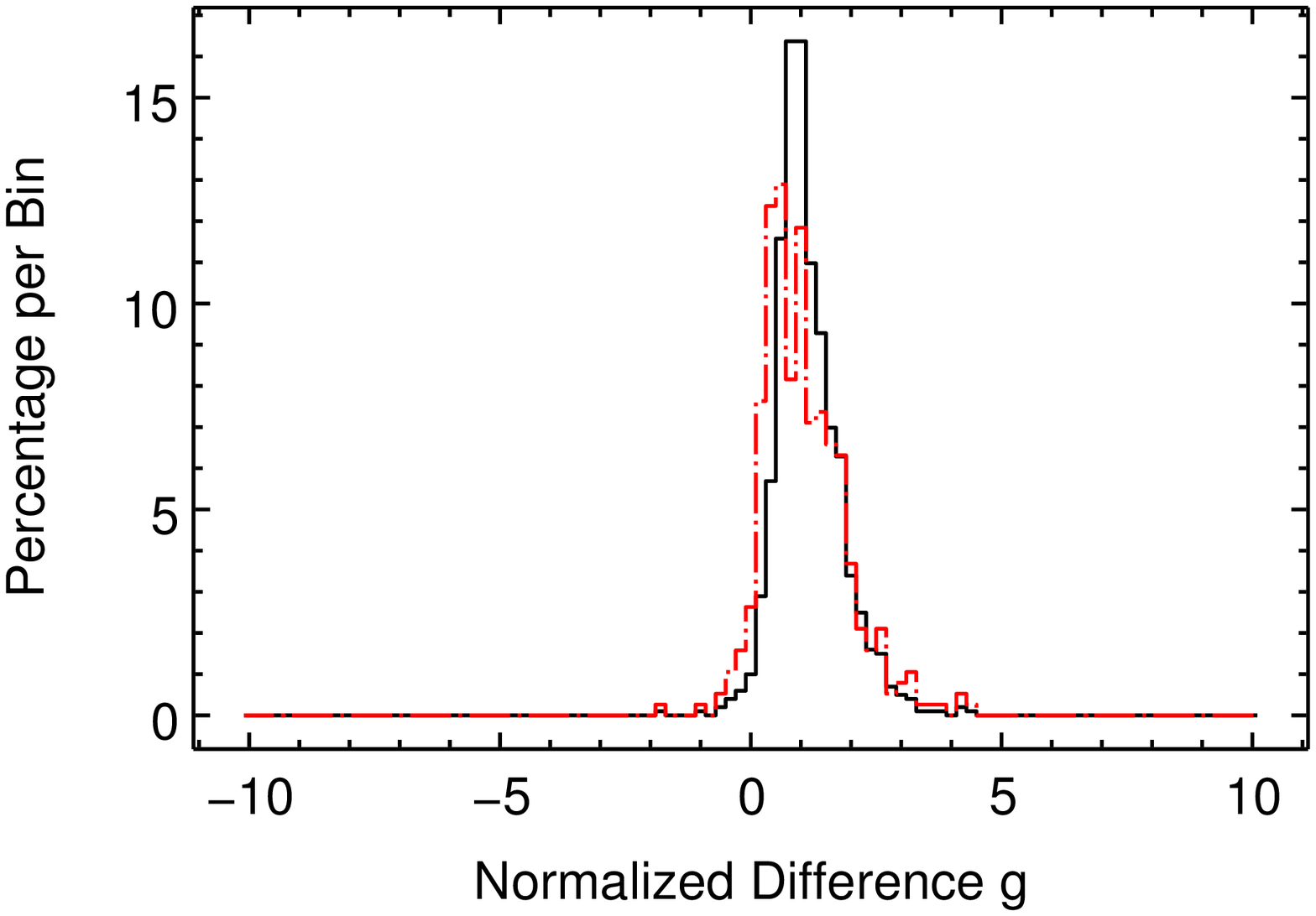}
\end{center}
\caption{\label{plfitparams} Distribution of normalized differences between
  fitted and simulated spectral parameters for sources with more than
  150 (black) and 500 (red, dashed) net \bband band counts:
 (top) power-law slope for 3,455 sources (black) and 802 sources (red, dashed);
 (bottom) $N_H$ for 1,002 sources (black) and 380 sources (red, dashed).}
\end{figure}

In Fig.~\ref{fig:cat_flux_color} we show further results for real
catalog sources, both when defining the colors via the aperture photon
fluxes and as calculated via the application of the \citet{park:06a}
algorithm.  The catalog hardness histograms have peaks comparable to
those of the powerlaw simulations, albeit with histogram tails that
extend to both harder and softer colors.  For hardnesses calculated
directly from aperture photon fluxes, both the medium vs. soft
histogram and the hard vs. medium histogram have local peaks at a
hardness ratio of 0.  These peaks are due to sources that were
detected in only the hard band, or only in the soft band,
respectively.  As the Bayesian algorithm of \citet{park:06a} is
specifically designed to properly handle cases with zero counts in a
given band, these local peaks are smoothed out when applying this
algorithm, as can be seen in Fig.~\ref{fig:cat_flux_color}.

\section{Spectral Fits}
\label{specfits}
For sources with more than 150 net counts in the \bband band, the \CSC attempts to
fit the observed counts spectrum with both absorbed power-law and absorbed
blackbody spectral models. We use the simulated spectra provided as part of
our point-source simulations to characterize the results of \CSCA model spectral
fits. We compare integrated \bband band model fluxes with input \bband band fluxes,
using a subset of simulated sources for which aperture photometry yields more than 150 b
band counts ({\it src\_cnts\_aper\_b}), and for which successful spectral model fits were
obtained . A total of 3,455 sources were used for power-law fits, and 2,897
sources for blackbody fits. Since the \CSCA reports integrated model fluxes as energy fluxes,
we convert input simulated photon fluxes to energy fluxes using the known
spectral parameters described in Section~\ref{pointsourcesims}. We used
conversion factors of $2.81\times10^{-9}$ and $6.64\times10^{-9}$ ergs~
photon$^{-1}$ for power-law and blackbody spectra, respectively. Our results
are shown in Figs.~\ref{plfitflux} and \ref{bbfitflux}, and are in
general similar to the results shown in Figs.~\ref{plflux} and \ref{bbflux},
albeit with many fewer sources. In particular, the systematic flux
overestimate for faint sources ($<\sim1-2\times10^{-14}$erg~cm$^{-2}$~s$^{-1}$) at
large off-axis angle is evident in the spectral model fits as well.

\begin{figure}
\epsscale{1.0}
\plotone{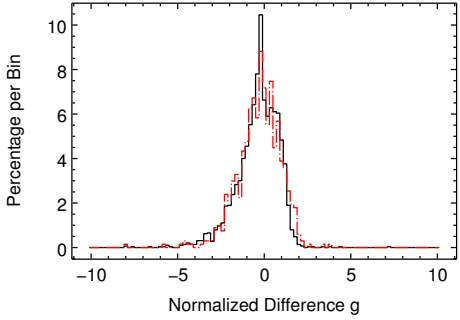}
\caption{\label{bbfitparams} Distribution of normalized differences between
  fitted values and simulated values for blackbody temperature $kT$. Black
  histograms refer to the entire sample of 2,897 sources. Red dashed histograms refer
  to the restricted sample of 669 sources with more than 500 net \bband source counts.}
\end{figure}

We compare fitted spectral parameters $\Gamma$, $kT$, and $N_{H}$, to input
spectral parameters for the corresponding model simulations, using normalized differences like
those defined in Section~\ref{sec:photometry}; we define $f=\Gamma_{fit}$ and $p=1.7$
for $\Gamma=1.7$ power-law spectra, $f=kT_{fit}$ and $p=3.0$ for $kT=3.0$ blackbody spectra,
and $f=N_{H,fit}$ and $p=3.0\times10^{20}$~cm$^{-2}$ for $N_{H}$ for both
models. Our results are shown in Figs.~\ref{plfitparams} and
\ref{bbfitparams}. For power-law fits, we find a median $\Gamma$ of 1.724 for
the 3,455 sources in our sample, with $\sim96$\% with normalized difference
$|g|<2$. If we restrict the sample to sources with more than 500 net counts, 
we find a median $\Gamma$ of 1.718 for
the 802 sources in the sample, with $\sim93$\% with $|g|<2$.  
For blackbody
fits, we find a median $kT = 2.90$~keV for 2,897 sources with more than 150 net counts,
and a median $kT = 2.96$~keV for 669 sources with more than 500 net counts. In
both cases, $\sim92$\% had $|g|<2$. We note that for both power-law and
blackbody models, the fitted spectra are slightly softer than the input
spectra. This result is expected, since no energy-dependent aperture
corrections are performed in spectral model fits. For the power-law fits, the
median values of $\Gamma$ are consistent with the softening of $0.03-0.05$ in
spectral index estimated in Section~3.9 of \citet{evans09}. 

For sources with simulated power-law spectra, fits converged to valid
values of both $N_H$ and its lower confidence bound 
for only 1,002 sources in the full
sample and for only 380 sources in the higher net count sample. For the
remainder of the sources, the fitting procedure encountered the lower
bound of the search region for $N_H$ ($1.0\times10^{15}$~cm$^{-2}$)
before encountering either the best-fit value or the lower
confidence bound. In many cases, neither were included in the parameter search region.  
We excluded these sources from analysis of the $N_H$ distributions. The resulting
distributions were skewed for both net count samples, as shown in panel (b)
of Fig.~\ref{plfitparams}. For the full sample, the median 
$N_H$~=~$1.2\times10^{21}$~cm$^{-2}$ with $\sim92$\% having $|g|<2$.
For the higher net count sample, the median $N_H=6.7\times10^{20}$~cm$^{-2}$ with
$\sim90$\% having $|g|<2$. We note that most ($\sim95\%$) sources in
the full sample had fewer than 1000 net counts and conclude that $N_H$
is poorly determined in the
\CSCA fits in this count range.
We do not cite a
result for $N_H$ for sources with simulated black-body spectra since most fits
were unable to converge to valid 
best-fit values or confidence bounds in the
range of parameter space used in the fitting routines. We attribute the
additional insensitivity of the fitting statistic to $N_H$ to
the relatively high temperature of 3 keV used to simulate the blackbody
spectra.

\begin{figure}
\begin{center}
\includegraphics[angle=90.0,width=0.4\textwidth]{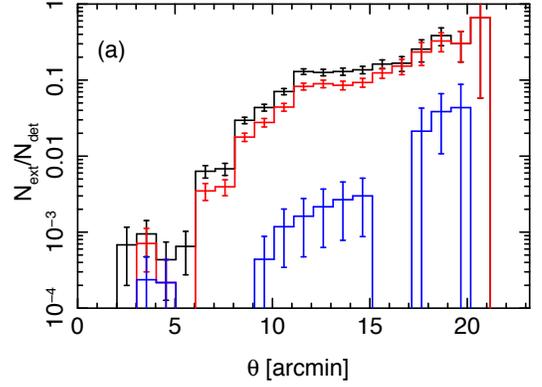}
\includegraphics[angle=90.0,width=0.4\textwidth]{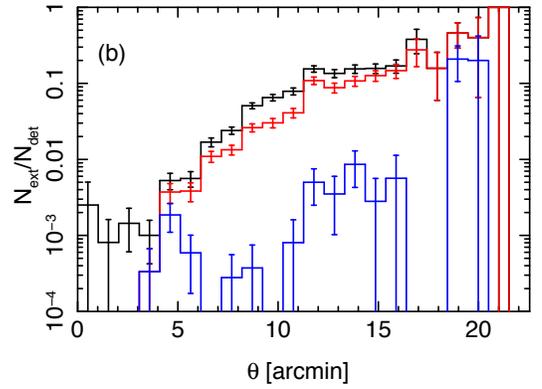}
\end{center}
\caption{\label{extentfig}
Fraction of simulated (a) powerlaw and (b) blackbody
point sources erroneously marked as extended in the \bband band
as a function of off-axis angle, $\theta$,
The black (top) histogram includes sources with
{\tt (extent\_code\&0x10) != 0}.
The red (middle) histogram includes sources with
{\tt (extent\_code\&0x10) != 0}, {\tt pileup\_warning}$< 0.01$,
and {\tt (conf\_code\&0x3) = 0}.
The blue (bottom) histogram includes sources with
{\tt (extent\_code\&0x10) != 0}, {\tt pileup\_warning}$< 0.01$,
and {\tt (conf\_code\&0xf) = 0}.
}
\end{figure}

\section{Source Extent}
\label{extent}
\newcommand{\falseextent}{f_\mathrm{fx}}

The raw extent of \CSC sources is parameterized by elliptical Gaussian sigma
values (\verb|mjr_axis_raw_b|, \verb|mnr_axis_raw_b|). For each \CSCA source, a
corresponding raw \psf elliptical Gaussian (\verb|psf_mjr_axis_raw_b|,
\verb|psf_mnr_axis_raw_b|) is derived by processing an SAOSAC simulation using
the same software.  For robust comparisons of raw source size (RSS), it is
convenient to define  the RSS as $a \equiv \left(\sigma_1^2 +
\sigma_2^2\right)^{1/2}/\sqrt{2}$, where $\sigma_i$ are the elliptical
Gaussian semi-axes.  \verb|extent_code| bits are set when the raw source size
exceeds the \psf size by a statistically significant amount within the
corresponding spectral band.

The method used to derive the elliptical Gaussian size parameters works well
for isolated sources embedded in relatively smooth background emission, but it
performs less reliably when the density of sources is high enough that source
regions overlap. The ellipse derived for a confused point source may not give
an accurate measure of the source size.  For each catalog source,
\verb|conf_code| indicates the nature of the overlap with nearby sources. For
example, \verb|(conf_code|\&\verb|0x3) = 0|, indicates that the source
detection region overlaps no other source detection region.
\verb|(conf_code|\&\verb|0xf) = 0|, indicates that the source detection region
overlaps no other region and the background region overlaps no other source
detection region.

Complicated image morphologies that arise from photon pileup in bright sources
may also confuse automated source extent measurements. The associated
\verb|pileup_warning| value may be used to gauge the importance of photon
pileup for a given source.

We define the {\it false extent} fraction, $\falseextent$, as the fraction of
detected point sources that are erroneously identified as extended because of
source confusion, or photon pileup, or any other reason such as a flaw in the
method used. We used the \marx point source simulations described in \S
\ref{pointsourcesims} to estimate $\falseextent$ as a function of off-axis
angle. Because the \marxh simulated sources are known to be point sources, any
non-zero \verb|extent_code| bit is, by definition, erroneous.
Fig.~\ref{extentfig} shows the \bband band false extent fraction as a
function of off-axis angle for powerlaw and blackbody sources.  The
black curve shows the false extent fraction based solely on the
\verb|extent_code| determined from the measured raw sizes of source
and \psf and the associated uncertainties. The red and blue curves in
Fig.~\ref{extentfig} show that, by modifying the source extent
criterion to exclude confused and piled sources, one can greatly
reduce the false extent fraction.  Source confusion is the most common
source of error because bright piled-up sources are relatively rare. 

\begin{figure}
\begin{center}
\includegraphics[angle=90.0,width=0.4\textwidth]{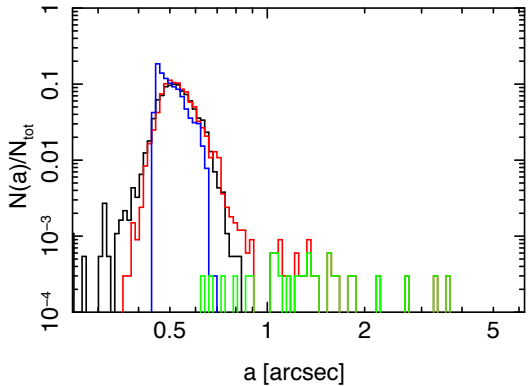}
\end{center}
\caption{\label{srcsizefig1}
Size distribution of power-law sources detected with $\theta <2.5^{\prime}$.
The histograms include only sources that have {\tt src\_cnts\_aper\_b > 25},
{\tt pileup\_warning < 0.01}, and {\tt (conf\_code\&0xf) = 0}.
The black curve shows 1,850 \marxh simulated point sources. The blue curve
shows 3,339 SAOSAC-simulated point sources.  The red curve shows 3,339 \CSCA
catalog sources; 33 of the selected \CSCA sources have $a>0.85^{{\prime}{\prime}}$.
The green curve shows \CSCA sources meeting the above criteria that
also have {\tt (extent\_code\&0x10) != 0}.
}
\end{figure}

Because the \marx and SAOSAC simulators have been tuned to closely approximate
the \chandra \psf, we expect close agreement between the point-source size
distribution derived from \marx and SAOSAC point-source simulations and the
size distribution derived from \CSCA point sources.  Furthermore, any extended
sources appearing in the \CSCA should appear as a tail extending above the
point-source size distribution.  Such extended sources should also be flagged
with one or more non-zero \verb|extent_code| bits.

Fig.~\ref{srcsizefig1} shows the distribution of RSS, $a$, among
\CSCA sources and \marxh\  and SAOSAC-simulated point-sources with off-axis angle
$\theta \le 2.5^{\prime}$. The \marx point-source distribution is broader than
the SAOSAC point-source distribtion because the \marx simulations sample much
fainter sources. In contrast, the SAOSAC sources are uniformly bright because
they were created primarily to provide an accurate measure of the \psf size.
The close agreement between the simulated point-source size distributions and
the observed \CSCA point-source size distribution confirms the accuracy of the
\marx and SAOSAC simulations.  A population of apparently extended \CSCA sources
is visible as tail extending to $a \approx 4^{{\prime}{\prime}}$.

A number of \bband band \CSCA sources with $\theta \lesssim 2.5^{\prime}$ are marked
as extended even though their raw source extent falls within the point source
size distribution.  For many of these sources, the \verb|extent_code| bit was
set erroneously because, for bright sources with $\theta \lesssim
3.5^{\prime}$, the uncertainty on the source size was underestimated, sometimes
falling below $0.1^{{\prime}{\prime}}$. As a result, some point sources were flagged as
extended even though the raw source size estimate exceeded the \psf size
estimate by $\lesssim 0.1^{{\prime}{\prime}}$. Imposing a minimum source size uncertainty
of $0.1^{{\prime}{\prime}}$, 379 \CSCA sources (81\% of which have $\theta < 2^{\prime}$ and
98\% of which have $\theta < 3.5^{\prime}$) would be reclassified from {\it
extended} to {\it point-source}. For $\theta \lesssim 4^{\prime}$, this change
in source size uncertainty eliminates most of the overlap between the size
distribution of point-sources and the size distribution of sources flagged as
extended.  We note that many of the affected sources also have
\verb|(conf_code|\&\verb|0xf) != 0| or \verb|pileup_warning|$>0.01$, making the
\verb|extent_code| value somewhat questionable for the reasons
discussed above. 

\begin{figure}
\begin{center}
\includegraphics[angle=90.0,width=0.4\textwidth]{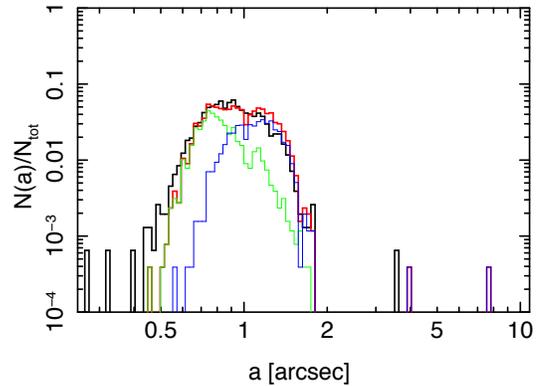}
\end{center}
\caption{\label{srcsizefig3}
Size distribution of power-law sources detected with $3.5^{\prime} < \theta
<4.5^{\prime}$.  The histograms include only sources that have {\tt
src\_cnts\_aper\_b > 25}, {\tt pileup\_warning < 0.01},
and {\tt (conf\_code\&0xf) = 0}. The black curve shows 1,543 \marxh simulated point
sources. The red curve shows 2,565 \CSCA catalog sources. The blue curve shows
\CSCA sources falling on \aciss.  The green curve shows \CSCA sources falling on
\acisi.
}
\end{figure}

At off-axis angles $\theta \gtrsim 4^{\prime}$, the \CSCA source extent
distribution appears consistent with that of the \marxh simulated point sources
(see Fig.~\ref{srcsizefig3}), suggesting that few genuinely extended sources
appear in the \CSCA catalog with $\theta > 4^{\prime}$. Additional
work is in progress to understand this effect.

For off-axis angles $3^{\prime} \le \theta \lesssim 10^{\prime}$, the point-source size
distribution is somewhat bimodal, consisting of a blend of two broad peaks
corresponding to sources detected on \acisi and on \aciss, respectively (see
Fig.~\ref{srcsizefig3} and Fig.~18 of \citealt{evans09}). The median imaging
\psf on \acisi is somewhat smaller than the median imaging \psf on \aciss
because the \acisi CCDs are positioned along the imaging focal surface, while
the \aciss CCDs are positioned along the Rowland torus of HETG.

\section{Variability}
\label{variability}

\begin{figure}
\begin{center}
\includegraphics[width=0.4\textwidth,angle=90]{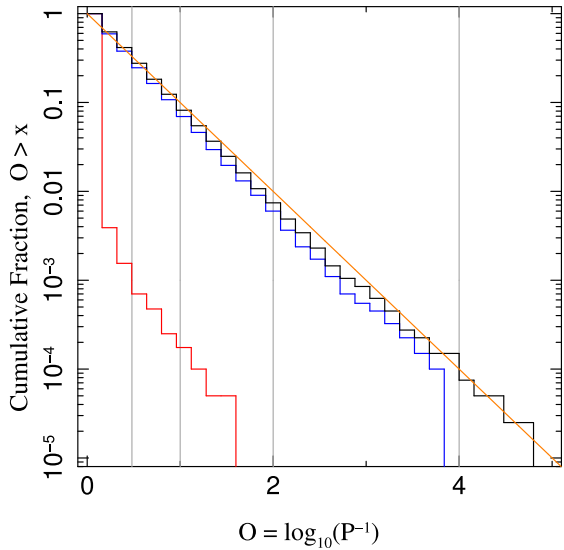}
\end{center}
\caption{\label{fig:white_noise_hists} Cumulative fraction of
  simulated white noise lightcurves (durations of 160\,ksec and mean
  rates of 0.032\,cps) detected with $O \equiv \log_{10} ( P^{-1} )$
  greater than the $x$-axis value.  $P$ is the probability that the
  lightcurve is consistent with a constant lightcurve.  Black line
  (top) is for the Kolmogorov-Smirnov test, blue line (middle) is for
  the Kuiper test, and red line is for the Gregory-Loredo test
  (bottom).  The straight orange line is $10^{-x}$.  Vertical grey
  lines correspond to the minimum $O$-values for which the \CSCA variability
  index (based upon the results of the Gregory-Loredo test) would be
  set to 5, 6, 7, or 8 (left to right).
}
\end{figure}

\begin{figure*}
\begin{center}
\includegraphics[width=\textwidth]{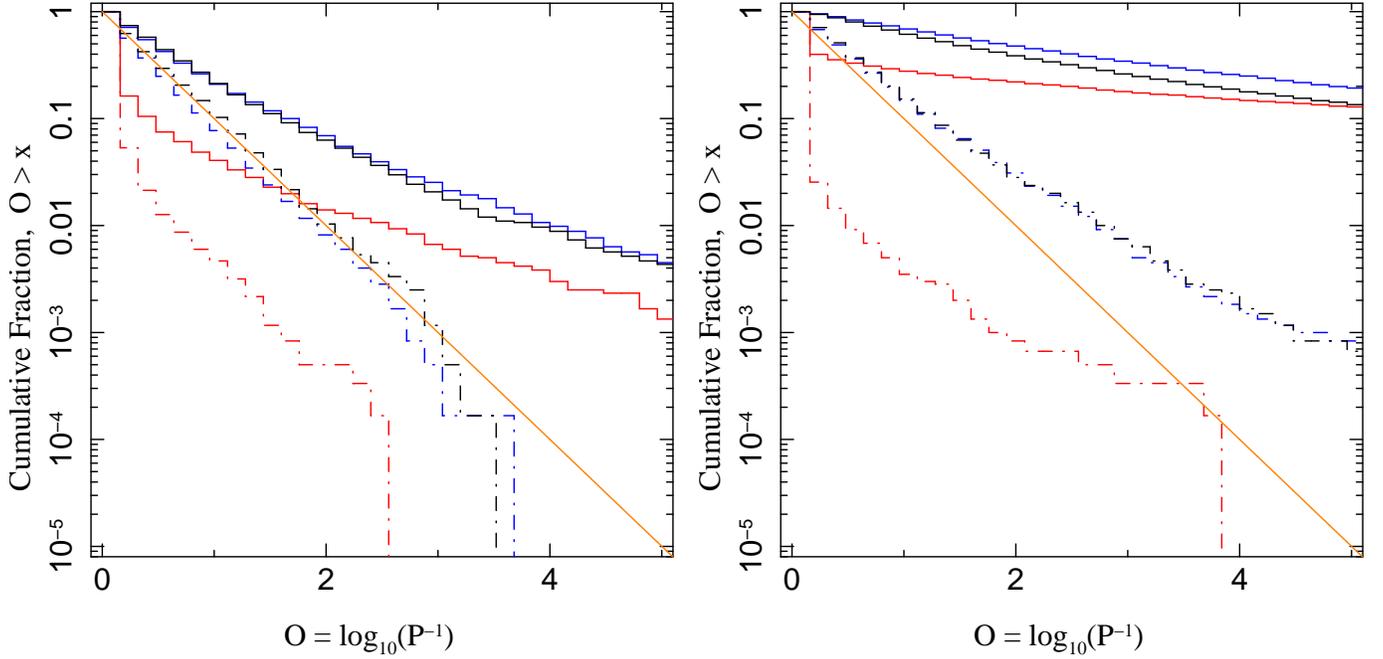}
\end{center}
\caption{\label{fig:var_hists} Cumulative fraction of simulated red
  noise lightcurves (durations of 50\,ksec) detected with $O \equiv
  \log_{10} ( P^{-1} )$ greater than the $x$-axis value.  $P$ is the
  probability that the lightcurve is consistent with a constant
  lightcurve.  Lightcurves used in the left figure have a mean rate of
  0.0032\,cps, while those used for the right have a mean rate of
  0.032\,cps.  For each, solid lines are for lightcurves with 30\%
  fractional rms, and dash-dot lines are for 7.5\% fractional rms.
  (Orange lines are $10^{-x}$.)  Black lines correspond to the
  Kolmogorov-Smirnov test, blue lines to the Kuiper test, and red
  lines to the Gregory-Loredo test.  }
\end{figure*}

\begin{figure*}
\begin{center}
\includegraphics[width=\textwidth]{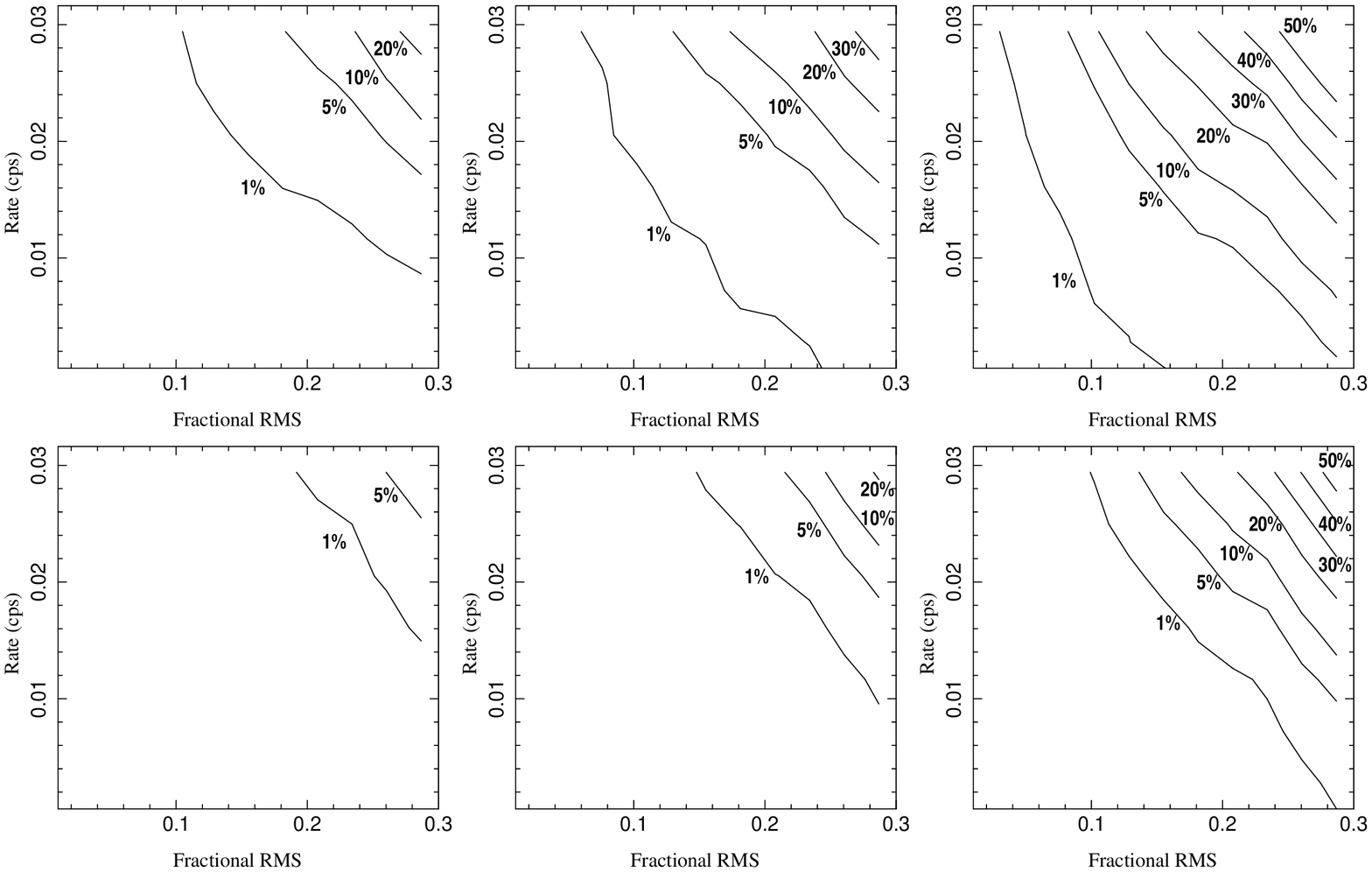}
\end{center}
\caption{\label{fig:var_contours} Contours for the fraction of
  simulated red noise lightcurves (as a function of simulated
  fractional rms and mean count rate) detected as variable with $O
  \equiv \log_{10} ( P^{-1} ) > 2$ (i.e., significantly variable at
  $>99\%$ confidence).  The top row corresponds to the results of the
Kuiper test, whereas the bottom row corresponds to the Gregory-Loredo
test.  From left to right, the durations of the lightcurves were
20\,ksec, 50\,ksec, and 160\,ksec.}
\end{figure*}

As described in \citet{evans09}, the \CSC utilizes three variability
tests: Kolmogorov-Smirnov, Kuiper, and Gregory-Loredo.  Results from
these tests are stored as a probability, $p$, that the lightcurve in a
given band for the indicated variability test is \emph{not} consistent
with being constant (i.e., pure counting noise, modulo source
visibility as described by the good time intervals and the
time-dependent fraction of the source region that falls on an active
portion of the detector).  For purposes of characterization, a more
useful probability is $P\equiv1-p$, which can be taken as the
probability that a constant lightcurve would have falsely indicated
the detected level of variability.  It is further convenient to take
the negative $\log_{10}$ of this quantity, i.e., define $O \equiv
\log_{10} (P^{-1})$.
This can be thought of being similar to the log of the odds ratio
that a variable lightcurve is a better description than a constant
one.  (Although the odds ratio is properly a Bayesian concept, and
hence applicable only to the Gregory-Loredo test, we define the
quantity $O$ for the Kolmogorov-Smirnov and Kuiper tests
via their frequentist probabilities $p$ as above so that we can more
easily compare results from the three tests.)
For much of the characterization work that follows, results are
presented in terms of this quantity $O$.  Note that even for a
``good'' variability test, a fraction, $f_P$, of lightcurves with a
constant mean rate should yield probabilities $P \le f_P$, or
equivalently, $O \ge \log_{10} ( f_P^{-1})$.

We first assess this expected property of the variability tests by
applying them to white noise simulations. For pure white noise
simulations, at least for the Kolmogorov-Smirnov and Kuiper tests, we
expect that the cumulative fraction of lightcurves with $O$ greater
than a given value, $x$, will follow $10^{-x}$.  Some deviations from
this relationship are expected for two reasons: First, we include a
simple model of pileup and assume that the pileup parameter
$\alpha=0.5$ (i.e., there is a $0.5^{(n-1)}$ probability that $n$
piled events will be detected as a single good event). This will tend
to suppress statistical fluctuations for the brighter lightcurves
\citep{davis:2001a}.  Second, we apply the lower count cutoff used
within the catalog by not including any lightcurves with fewer then
ten counts, and thus we are suppressing some range of inherent Poisson
variability (fluctuations to low counts from lightcurves with mean
counts just above the threshold, and fluctuations to high counts from
lightcurves with mean counts just below the threshold).

We simulate 40,000 lightcurves at each of seven different lengths
ranging from 1\,ksec to 160\,ksec and 8 different mean rates ranging
from 5.6e-4\,cps to 3.2e-2\,cps, for a total of 2,240,000
simulations. Histograms of the test results for the longest, brightest
lightcurves are presented in Fig.~\ref{fig:white_noise_hists},
although results for lightcurves of different lengths and mean rates
are comparable.  We find that for the most part, the
Kolmogorov-Smirnov and Kuiper tests yield the expected results for
the white noise lightcurve.  That is, the cumulative fraction of
simulated lightcurves with test results indicating variability
decreases with the significance level of the results.  Given that
Fig.~\ref{fig:white_noise_hists} represent 40,000 lightcurves, we find
as expected $\approx 400$ simulations that (falsely) indicate
variability at $\ge 99.9\%$ confidence.  Note, however, that the
Kolmogorov-Smirnov test and especially the Kuiper test each show a
small deficit of lightcurves with high variability significance levels.
We attribute this primarily to the effect of pileup on the generated
lightcurves.  These deficits are small, however, and we find that the
usual notion of significance levels applies well to these simulated
lightcurves when using the Kolmogorov-Smirnov and Kuiper tests.

The Gregory-Loredo test assigns even fewer white noise lightcurves to
formally significant statistic levels.  It is important to remember,
however, that the Gregory-Loredo test is answering a more restrictive
question.  Rather than asking the simple question, ``Is this
lightcurve consistent with a constant rate?'', it is instead asking,
``Is a uniformly binned lightcurve with multiple time bins a better
description than a single bin, constant rate lightcurve?''.  The
Gregory-Loredo test, for example, is not well-suited for discovering a
single, short flare interspersed in an otherwise steady lightcurve.
We find that the Gregory-Loredo test (which, again, is the basis for
the \CSCA tabulated variability indices) yields fewer false positives;
however, as we show below, it is also less sensitive to real
variability.  The Gregory-Loredo test is therefore a somewhat more
conservative measure of variability than either the
Kolmogorov-Smirnov or Kuiper tests.

We next turn to the question of sensitivity to real lightcurve
variability.  We simulated red noise lightcurves with the same lengths
and mean rates as for the white noise simulations; however, we further
considered a range of 12 fractional rms levels, ranging from 1\% to
30\%.  We performed 6,000 simulations for each combination of
lightcurve length, mean rate, and fractional rms, yielding a total of
4,032,000 simulations.  The cumulative fractions of simulated
lightcurves above a given significance threshold, for a subset of
simulated lightcurve lengths, rates, and fractional rms values, are
shown in Fig.~\ref{fig:var_hists}.  The variability tests performed on
these simulations -- for lightcurves that are sufficiently bright,
long, and/or variable -- clearly indicate variability above and beyond
the expectations of pure white noise.

To further quantify the meaning of ``sufficiently bright, long, and/or
variable'', in Fig.~\ref{fig:var_contours} we present what essentially
amount to ``variability detection probability'' contours as a function
of rms variability ($x$-axis) and mean lightcurve rate ($y$-axis) for
a variety of lightcurve lengths (individual panels).  For example,
here we choose as a ``significant'' detection threshold a variability
test value of $O\ge2$.  The calculated fraction of simulated
lightcurves that yield a variability significance above this value is
a measure of the sensitivity of the tests for these particular types
of lightcurves\footnote{The simulations create lightcurves with a
  \emph{mean} power spectral density profile that is $\propto f^{-1}$.
  Any single realization of this mean power spectrum profile can yield
  lightcurves with properties significantly different than the average
  profile and yield low variabilities for that specific instance.
  These inherent lightcurve variations are convolved with the
  sensitivities of the tests themselves to yield the final detection
  probabilities.}.  

In general we see that the Kuiper test is more sensitive than the
Gregory-Loredo test.  (The Kolmogorov-Smirnov test yields contours
similar to the Kuiper test.)  Not unexpectedly, the brighter, more
highly variable, and longer the lightcurve, the more sensitive the
tests.  Ideally, for a set of truly variable, well-observed
lightcurves and a chosen 
threshold for the value of 
$O = \log_{10} ( P^{-1} )$, we hope to find that the fraction, $f_{\rm
  lc}$, of lightcurves \emph{exceeding} this threshold to be $f_{\rm
  lc}\gg P = 10^{-O}$. For many realistic parameter regimes, however,
$<10\%$ of the simulated variable lightcurves are in fact detected as
being variable with $O >2$ (or equivalently, $P < 10^{-2}$).  This is
to be borne in mind when considering the catalog results which we
discuss below.

\begin{figure}
\begin{center}
\includegraphics[width=0.38\textwidth]{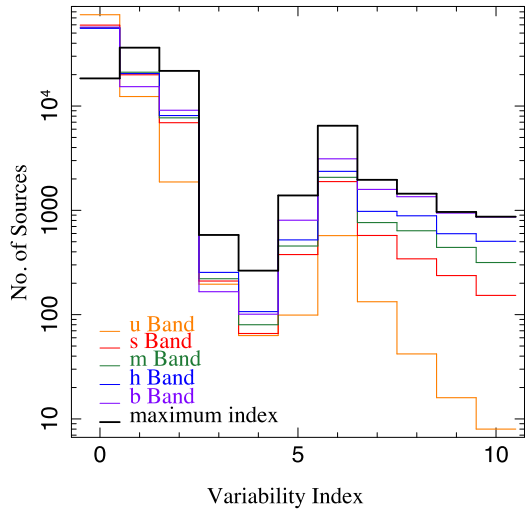}
\end{center}
\caption{\label{fig:cat_vindex_hists} Histograms of variability results from
  the \CSCA, for different energy bands, in terms of the variability
  index, excluding sources that 
  dither across a chip edge.  Orange, red, green, blue, and purple
  lines represent 
  the \uband, \sband, \mband, \hband, and \bband bands, respectively.
  The thick black line is the
  maximum variability index from the five bands.} 
\end{figure}

Results from applying the variability tests to \CSCA sources are shown
in Fig.~\ref{fig:cat_vindex_hists}.  Specifically, we show histograms
of the variability indices (derived from the Gregory-Loredo test;
\citealt{evans09}) in each of the \acis\ energy bands used in the
catalog.  Note that here we have excluded any source that dithers over
a chip edge\footnote{Corrections are made in the variability tests for
  the fraction of source area that is on a chip at any given moment.
  However, in release 1.0.1 of the \CSCA there is a programming error
  that affects any near-edge source that dithers onto a chip that was
  either turned off or was otherwise excluded from processing.
  Although such sources are a small minority of all near-edge sources,
  they are difficult to automatically identify in downloads of the
  source properties.  Therefore, unless otherwise noted, the results
  shown here exclude all sources that dither over a chip edge.}.  Of
the over 90,000 sources examined, nearly 13\% have a maximum
variability index $\ge 6$, and nearly 6\% have a maximum variability
index $\ge 7$.  These two variability indices represent, respectively,
$>90\%$ and $>99\%$ confidence that the source is better described by
a uniformly binned, variable lightcurve rather than by a white noise
lightcurve.  The \bband band shows the most highly
significant variability detections, most likely due to the increased
counting statistics available for this band.  Otherwise, detection
significance tends to decrease from the hardest \hband to the
softest \uband bands.  This is likely a combination of
detector properties (\acisi has very little sensitivity in the \uband
band and has reduced sensitivity in the \sband band compared to
\aciss), observational properties (e.g., the soft energy bands are
easily obscured by interstellar absorption), and intrinsic source
properties.

We next turn to the significances as determined by the variability
tests.  Examining the three different test results in the \sband,
\mband, \hband, and \bband energy bands individually, we find that
between 4--16\% of the lightcurves have $O \ge 2$, and 1--7\% of the
lightcurves have $O \ge 3$ (again, roughly corresponding to the
$>90\%$ and $>99\%$ confidence levels for significant variability,
respectively).  Within each energy band, the lower end of the
percentage range is for the Gregory-Loredo test (which again, is
asking a more stringent question than merely is the lightcurve
variable), while for all tests the soft band shows the smallest
percentage of significantly variable lightcurves, consistent with the
results of the catalog variability indices discussed above.

\begin{figure}
\begin{center}
\includegraphics[angle=90.0,width=0.4\textwidth]{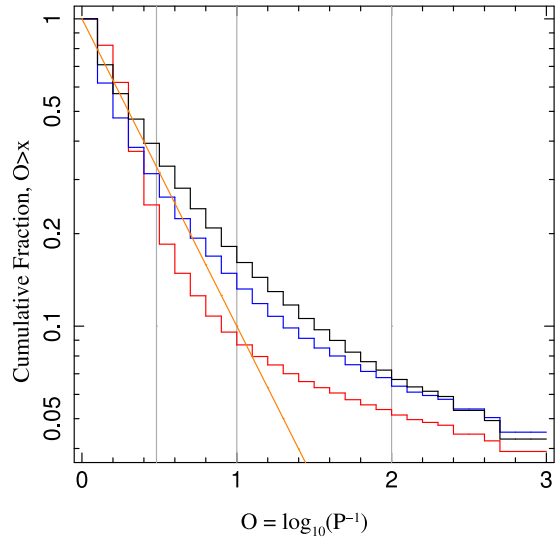}
\end{center}
\caption{\label{fig:cat_test_hists} Cumulative fraction of sources
  from the \CSCA (excluding sources that dither across a chip edge)
  that exceed a given variability significance (expressed as 
  $O = \log_{10} (P^{-1}) = - \log_{10} (1-p)$) 
  for the three variability tests performed in the
  (\bband) band.  Black histogram (top) is for the
  Kolmogorov-Smirnov test, blue histogram (middle) is for the Kuiper
  test, and red histogram (bottom) is for the Gregory-Loredo test.
  The orange straight line is the expectation for constant rate lightcurves,
  subject only to Poisson noise.  The grey vertical lines are the boundaries
  for the catalog variability indices (based upon the Gregory-Loredo
  test) 5, 6, and 7.}
\end{figure}

At the above respective significance levels, we expect that $<10\%$
and $<1\%$ of an ensemble of white noise lightcurves would show
comparably significant results.  Thus we see that up to approximately
5--6\% of the \CSCA sources (i.e., the excess above the $<1\%$ of
sources we expect to have $O>3$) are detected as being truly variable.
This is to be compared to, for example, the $<1\%$ of detections
(2,307/246,897) classified as variable in the 2XMM catalog
\citep{watson:2009a}.  In practice, for the \CSCA as a whole a
significant population of variable sources begins to appear at
variability indices $\ge 5$ and variability test 
   values
$O>1$.  This is illustrated in Fig.~\ref{fig:cat_test_hists}, which
shows the \CSCA variability test results for the \bband 
band.  Here we show the cumulative fraction of sources with
$O=\log_{10}(P^{-1})$ greater than a given value for each of the three
tests. This is to be compared to the white noise expectation that the
curves follow $10^{-x}$.  Excesses above this line represent
populations of significantly variable sources.  

In practice, one would identify variability in a subset of catalog
sources by choosing a 
   threshold value of $O$.  
Sources with $O$ exceeding this threshold would be identified as variable.  A low
threshold would yield a larger number of variable sources, but also a
larger fraction of ``false positives''.  On the other hand, choosing
very high test significances for the threshold will reduce the
number of flagged sources.  For the catalog as a whole, choosing
$O\ge2$ in either the Kolmogorov-Smirnov or Kuiper test, or nearly
equivalently\footnote{For the \bband band, sources with a variability
  index of 7 have a mean value of $O=2.4$ for the Kuiper test and
  $O=2.3$ for the Kolmogorov-Smirnov test.} a variability index $\ge
7$, maximizes the difference between the cumulative histograms for the
detected and white noise significances.  Approximately 6\% of the
sources will be flagged as variable, of which $\approx 17\%$ are
likely false positives (i.e., 1/6, as we expect 1\% of non-variable
sources to achieve such high test significance values).  Given that
the Kolmogorov-Smirnov and Kuiper tests have very well-characterized
properties for white noise lightcurves, those test results can be used
as a guide for assessing variability in any sub-populations taken from
the catalog.  Those tests specifically should allow users to choose
their own optimization of number of variable sources vs. fraction of
false positives.  The Gregory-Loredo test, having less
well-characterized white noise properties, is less well-suited for
that task; however, its chief advantage lies in the fact that it also
provides an estimate of the lightcurve which can be used in more
sophisticated analyses.

\begin{figure}
\begin{center}
\includegraphics[width=0.4\textwidth]{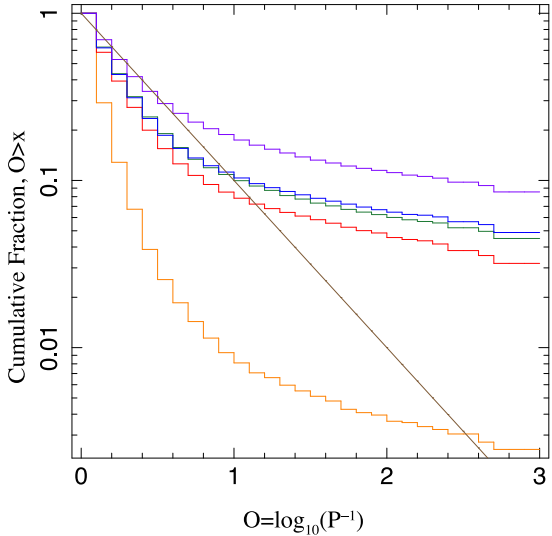}
\end{center}
\caption{\label{fig:master_var} Cumulative fraction of
  \CSCA v.1.0.1 master sources (comprised of two or more individual
  observations) detected with inter-observation variability above a
  given value of $O \equiv \log_{10} ( P^{-1} )$,
  greater than the $x$-axis value.  Bottom line (orange) is for
  the \uband band, followed by the \sband (red), \mband
  (green), \hband (blue), and \bband (purple) bands.  The straight
  line (brown) is $10^{-x}$, and again is the expectation for random
  noise fluctuations.  }
\end{figure}

We separately have analyzed the variability from catalog sources that
dither over a chip edge (by selecting the approximately 38,000 sources
with {\tt edge\_code} or {\tt multi\_chip\_code} $>0$).  To minimize
issues arising from the programming error related to sources dithering
onto an off or excluded chip, we did not include any sources from
ObsIDs with an excluded chip.  (A list of such ObsIds is maintained on
the \CSCA website.)  The results are very similar to the above.  17\%
of those sources have a maximum variability index $\ge 6$, and 7\%
have a maximum variability index $\ge 7$.  Examining the three
different test results in four energy bands separately, we find that
between 5--17\% of the lightcurves have $O \ge 2$, and 2--7\% of the
lightcurves have $O \ge 3$.  These percentages are slightly higher
than those quoted above, but not dramatically so.  There is likely
some additional false variability associated with dithering over the
edge, but this does not dominate the results from these sources if one
choose a test threshold of $O=2$.

Although we have not performed simulations to assess the sensitivity
of our procedures for detecting \emph{inter-}observation variability,
as for the intra-observation variability tests discussed above we have
conducted a preliminary assessment of the actual \CSCA v.1.0.1
results.  The \CSCA includes master source variability probabilities,
{\tt var\_inter\_prob\_*}, that represent the probability that the
multiple observations that comprise a given master source
are \emph{not} consistent with a constant flux in a given energy band.
To be consistent with our prior discussion of intra-observation
variability, we again convert these probabilities, $p$, into a
   quantity similar to a 
logarithmic odds ratio, $O \equiv \log_{10}(1-p)$.  We again consider
the cumulative fraction of sources above a given value, $O$.  Again,
even for non-varying sources, we expect by random noise for 10\% to
have $O\ge1$, 1\% to have $O\ge2$, etc.  Results for master sources
comprised of two or more individual observations are presented in
Fig.~\ref{fig:master_var}.

The selection of master sources comprised of two or more individual
observations (necessary for the definition of inter-observation
variability) limits the selection to 17,538 unique master source IDs.
It should be noted, however, that although there are multiple
observations for each of these master sources, each energy band is not
necessarily significantly detected in each individual observation.
This is reflected in Fig.~\ref{fig:master_var}, where the \uband band
is seen to be skewed towards extremely low inter-observation
variability significance.  This is unsurprising as the \uband band
flux might have been significantly detected in an \aciss observation,
yet remain undetected in an \acisi observation.  In general, we see
that the harder bands, and especially the \bband band, follow more
closely the expected $10^{-x}$ behavior for low values of $O$.

We see, however, that all energy bands show a tail of larger $O$
values that represent the significant detection of inter-observation
variability.  This tail is most pronounced for the \bband band, where
$\approx 20\%$ of sources have $O\ge1$, and 10\% of sources have
$O\ge2$.  Thus, approximately 10\% of all master sources comprised of
multiple observations show significant inter-observation variability.
Furthermore, choosing a selection critereon of {\tt
var\_inter\_prob}$\ge0.99$ identifies these sources, with only
$\lesssim 10\%$ of them being ``false positives''.

\section{Conclusions}
\label{conclusions}

The \CSC is intended to be a general resource for
astronomers at all wavelengths. It differs from the many excellent
\chandra catalogs derived as part of specific scientific programs in
that its data selection and analysis procedures are not optimized for
any particular scientific goal. With few exceptions, data from all
detectors active in each observation are included, and data from all
observations are processed in a uniform manner with a uniformly
defined set of source properties. The statistical characterization
studies we present here are based on extensive simulations and
comparisons to other catalogs, and illuminate the differences between
the \CSCA and other \chandra catalogs.

The first release of the \CSC includes a large fraction of all \chandra
\acis non-grating observations made in the first eight years of the
\chandra mission. Significant characterization results include the following.
\begin{itemize}
\item 
The catalog contains $\sim94,700$ distinct X-ray sources from
$\sim3,900$ separate \acis observations.
\item
The total sky coverage is
$\sim320$ deg.$^2$ for sources with a 0.5--7.0\,keV photon flux greater
than $\sim4\times10^{-5}$\,ph\,cm$^{-2}$\,s$^{-1}$.
\item
Detection
efficiencies are:
\begin{itemize}
\item 
typically near $\sim100$\% for sources within
$\sim5^{\prime}$ of the aimpoint and brighter than
$\sim1$--$3\times10^{-6}$\,ph\,cm$^{-2}$\,s$^{-1}$, depending on
exposure, and 
\item
$\sim50$\% or better for sources between $\sim5$--$10^{\prime}$ off axis.
\end{itemize}
\item
False
source detections appear to cluster near chip edges and the boundaries
between back- and front-illuminated chips, but the false source rate
is appreciable only for observations with exposures longer than
$\sim$50 ksec. 
\item
Fewer than $\sim1$\% of the sources in
the \CSCA are spurious.
\item
Average positional errors of \CSCA sources range
from $\sim0.2^{{\prime}{\prime}}$ on-axis to
$\sim4^{{\prime}{\prime}}$ at $\sim14^{\prime}$ off-axis.
\item
Systematic errors in photon fluxes inlcude an overestimate of a factor
of $\lesssim2$ for sources fainter than
$\sim3\times10^{-6}$\,ph\,cm$^{-2}$\,s$^{-1}$
and at off-axis angles $\theta \gtrsim10^{\prime}$, due at
least in part to an uncorrected Eddington bias when detection
efficiency is low. Additional systematic errors at higher fluxes
include both underestimates and overestimates of $\sim10-30\%$,
depending on energy band and source spectrum, and are attributed to
the use of a monochromatic effective area in computing
fluxes. Systematic errors in \uband band fluxes can be
$\gtrsim30\%$, for some source spectra.
\item Extended sources with sizes of a few arcseconds can be detected
within $\sim2.5^{\prime}$ of observation aimpoints; further work
is required to fully characterize \CSCA extent capabilities farther off-axis. 
\item  Choosing a 99\% confidence level for source variability (using
either the Kuiper or Kolmogorov-Smirnov tests), 6\% of all \CSCA
sources are found to be significantly variable.  Less than 1/6 of
these detections are expected to be false positives.
\item Approximately 10\% of all master sources comprised of multiple
observations show significant inter-observation variability.  Less
than 10\% of these detections are expected to be false positives.
\end{itemize}

Results presented here apply to the Release 1.0.1 of the \CSC. However,
they should also apply to \acis \CSCA sources in incremental Release 1.1,
which was made public in August, 2010. \acis analysis
procedures do not, in general, differ between releases 1.0.1 and
1.1. The latter does, however, include \hrci data, and although \hrci
analysis procedures are not different, its different detector
characteristics merit additional characterization. Additional \hrci
characterization results will be presented when available.

\acknowledgments
We wish to thank the anonymous referee for a very careful and detailed
review of the manuscript.

The characterization analysis made extensive use of the CIAO and ChIPS
software packages, developed by the Chandra X-ray Center, the 
SAOImage DS9 imager, developed by the Smithsonian Astrophysical
Observatory, the ISIS software package, developed by John
Houck of the 
MIT Kavli Institute for Astrophysics and Space Research,
the S-Lang software package, developed by John Davis of the 
MIT Kavli Institute for Astrophysics and Space Research,
and the TOPCAT (www.starlink.ac.uk/topcat) and STILTS
(www.starlink.ac.uk/stilts) software packages, developed by Mark
Taylor of Bristol University, UK.
  
The authors acknowledge the support of the National Aeronautics and 
Space Administration via grants NAS 8-03060 and SV3-73016. 

\appendix
\section{A Comparison of the MARX and SAOTrace PSFs}
\label{marx}

\begin{figure*}
\begin{center}
\includegraphics[width=6.2in]{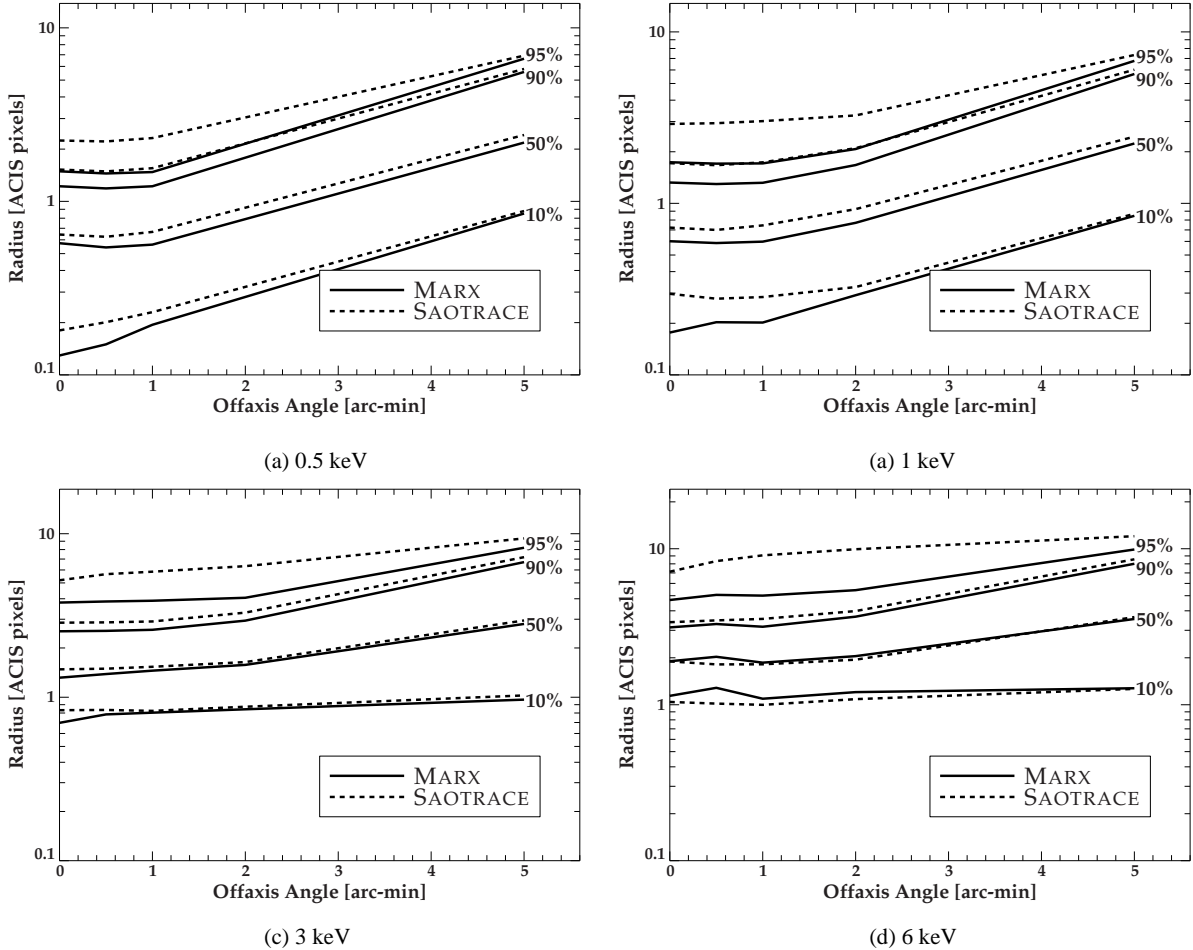}
\end{center}
\caption{\label{fig:acisipsf} The \acisi encircled energy radius at
the 10, 50, 90, and 95 percent levels as
a function of off-axis angle for various energies}
\end{figure*}

\begin{figure*}
\begin{center}
\includegraphics[width=6in]{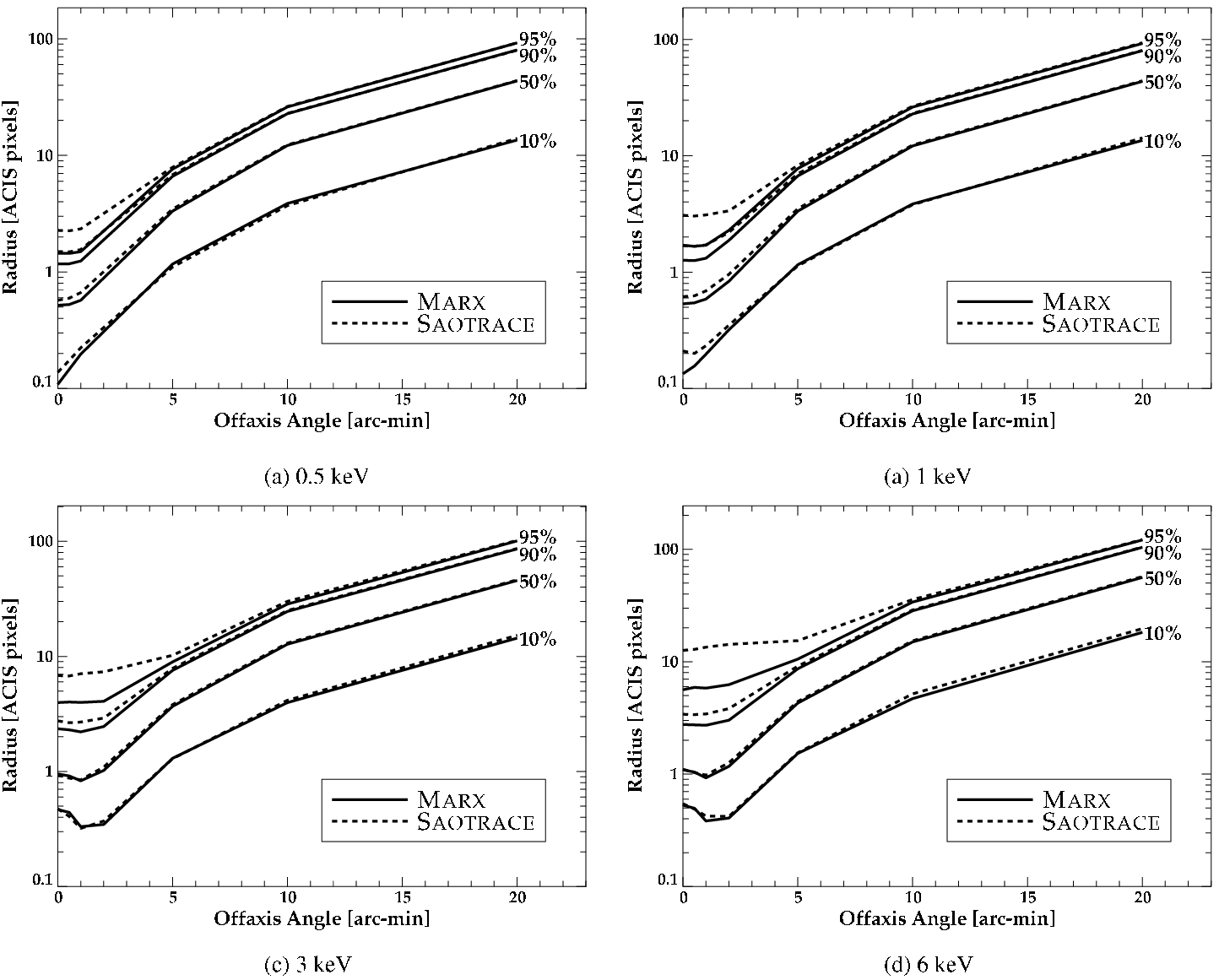}
\end{center}
\caption{\label{fig:acisspsf} The \aciss encircled energy radius at
the 10, 50, 90, and 95 percent levels as
a function of off-axis angle for various energies.}
\end{figure*}

\begin{figure*}
\begin{center}
\includegraphics[width=6in]{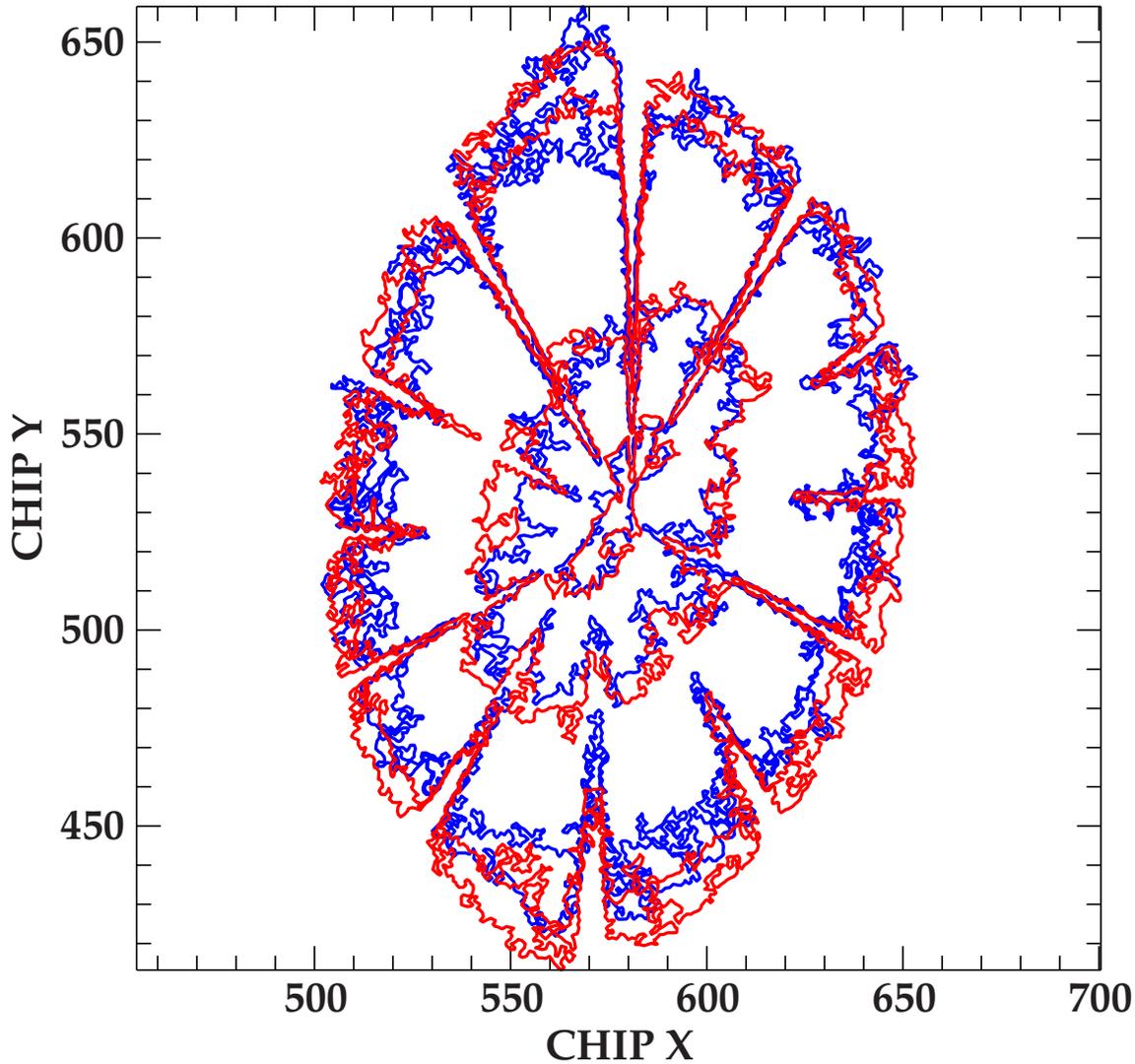}
\end{center}
\vspace{-0.4cm}
\caption{\label{fig:psfoffaxis} The 10, 50, 90, and 95 percent \marx
 (blue) and \saotrace (red, or halftone) encircled energy contours 
 for a 3 keV source 20'
off-axis on \aciss.}
\end{figure*}

\begin{figure}
\begin{center}
\includegraphics[width=5in]{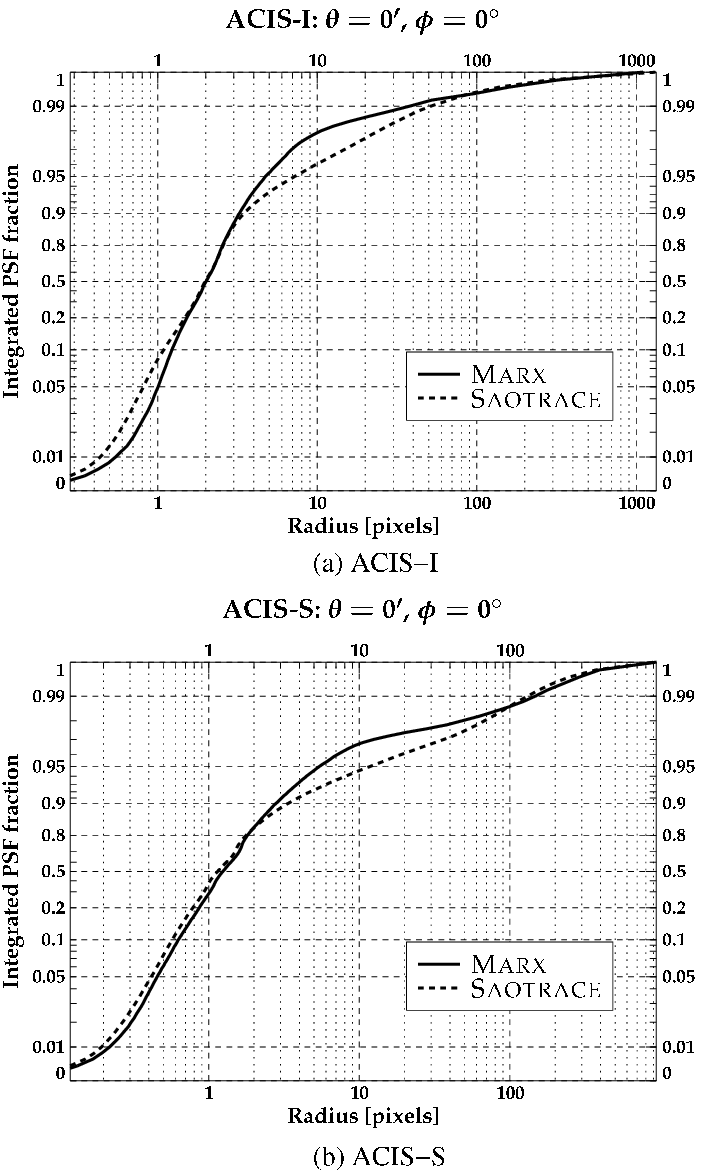}
\end{center}
\caption{\label{fig:acispsf} The encircled energy as a function of
radius for an on-axis source on the \acisi and \aciss arrays.  The
source spectrum was assumed to be an absorbed powerlaw with an
absorbing column of $10^{21}$ cm$^{-2}$ and index of $1.7$.}
\end{figure}

\marx\footnote{http://space.mit.edu/cxc/marx/} (Model of AXAF response
to X-rays) is a suite of  programs designed to simulate the on-orbit
response of the \chandra 
X-ray Observatory.  It was used for release 1.0.1 of the catalog to
characterize the detection efficiency, flux accuracy, and relative
astrometry via point sources simulated at various off-axis angles,
energies, and instrument configurations.  To better understand the
accuracy of the characterization, it is important to know how well the
\marx Point Spread Function (\psf) approximates that of the
telescope.  It is far beyond the scope of this work to make a direct
comparison of the simulated \marx \psf to that of actual flight data.
Instead, we compare the \marx \psf to that of the high-fidelity High
Resolution Mirror Assembly (\hrma) ray-trace program \saotrace, which
has undergone extensive pre-flight and post-flight calibration.  

The shape of the {\em observed} \psf is a complicated non-linear
function that depends upon a number of variables including off-axis
angle, energy, instrument configuration, detection mode, and source
flux.  Since incident photons first interact with the \chandra \hrma
before arriving at the detector, the observed \psf is a convolution of
a \hrma \psf and detector \psf.  The detector \psf consists of an
astigmatic component caused by deviations of the detector geometry
from that of the ideal focal surface, a component due to the use of
finite size detector pixels, and an intrinsic component that arises
from the interaction of the photon with the detector.  With the
exception of the latter, the former two components are purely
geometrical and are handled in a straight forward manner by the
\marx raytrace.  Positional uncertainties from the physical
interaction of the photon with the detector are handled statistically
by assuming an additional gaussian blur when \marx constructs event
coordinates. 

The \hrma \psf may be broken into two parts.  The first is a component
that dominates the core of the \psf and is a consequence of
misalignments and low spatial frequency deviations from the perfect
type-I Wolter geometry.  The second part gives rise to the scattering
wings of the \psf and is caused by high frequency surface errors or
microroughness.  In principle, given a detailed geometric model of the
mirror, the core of the \psf could be simulated via ray-tracing.
However, \marx lacks the detailed geometric details of the \hrma but
instead assumes perfect type-I Wolter geometry for each of the mirror
shells and takes into account misalignments between them.  \marx
models the low spatial frequency deviations from the ideal Wolter-I
geometry by rotating the surface normal at the intersection point of a
ray about a random direction perpendicular to the normal by a small
angle chosen from a gaussian distribution.  The scattering wings of
the \hrma \psf are treated statistically by \marx using a
parametrization developed by \citet{lvs1989} of the
\citet{beckmann1963} scattering model.

The encircled energies of the \marx and \saotrace \acisi \psfs as a
function of off-axis angle at various energies are shown in
Fig.~\ref{fig:acisipsf}; the corresponding \psfs for \aciss are
depicted in Fig.~\ref{fig:acisspsf}.  From these plots one can see
that beyond about $5^\prime$ off-axis, the \marx and \saotrace \psfs
agree quite well. This agreement can also be seen in
Fig.~\ref{fig:psfoffaxis},  which shows 2d encircled energy contours
for a $20^\prime$ off-axis source.  
Fig.~\ref{fig:acispsf} shows that on-axis, the encircled energies of
the \marx and \saotrace \psfs agree out to about 90 percent of the
integrated flux, but differ in the scattering wings. 

The fact that the \marx and \saotrace \psfs agree far off-axis, but
disagree near on-axis in the wings should not be surprising.  The
various statistical parameters that \marx uses to characterize the
\psf were tuned to match the High Efficiency Transmission Grating
Spectrometer's (\hetgs) on-axis Line Spread Function (\lsf) as
determined through \hetgs observations of Capella
\citep{canizares2005}.  Due to the lack of adequate counts in the
wings of the \lsf, only the parameters influencing the \psf core could
be determined with sufficient resolution.  The use of the \hetgs for
this purpose is a reflection of the fact that \marx started out as a
simulator for the \hetgs.  In contrast, the on-axis \saotrace \psf was
compared to \hrci observations of Ar-Lac \cite{diabTBD}, where the
residuals in the core of \psf were estimated to be less than 10
percent.  The wings of the \saotrace \psf were accessed using the
zeroth order \hetgs data from a 50\,ksec observation of Her X-1.  From
this observation, the uncertainties in the flux of the \saotrace wings
were estimated to be at least 30--50\% \citep[see the discussion
  of][]{xiang:2009a}.  

For near on-axis sources, the relative positional accuracy in the sky
tangent plane system between the \marx and \saotrace \psfs was
determined by comparing the tangent plane locations of the centroids
of their \psfs.  For such cases, we found \marx to be consistent with
\saotrace to subpixel accuracy.

Centroiding was less useful for far off-axis sources where the
distortions in the core of the \psf become quite noticeable.  In this
situation, the intersection of the shadows caused by the \hrma support
struts as seen in the sky tangent plane coordinate system was used to
determine the source position.  The astigmatic effects associated with
the different path lengths of rays from the \hrma to the detector
surface mean that the strut shadows may not have a common intersection
point in the sky and detector coordinate systems.  This is
particularly noticeable for the \aciss detector planes, which were
designed to approximate the Rowland surface of the \hetgs causing them
to be offset from the imaging focal surface. The accuracy of this
method was estimated to be less than 2 arc-seconds for sources 25
arc-minutes off-axis.

The previous technique was used to compare the \marx \psf to that of
the \chandra observation (OBSID 1068) of LMC X-1, observed 24.8
arc-minutes off-axis.  A Level 2 event file was created using \ciao 4.2
and loaded into SAOImage {\tt ds9} \cite{ds9} to view the (binned) source
events in the sky tangent plane system.  Using the intersection of the
strut shadows as described above, the source was estimated to have a
right ascension of 84.9115$\pm$0.0002 degrees and a declination of
-69.74335$\pm$0.00028 degrees.  These values were used to
to specify the source position for a \marx point source simulation of
OBSID 1068.  The resulting \marx event file and the \chandra
observation level 2 event file were loaded into SAOImage ds9 to
visually compare the observed and simulated \psfs by ``blinking'' one
against the other.  As expected, qualitative differences were seen in
the core of the \psf but the positions of the support strut shadows
were nearly on top of one another with a registration uncertainty
estimated to be less than two sky tangent plane pixels, which is
consistent with the uncertainties in the source position estimated using
the support struts.


\end{document}